\documentclass[showpacs,twocolumn,aps,prl,superscriptaddress,longbibliography, floatfix]{revtex4-2}
\usepackage{amsmath,amssymb,amsfonts,bm,graphicx,mathtools,braket}
\usepackage{graphicx}
\usepackage[caption=false]{subfig}
\usepackage{ amssymb }
\usepackage{placeins}
\usepackage{dcolumn}
\usepackage{bm}
\usepackage{times}{\Large }
\usepackage{xcolor}
\usepackage{mathrsfs}
\usepackage{hyperref}
\hypersetup{colorlinks=true,linkcolor = blue,
            urlcolor  = blue,
            citecolor = blue,
            anchorcolor = blue, linktocpage}

\newcommand{\equa}[1]{Eq.~\eqref{#1}} 
\newcommand{\fig}[1]{Fig.~\ref{#1}}

\newcommand{\violet}[1]{\textcolor{violet}{#1}}   
\renewcommand{\violet}[1]{}

\newcommand{\nonum}{\nonumber \\}

\newcommand{\apphref}[2]{Appendix~\hyperref[#1]{#2}}

\newcommand{\hidehide}[1]{}                 

\begin{document}

\title{Dynamical Tunneling Blockade in Layer-Pseudospin Ferromagnets}
	\author{Lingxian Kong}
	\affiliation{State Key Laboratory
		of Low-Dimensional Quantum Physics and Department of Physics, Tsinghua University, Beijing 100084, China}
	\affiliation{International Center for Quantum Materials, School of Physics, Peking University, Beijing 100871, China}
    \affiliation{Department of Physics, KTH Royal Institute of Technology, SE-106 91 Stockholm, Sweden}
    \author{Yuta Murakami}
    \affiliation{Institute for Materials Research, Tohoku University, Sendai 980-8577, Japan}
    \affiliation{RIKEN Center for Emergent Matter Science (CEMS), Wako 351-0198, Japan}

	\author{Ryuichi Shindou}
	\affiliation{International Center for Quantum Materials, School of Physics, Peking University, Beijing 100871, China}%
	
	\author{Zhiyuan Sun}%
	\email{Corresponding author. zysun@tsinghua.edu.cn}
	\affiliation{State Key Laboratory
		of Low-Dimensional Quantum Physics and Department of Physics, Tsinghua University, Beijing 100084, China}%
	\affiliation{Frontier Science Center for Quantum Information, Beijing 100084, China}
	
	\date{\today}
	\begin{abstract}

Double-layer systems with narrow or flat electronic bands may
exhibit spontaneous interlayer coherence, described by a layer-pseudospin ferromagnetic order parameter. 
In modern realizations of layer-pseudospin ferromagnets, the electrons carry additional valley degeneracies, endowing each layer with a valley-pseudospin degree of freedom. 
We study their nonequilibrium steady states driven by a static interlayer voltage bias in the presence of interlayer tunneling. While in the low-bias regime, the system is in a static state with parallel valley pseudospins between the two layers, we show that in the high-bias regime, 
the dynamical precession of the layer-pseudospin stabilizes the valley pseudospins in the antiparallel configuration, strongly suppressing the interlayer tunneling. 
This dynamical tunneling blockade leads to an absence of ac current and a sharp drop of the dc tunneling current to a tiny value, in stark contrast to conventional quantum Hall double layers. 
\end{abstract}

    \maketitle

  	\textit{Introduction.---}Double-layer systems with narrow or flat electronic bands, such as quantum Hall double layers, have  been  versatile platforms for realizing correlated electronic phases~\cite{Halperin1983, Zhang1994, West1994, Zhang1995, Yoshioka1996, Yang2001,
  		Hasebe2002, MacDonald2004, Eisenstein2014, Lee2025, rbvy-4rwd}. 
  	Among them,  a prominent example is the layer-pseudospin ferromagnetic state favored by the exchange interaction~\cite{Zhang1994,Zhang1995,Stern2000,Hasebe2002}, also interpreted as a condensate of interlayer excitons~\cite{MacDonald1996,Eisenstein2003,Eisenstein2014}.
  	It has been observed in the interlayer-tunneling conductance~\cite{West2000, West2001, Tutuc2018}, 
  	counterflow resistance~\cite{West2004, Huse2004, Dean2017, Dean2022, Li2026}, 
  	and drag effects~\cite{West2002, West2012, Kim2017}. 
  	Typical material platforms include double quantum wells in GaAs structures~\cite{West2000, West2001, West2019, West2004, Huse2004, West2002, West2012} and graphene double layers~\cite{Dean2017, Kim2017, Dean2022,Li2026, Kim2023} in the quantum Hall regime, and, more recently, double moiré-bilayer systems without a magnetic field~\cite{Vishwanath2021, MacDonald2022, Mak2023}.

\begin{figure}[!t]
    \centering
	\includegraphics[width=0.95\linewidth]{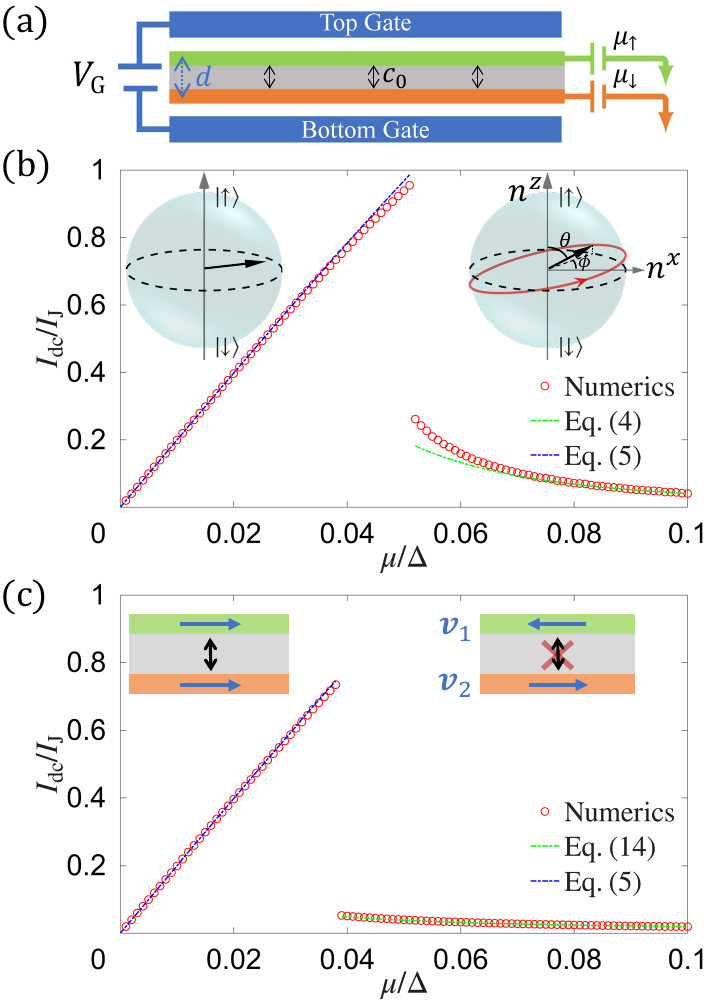} 
    \caption{
(a) Double-layer device with the two layers connected to leads at chemical potentials $\mu_{\uparrow}$ and $\mu_{\downarrow}$ and separated by a tunneling barrier of thickness $d$. 
The gate voltage $V_{\rm G}$ applies an out-of-plane electric field.
(b) 
The dc interlayer current density $I_{\rm dc}$ (in units of  $I_{\rm J}$) as the bias voltage $\mu$ is swept upward  for systems without valley degeneracy.
Red circles are numerical results while green and blue curves are analytical results. 
The low-bias state  has  a static layer pseudospin $\bm n$ (black arrow in the left inset, the Bloch sphere of $\bm n$ with the north/south pole corresponding to the macroscopic state $\ket{\uparrow}$/$\ket{\downarrow}$ and the dashed circle being the equator at $n^z=0$),
 while the high-bias state has a precessing $\bm n$ along the red orbit on the Bloch sphere (right inset) accompanied by both ac and dc currents. 
The parameters are  $c_0=0.005\Delta$, $\beta=0.1$, and $\Gamma=0.2$.
For $\Delta D_Q(\Delta)\sim1\,\mathrm{meV}$ and $a\sim10\,\mathrm{nm}$, one has $I_{\rm J}\sim10^4\,\mathrm{nA}/\mu\mathrm{m}^2$.
(c) Same as (b) but with a valley degeneracy in each layer. 
The low-bias branch has parallel valley pseudospins $\bm v_1$ and $\bm v_2$ (blue arrows in the left inset) allowing interlayer tunneling (black double-headed arrow). 
At high bias, they become nearly antiparallel, producing a dynamical tunneling blockade (red cross in the right inset) with zero ac current and strongly suppressed dc current.
}
    \label{fig1}
    \end{figure}

A typical device consists of two electronic layers separated by an insulating barrier of thickness $d$, as shown in \fig{fig1}(a). 
The two layers are separately connected to metallic leads held at chemical potentials $\mu_{\uparrow}$ and $\mu_{\downarrow}$~\cite{Mak2019, Shan2021, Wang2023b, Mak2025, Wang2025, Shan2026}, respectively. 
These leads act as reservoirs that exchange electrons with each layer. 
In addition, a gate voltage $V_{\rm G}$ between the top and bottom gates generates an electric field $E_{\rm G}$ along the $z$ direction, which shifts the relative single-particle energy between the two layers~\cite{Kis2019, Mak2022, Wang2022, Schaibley2022, Gunnink2026}. 
This setup allows the interlayer voltage bias and  energy offset to be controlled independently.
It has been understood that together with a weak interlayer tunneling, the interlayer electronic order enables coherent interlayer tunneling similar to the Josephson effect~\cite{Iwazaki1993, Zee1993, West2000, West2001, Radzihovsky2001, Ma2001, Wilczek2001, MacDonald2001, MacDonald2005, vonKlitzing2008, West2013, Millis2021a}.
An interlayer voltage bias $\mu = \mu_\uparrow-\mu_\downarrow$ results in an ac Josephson effect where the phase of the order parameter constantly evolves in time~\cite{Iwazaki1993, Zee1993, Wilczek2001, MacDonald2005}.
This phase-precession picture follows when the layer pseudospin is the only active order parameter.
However, in  modern platforms such as graphene-based quantum Hall double layers~\cite{Dean2017, Kim2017, Dean2022,Li2026} and double moiré bilayers~\cite{MacDonald2022, Mak2023, Wang2024},
the order parameter also possesses internal valley degrees of freedom in each layer~\cite{Jin2026, Wang2026}. 
How these additional valley degrees of freedom affect the interlayer Josephson dynamics remains a mystery.

In this Letter, we study the nonequilibrium steady states of layer-pseudospin ferromagnets driven by the interlayer voltage bias $\mu$, emphasizing  the effects of valley degrees of freedom. 
We describe the intertwined dynamics of the layer pseudospin and the valley pseudospins using generalized Landau-Lifshitz-Gilbert (LLG) equations derived from microscopic theory. 
The valleyless case, where the layer pseudospin is the only active degree of freedom, serves as a baseline. At low bias, the system is in a static state with a dc tunneling current, where the layer-pseudospin stays along a fixed direction set by the bias [left inset of Fig.~\ref{fig1}(b)].
Above a critical bias $\mu_{\rm c}$,  the layer pseudospin is driven by the bias to precess continuously following an orbit on the Bloch sphere [right inset of Fig.~\ref{fig1}(b)], leading to the ac Josephson effect accompanied by  an ac interlayer tunneling current and a weak dc current. 

If the order parameter additionally carries a valley-pseudospin from each layer, 
the low-bias state still has a static layer pseudospin with the two valley-pseudospins locked to be parallel, see the left inset of Fig.~\ref{fig1}(c). Interestingly, in the high-bias state, the dynamical orbit of the  layer-pseudospin becomes unstable: it generates an effective force that drives the valley-pseudospins toward the antiparallel configuration and stabilizes them there, see the right inset of Fig.~\ref{fig1}(c). Because interlayer tunneling relies on the overlap between the valley spinors, it is strongly suppressed in this state, which we refer to as a \textit{dynamical tunneling blockade}. Consequently, as the system enters this blockade regime after the bias voltage is tuned beyond the critical value, the dc current drops sharply while  the ac current stays zero, eliminating the ac Josephson effect.

\textit{Systems without valley degeneracy.---} 
We start with the simplest setting, in which each layer hosts a single flat or narrow band without valley degeneracy. 
This can be realized, for example, in GaAs double quantum wells in the quantum Hall regime~\cite{West2000, West2001}. 
The layer degree of freedom of electrons can be regarded as a pseudospin-$1/2$.
Interactions may induce ferromagnetism in the layer-pseudospin, as the polarized state increases the inter-electron distance in the many-body wave function so that the interaction energy is lowered~\cite{Zhang1995}.
The part of the microscopic model (see Appendix~\hyperref[appe:microscopic]{A}) that is $SU(2)$ symmetric in layer-pseudospin space yields  the mean-field Hamiltonian 
$H^{\rm{MF}}_{ \bm{k}}=\epsilon_{\bm k} \sigma_0-\Delta (\bm{n}\cdot\bm{\sigma})$
at momentum $\bm k$, 
where $\sigma_0$ is the identity matrix in the layer space, $\epsilon_{\bm k} $ is the band dispersion,
$\Delta$ is the order-parameter amplitude that leads to spontaneous splitting of the two bands,
the unit vector $\bm{n}=(\sin\theta\cos\phi,\sin\theta\sin\phi,\cos\theta)$ specifies the macroscopic layer-pseudospin,
and $\bm{\sigma}=(\sigma_x,\sigma_y,\sigma_z)$ are Pauli matrices in layer-pseudospin space,  see Fig.~\ref{fig1}(b).

In the absence of an interlayer bias ($\mu=0$), the low-energy dynamics of the layer-pseudospin is governed by the LLG equation:
\begin{align}
\frac{d{\bm n}}{dt}
=
-\bm n\times \Big[ 
\Gamma\frac{d{\bm n}}{dt}
+
2 \frac{\partial E(\bm n)}{\partial\bm n} \Big], 
\label{LLG_equili}
\end{align}
where $\Gamma$ is the dimensionless damping coefficient arising from 
exchanging particles with the leads~\cite{supplementary}. 
In the dissipationless limit, the pseudospin precesses along a constant-energy contour on the unit sphere.
With a nonzero damping $\Gamma$, it precesses while relaxing to the lowest-energy direction. The orientation-dependent energy $E(\bm n)=E_{\rm a}+E_{\rm p}+E_{\rm t}$ contains an easy-plane term $2E_{\rm a}=\Delta \beta (n^z)^2$ originating from the difference between intra- and interlayer Coulomb interactions,
a layer-polarization term $2E_{\rm p}=  n^z E_{\rm G} d $ induced by the vertical gate field, 
and an interlayer Josephson energy 
$E_{\rm t}=-c_0\sin\theta\cos\phi$ arising from interlayer tunneling of strength $c_0$ [black arrows in Fig.~\ref{fig1}(a)]. 
The Josephson term explicitly breaks the $U(1)$ symmetry of varying the relative phase $\phi$ between the layers and pins the equilibrium orientation to $\phi=0$. 
With this normalization,  the energy per electron is $E_{\rm per}=E(\bm n)D_2(\Delta)$, where the dimensionless  factor is 
$D_Q(\Delta)\equiv \int \rho(\varepsilon)\bigl[n_{\rm F}(\varepsilon-(Q-1)\Delta)-n_{\rm F}(\varepsilon+\Delta)\bigr]\,d\varepsilon/(2\pi)$, 
with $Q=2$ for the present system  and $Q=4$ for the system with valley degeneracy to be discussed below. 
Here $\rho$ is the electronic density of states per electron and  $n_{\rm F}$ is  the fermionic occupation function~\cite{supplementary}.
Equation~\eqref{LLG_equili} neglects fluctuations of the order-parameter amplitude $\Delta$, which give only subleading corrections as the low-energy  dynamics is dominated by the pseudospin orientation~\cite{supplementary}. 
Note that the LLG dynamics applies universally as long as the original electronic bands have the same dispersion, being flat or not.

A nonzero interlayer bias  $\mu$ turns the two leads into a nonequilibrium reservoir.
It drives the pseudospin dynamics through the following substitution in the damping term in \equa{LLG_equili}: 
\begin{align}
\Gamma \frac{d\bm{n}}{dt} \to 
\Gamma \left(\frac{d{\bm n}}{dt}-\mu\hat{\bm z}\times\bm n \right),
\label{Substitute}
\end{align}
see Ref.~\cite{supplementary} for a derivation using the Keldysh path integral. Physically, this nonequilibrium damping tends to relax the layer-pseudospin toward a precessing state satisfying $d\bm n/dt=\mu\hat{\bm z}\times\bm n$.
This is the essence of the ac Josephson effect: the interlayer phase winds at the rate $d\phi/dt=\mu$, in analogy to dynamical  condensates of interlayer excitons~\cite{Millis2024a, Millis2024b}.
Experimentally, the interlayer-coherent state at total filling
$\nu_{\rm T}=1$ 
is commonly studied in the density-balanced
configuration, corresponding to half filling in each layer,
$\nu_{\uparrow}=\nu_{\downarrow}=1/2$
~\cite{West2000,West2001,West2019,Wegscheider2009,West2013}. 
Accordingly, we tune the gate voltage to $E_{\rm G}d=\mu$ 
so that the system is layer-balanced when the interlayer tunneling is absent.
In this case, the
biased LLG equation admits the steady-state solution $\phi(t)=\phi_0+\mu t$ and $n^z=0$ on the equators of the Bloch spheres in \fig{fig1}(b), where $\phi_0$ is an arbitrary initial phase. 

Via the Josephson energy $E_{\rm t}$, nonzero interlayer tunneling distorts the dynamical orbit into the 
red curve $(\phi=\phi_0+\mu t+\delta\phi(t),\, n^z(t))$ in the right inset of \fig{fig1}(b).
The dynamical orbit is accompanied by the interlayer tunneling current density:
    \begin{align}
		I(t) = I_{\rm J} \sin\theta \sin \phi ,
		\label{J_2}
	\end{align}
where $I_{\rm J}\equiv c_0 D_Q e/(a^2\hbar)$ is the ``Josephson critical current density'' and  $1/a^2$ is the electronic density (for quantum Hall systems, $a$ is the magnetic length). 
At leading order in $c_0$, \equa{J_2} gives an ac interlayer current at frequency $\mu$. 
On top of it, the oscillating Josephson term induces a dynamical phase
shift $\delta\phi(t)=\mathcal{O}(c_0)$~\cite{Millis2024a,Diehl2025},
which generates a dc tunneling current at second order in $c_0$, see Appendix~\hyperref[appe:varphi]{F}.
Time-averaging \equa{J_2} with $\phi(t)=\phi_0+\mu t+\delta\phi(t)$ yields
\begin{align}
I_{\rm dc} \approx
I_{\rm J}
\frac{c_0 \Gamma}{\mu}
\frac{4\beta^2\Delta^2+(1+\Gamma^2)\mu^2}
{4\Gamma^2\beta^2\Delta^2+(1+\Gamma^2)^2\mu^2}.
\label{j_dc_2}
\end{align}
Note that the fluctuation $n^z=\mathcal{O}(c_0)$ does not contribute to the dc current at $\mathcal{O}(c_0^2)$.

If the bias is smaller than a critical value $\mu_{\rm c}=2c_0/\Gamma$, the modified LLG equation admits a static solution,  $d\bm{n}/dt=0$, which to leading order in $c_0$ reads~\cite{supplementary}
\begin{align}
    \phi \approx \arcsin(\mu/\mu_{\rm c}), \ \ \ 
     n^z  \approx  -\mu/(2\beta\Delta).
    \label{sol_static}
\end{align}
In this state, the ac interlayer current vanishes, while the dc current density is given by \equa{J_2} with \equa{sol_static}. 
Note that for $|\mu|<\mu_{\rm c}$,
both static and dynamical states may exist at the mean-field level, depending on  initial conditions.
As the bias voltage is swept upward, the resulting $I_{\rm dc}$-$\mu$ characteristic is shown in \fig{fig1}(b).
Within the static state, the dc current $I_{\rm dc} \approx I_{\rm J} \mu/\mu_{\rm c}$ increases approximately linearly with $\mu$. 
Above $\mu_{\rm c}$, the system undergoes a sharp transition into the dynamical phase, after which the dc current decreases with $\mu$ according to \equa{j_dc_2}, giving rise to negative differential conductance. 
Within each phase, the analytical formulae agree well with the numerically exact results in \fig{fig1}(b).   

Qualitatively similar current-voltage characteristics have been reported in experiments~\cite{West2000, West2001, Eisenstein2003, Spielman2004,  vonKlitzing2008, Wegscheider2009,West2013}  and theoretical studies~\cite{Rosenow2011,MacDonald2014}.
The static and dynamical phases are analogous to those of a current-biased Josephson junction described by the resistively and capacitively shunted junction (RCSJ) model~\cite{Tinkham2004a}. This analogy and the associated steady-state regimes have been explored in earlier phenomenological descriptions of interlayer tunneling in quantum Hall double layers~\cite{Zee1993, Spielman2004, Revishvili2006, MacDonald2014}. 
The derivation of the generalized LLG equations here using the Keldysh path-integral approach~\cite{supplementary} provides a microscopic and systematic framework for studying these nonequilibrium devices.
Compared with the reduced pendulum
description that yields the $I_{\mathrm{dc}}\propto\mu^{-3}$ scaling at high bias~\cite{Revishvili2006}, 
\equa{j_dc_2} from the full LLG approach
predicts, for weak damping,
that the current decay slows from $\mu^{-3}$ to $\mu^{-1}$
as the bias increases through $\mu\sim2\beta\Delta$.

\textit{Systems with valley degeneracy.---}
We now move on to the case where electrons carry additional valley degeneracies.
This applies to the modern platforms for layer-pseudospin ferromagnetism such as graphene double layers in the quantum Hall regime~\cite{Dean2017, Kim2017, Dean2022,Li2026} and double moir\'e bilayers~\cite{MacDonald2022, Mak2023, Wang2024}, where each layer has two degenerate valleys~\cite{Jin2026, Wang2026}. 
It turns out that the valley degeneracy leads to qualitatively different dynamical states in biased devices.

With two valleys in each layer, the double-layer system contains four electronic bands. We consider an interaction-induced layer- and valley-pseudospin ferromagnet at total filling $\nu_{\rm T}=1$ within the four active bands, with balanced layer fillings $\nu_\uparrow=\nu_\downarrow=1/2$~\cite{Kim2017, Dean2022, Wang2024, Li2026}. 
The $SU(4)$-symmetric part of the model (see Appendix~\hyperref[appe:microscopic]{A}) yields the mean-field Hamiltonian $H^{\rm{MF}}_{ \bm{k}}=\epsilon_{\bm k} \sigma_0\otimes \tau_0 -\Delta (\bm{\lambda}\cdot\bm{\Lambda})$, 
where $\tau_0$ is the identity matrix in the valley space, $\Delta$ is the order-parameter amplitude, $\bm{\Lambda}$ denotes the 15 generators of the group $SU(4)$, 
and $\bm{\lambda}$ specifies the order-parameter orientation. 
It is convenient to represent the order parameter by $\bm{\lambda}\cdot\bm{\Lambda} = 4|\psi\rangle \langle \psi|-\sigma_0\otimes \tau_0$ where
\begin{align}
\ket{\psi}
&= e^{-i\phi/2}
\cos\frac{\theta}{2}\,
\ket{\uparrow}\otimes\ket{1}
+
e^{i\phi/2}
\sin\frac{\theta}{2}\,
\ket{\downarrow}\otimes\ket{2}
\label{four_spinor}
\end{align}
is the single-electron wave function that has the lowest energy in the four-dimensional layer-valley flavor space at each $\bm{k}$. 
Here $\ket{\uparrow}$ and $\ket{\downarrow}$ are top-layer and bottom-layer states while $|1\rangle$ and $|2\rangle$ are valley spinors. 
Therefore, the order parameter  is parametrized by three unit vectors~\cite{Goerbig2021}: 
the layer-pseudospin vector $\bm{n}\equiv(\sin\theta\cos\phi,\sin\theta\sin\phi,\cos\theta)$  and
the valley-pseudospin vectors $\bm{v}_{1}=\langle 1| \bm{\tau} |1\rangle$ in the top layer and $\bm{v}_{2}=\langle 2| \bm{\tau} |2\rangle$ in the bottom layer 
[blue arrows in the insets of \fig{fig1}(c)], where $\bm{\tau}=(\tau_x, \tau_y, \tau_z)$ are the Pauli matrices in the valley space. We choose the gauge where $\langle1|2\rangle$ is real and nonnegative, absorbing its complex phase into $\phi$.

The dynamics of the order parameter is governed by a set of coupled LLG equations for the three vectors. The layer-pseudospin  obeys
\begin{align}
	 \mathcal{D}_t {\bm n} = -\bm{n} \times \left[ \Gamma (\mathcal{D}_t {\bm n} -\mu \hat{\bm z}\times\bm n) + 2\frac{\partial E}{\partial\bm n} \right],
	\label{LLG_eff}
\end{align}
while the two valley-pseudospins obey
\begin{equation}
    \begin{aligned}
\frac{d {\bm v}_1}{dt}
&=
-\bm v_1\times\left(
\Gamma\frac{d {\bm v}_1}{dt}
+\frac{4}{1 + n^z}
\mathcal D_{\bm v_1} E
\right),
\\
\frac{d {\bm v}_2}{dt}
&=
-\bm v_2\times\left(
\Gamma\frac{d {\bm v}_2}{dt}
+\frac{4}{1 - n^z}
\mathcal D_{\bm v_2} E
\right).
\end{aligned}
\label{eq:Q4_LLG_v}
\end{equation}
Here $\mathcal{D}_t {\bm n} \equiv 
(d/dt + a_t \hat{\bm z} \times) \bm{n}$
is a covariant time derivative with $a_t=\bm a\cdot d(\bm v_1+\bm v_2)/dt$ being an effective gauge potential generated by the valley dynamics~\cite{Refael2009}, where $\bm a\equiv(\bm v_1\times\bm v_2)/[2(1+\bm v_1\cdot\bm v_2)]$ represents the relative Berry connection of the two valley pseudospins. 
The covariant valley derivatives are defined as 
$\mathcal D_{\bm v_i} \equiv \partial_{\bm v_i}-\bm a\,\partial_\phi$.  In a form similar to \equa{Substitute}, the bias $\mu$ couples directly only to the layer-pseudospin, which is the interlayer order.
Note that Eqs.~\eqref{LLG_eff} and \eqref{eq:Q4_LLG_v} have a compact form in terms of the quantum geometry defined on the order-parameter manifold, see  Appendix~\hyperref[appe:EoM]{B} and the microscopic derivation in  Ref.~\cite{supplementary}. 

As in the previous case, the orientation-dependent energy is $E(\bm{n},\bm{v}_1,\bm{v}_2)=E_{\rm a}+E_{\rm p}+E_{\rm t}$. 
The easy-plane anisotropy and layer-polarization energies retain their previous forms: $2E_{\rm a}=\beta\Delta(n^z)^2$ and $2E_{\rm p}=n^z E_{\rm G}d$, whereas the interlayer tunneling energy $E_{\rm t}$ additionally depends on the valley-pseudospin orientations. To determine this dependence, we write the interlayer tunneling Hamiltonian as
\begin{align}
    H_{\rm t}
    =
    -c_0
    \sum_{\bm k; \alpha,\beta=K,K'}
    a^{\dagger}_{\uparrow\alpha,\bm k}
    X^{\alpha\beta}
    a_{\downarrow\beta,\bm k}
    + \mathrm{h.c.},
    \label{tunneling_H_Q4}
\end{align}
where $a_{\uparrow/\downarrow,\alpha,\bm k}$ annihilates an electron with momentum $\bm k$ in the $\uparrow/\downarrow$ layer and valley $\alpha=K,K'$. 
By independent valley-basis rotations in the two layers, one can always bring $X$ to a diagonal form with nonnegative diagonal entries. For simplicity, we take valley-conserving tunneling: $X=\tau_0$.
The consequences of valley-nonconserving tunneling are discussed in Appendix~\hyperref[appe:unequal_tunneling]{E} and are detailed in Ref.~\cite{supplementary}. 
The interlayer tunneling  then gives the Josephson energy
\begin{align}
    E_{\rm t}
    =
    -c_0 \cos\frac{\Theta}{2} \sin\theta \cos \phi 
    \,,
    \label{E_tunnel}
\end{align}
where $\Theta\in[0,\pi]$ is the relative angle between the two valley-pseudospin vectors $\bm{v}_1$ and $\bm{v}_2$ with $\cos(\Theta/2) = \langle 1|2\rangle$. Correspondingly, the current in \equa{J_2} is modified to $I(t) = I_{\rm J} \cos\frac{\Theta}{2} \sin\theta \sin \phi$.
In the present model, $E_{\rm t}$ is the only contribution to the energy that depends on the valley-pseudospin orientations. 
This weak potential leads to valley-pseudospin motion via the last terms in \equa{eq:Q4_LLG_v}. 

In equilibrium with zero voltage bias, $\mu=0$, the tunneling energy is minimized when the valley-pseudospin vectors in the two layers are parallel, $\bm{v}_1 = \bm{v}_2$ ($\Theta=0$), and the layer-pseudospin phase is locked to $\phi=0$. 
At sufficiently low bias, Eq.~\eqref{eq:Q4_LLG_v} supports a static parallel valley configuration, $\bm{v}_1=\bm{v}_2$. In this configuration, one has $\bm{a}=0$, $\Theta=0$, 
and \equa{LLG_eff} reduces to that for the valleyless system. 
The coupled system therefore admits the same static solution as in \equa{sol_static}, accompanied by parallel valley pseudospins and the dc tunneling current, see the left inset of \fig{fig1}(c).

The state at high bias is qualitatively different from that in the valleyless case. 
To obtain it,  a natural starting point is the precession solution $(\phi,\theta)=(\mu t+\phi_0,\pi/2)$ with the valley pseudospins being static, which is exact in the absence of interlayer tunneling, $c_0=0$.
During the precession, the Josephson energy $E_{\rm t}$ alternately favors parallel valley pseudospins ($\Theta=0$) when $\cos\phi>0$
and antiparallel valley pseudospins ($\Theta=\pi$) when $\cos\phi<0$.
It gives no preference for a valley-pseudospin configuration  since the time average $\langle E_{\rm t} \rangle_t$ over a period is zero. 
However, the tunneling term distorts the orbit by introducing a perturbative correction, giving $\phi=\mu t+\phi_0+\mathcal{O}(c_0 \cos\frac{\Theta}{2})$, as in the right inset of \fig{fig1}(b). 
One may get a glimpse of the preferred valley pseudospin configuration in the dissipationless limit, where energy conservation implies slower phase rotation in regions of higher Josephson 
energy.
The phase therefore spends more time in regions with $\cos\phi<0$, giving a positive Josephson energy $\langle E_{\rm t} \rangle_t \propto \left(c_0 \cos\frac{\Theta}{2} \right)^2$ after time averaging. This energy penalizes the state with nonzero overlap between the two valley spinors, thus favoring antiparallel valley pseudospins ($\Theta=\pi$).

To obtain the preferred valley configuration quantitatively, we derive  from \equa{eq:Q4_LLG_v} the dynamics of the relative valley angle:
\begin{align}
    \frac{d \Theta}{dt} &= -\frac{4c_0}{1+\Gamma^2}
\frac{\sin(\Theta/2)}{\sin\theta}
\left(\Gamma\cos\phi + n^z\sin\phi  \right).
\label{eq:Theta}
\end{align} 
We separate the slow evolution of $\Theta$ from its fast oscillation and denote its average over one precession period by $\bar{\Theta}\equiv\langle\Theta\rangle_t$. 
The unperturbed precession orbit $(\phi,\theta)=(\mu t+\phi_0,\pi/2)$ gives a vanishing cycle average in Eq.~\eqref{eq:Theta}.
Expanding the periodic corrections $(\delta\phi,\delta\theta,\delta\Theta)$ of the orbit to first order in $c_0$, 
and time averaging the right-hand side of Eq.~\eqref{eq:Theta} over one period yields the slow dynamics at order $c_0^2$:
\begin{align}
    \frac{1}{\Gamma}\frac{d\bar{\Theta}}{dt}
    &=
    -\frac{8}{1+\Gamma^2}\frac{\partial V_{\rm eff}}{\partial\bar{\Theta}},
    \nonumber\\
    V_{\rm eff}(\bar{\Theta})
    &=
    \frac{c_0^2\beta\Delta(1+\cos\bar{\Theta})}
    {\mu^2(1+\Gamma^2)^2+(2\beta\Delta\Gamma)^2}.
    \label{eq_ponder}
\end{align}
Here $V_{\rm eff}$ is the effective static potential felt by $\bar{\Theta}$, a ponderomotive potential~\cite{Sun2024}.
Evidently, $V_{\rm eff}$ favors the antiparallel configuration $\Theta = \pi$ and relaxes the  two valley pseudospins toward this state.
Because the two valley spinors  are orthogonal there, the effective tunneling strength $c_0 \langle 1 | 2\rangle$ is strongly suppressed, leading to the \textit{dynamical tunneling blockade}, see the right inset of Fig.~\ref{fig1}(c). 
The numerically exact evolution of $\Theta$ is shown in Fig.~\ref{fig_Theta} of Appendix~\hyperref[appe:Theta]{C}, which exhibits an oscillatory relaxation toward the near-antiparallel configuration, whose slow evolution is well captured by Eq.~\eqref{eq_ponder} within its regime of validity.

We now discuss the final stable state as $\Theta$ approaches $\pi$. In this regime,
the gauge connection $\bm a$ becomes large with $|\bm a|=\tfrac{1}{2}\tan(\Theta/2)\sim1/(\pi-\Theta)$, which allows the gauge potential $a_t$ generated by the valley dynamics to be enhanced. 
Since $a_t$ effectively changes the bias  for the layer-pseudospin to $\mu_{\rm eff}=\mu-a_t$, 
it may compensate the applied bias and stop the bias-driven precession. 
The perturbative solution used to derive Eq.~\eqref{eq_ponder} ceases to be valid then.
We therefore solve the coupled Eqs.~\eqref{LLG_eff} and \eqref{eq:Q4_LLG_v} directly in the near-antiparallel regime, obtaining a phase-locked state:
\begin{align}
    a_t &\approx\mu, \qquad
    \phi\approx\operatorname{sgn}(\mu)\frac{\pi}{2},   \nonumber\\
     \Theta&=\pi-\epsilon(\mu), \ \ 
    \epsilon(\mu)\approx
    \frac{4c_0\Gamma}{|\mu|(1+\Gamma^2)}
    \label{eq:blockade_static}
\end{align}
to leading order in $c_0$~\cite{supplementary}. 
It shows that the valley pseudospins are stabilized in a near-antiparallel configuration, with the relative angle offset from $\pi$ by the small value $\epsilon(\mu)$.
This stationary value agrees well with the numerically exact result, as shown in Fig.~\ref{fig_Theta} of Appendix~\hyperref[appe:Theta]{C}. 
The near-antiparallel valley configuration reduces the effective tunneling strength to $c_0\cos(\Theta/2)\approx c_0\epsilon(\mu)/2$.
Note that although $\phi$ and $\Theta$ are stationary, the valley pseudospins continue to
precess together, generating the gauge potential $a_t\approx\mu$ that
compensates the bias.
The resulting tunneling current from \equa{E_tunnel} is therefore purely dc:
\begin{align}
    I_{\rm dc} 
    \approx
    I_{\rm J} \frac{c_0}{\mu }  
  \frac{2\Gamma}{1+\Gamma^2},
    \label{eq:blockade_current}
\end{align}
to leading nonvanishing order in $c_0$. Note that the dc current here has a physical origin distinct from that of Eq.~\eqref{j_dc_2} in the valleyless case. The latter is the time-averaged dc component of the dynamical current in a precessing state.

Taken together, these results reveal a bias-driven nonequilibrium phase transition from parallel to (nearly) antiparallel valley pseudospins. Unlike in the valleyless case, the steady states on both sides of the transition have static layer pseudospins. 
The $I_{\rm dc}$-$\mu$ characteristics in Fig.~\ref{fig1}(c) show excellent agreement between the analytical and numerical results. 
Compared with the valleyless case in Fig.~\ref{fig1}(b), the dc current exhibits a substantially larger drop at the transition, reflecting the strong suppression of the effective tunneling strength by the tunneling blockade. 
Note that the critical bias in Fig.~\ref{fig1}(c) is smaller than $2c_0/\Gamma$
because of an earlier onset of instability in the valley pseudospin configuration,
see Ref.~\cite{supplementary} for a detailed derivation from Eq.~\eqref{eq:Theta}. 
The blockade transition is directly confirmed by the numerical steady-state
solution ($\zeta=0$) in Fig.~\ref{fig:unequal_tunneling_Z}(a) of
Appendix~\hyperref[appe:unequal_tunneling]{E}, which shows an abrupt drop in the normalized tunneling amplitude $Z\equiv|\langle1|X|2\rangle|=\cos(\Theta/2)$.

\textit{Discussion.---} 
Experimentally, in addition to the sharper drop of the $I_{\rm dc}$-$\mu$ curve at the critical bias, a strong signature of the tunneling blockade phase is the complete absence of the coherent ac tunneling current. In the valleyless case, the ac current emits electromagnetic radiation which reflects the phase coherence (see details in Ref.~\cite{supplementary}) while the radiation is absent in systems with valley degeneracy. 
Note that in both cases, hysteresis exists because of coexisting steady-state solutions with the bias swept upward and downward, see \fig{fig:hysteresis} in Appendix~\hyperref[appe:hysteresis]{D}. 
Further, in double moiré bilayers with valley-contrasting Chern bands~\cite{MacDonald2019}, the reorganization of valley pseudospins across the dynamical tunneling blockade transition may produce 
abrupt changes in responses to valley-sensitive probes, potentially detectable through in-plane Hall transport~\cite{Park2023FQAH}, Kerr rotation~\cite{Lee2016Kerr,Volmer2023Kerr}, or magnetic circular dichroism~\cite{Cai2023FQAH}. 

While we have focused on the valley-conserving tunneling terms, the dynamical tunneling blockade transition remains robust for generic tunneling  in \equa{tunneling_H_Q4}. 
Direct numerical solutions of 
Eqs.~\eqref{LLG_eff} and \eqref{eq:Q4_LLG_v} with the corresponding Josephson term
reveal the same qualitative behavior: a sharp suppression of the effective tunneling amplitude $Z$ and the interlayer tunneling current as the bias is increased beyond the critical value, as shown in
Fig.~\ref{fig:unequal_tunneling_Z} of
Appendix~\hyperref[appe:unequal_tunneling]{E}. 
The tunneling blockade mechanism should also be robust against easy-axis or easy-plane anisotropies of the valley pseudospin, as it relies essentially on the dynamical breaking of degeneracy between parallel and antiparallel valley pseudospins, see details in Ref.~\cite{supplementary}.
Furthermore, the blockade may also apply to broader platforms such as dynamical interlayer exciton condensates~\cite{Millis2024a, Millis2024b, Golez2025} with internal spin/valley degeneracies, and Josephson junctions made of superconductors with multiple degenerate components~\cite{Khalaf2022,Han2025Chiral,Geier2026}. 
Finally, we note that the current work is restricted to a mean-field analysis, while fluctuations~\cite{Sieberer2016,Diehl2025} and disorder effects~\cite{MacDonald2001,Wilczek2001} in these nonequilibrium steady states warrant future study. 

\textit{Acknowledgments.---} 
This work is supported by the National Natural Science Foundation of China (Grants No. 12374291, No. 12421004 and No. 12474150), the Beijing Natural Science Foundation (Z240005), and the National Key Research and Development Program of China (2022YFA1204700 and 2024YFA1409000).
Y. M. was supported by JSPS KAKENHI Grant Nos. JP24H00191, JP25K07235, JP26K00646, and JP26H01281, Japan. L. K. acknowledges Roland Gustafssons Stiftelse för teoretisk fysik for travel support. 
		
\textit{Data Availability.---}
The data supporting the findings of this work are available on Zenodo at \href{https://doi.org/10.5281/zenodo.23289555}{doi:10.5281/zenodo.23289555}.

\bibliography{layer_magnet}
    
\newpage

\onecolumngrid
\begin{center}
{\large\textbf{End Matter}}
\end{center}
\twocolumngrid

\appendix	

\setcounter{equation}{0}
\renewcommand{\theequation}{A\arabic{equation}}
\emph{Appendix A: Microscopic model Hamiltonian.}
\phantomsection
\label{appe:microscopic} 
The double-layer Hamiltonian  reads $H =H_0 + V_{\rm x} + V_{\rm a} + H_{\rm p} + H_{\rm t}$ with
\begin{align}
H_0=&\sum_{q, \bm k} \epsilon_{\bm k} a^\dagger_{q,\bm k}a_{q,\bm k}
,
\label{eq:microscopic_H}
\\
V_{\rm x}=& \sum_{i\ne j} -J(\bm{r}_i - \bm{r}_j) \hat{\bm\Lambda}(\bm{r}_i) \cdot\hat{\bm\Lambda}(\bm{r}_j)
\label{eq:Vx}
,
\\
V_{\rm a} =& g_{\rm a} \sum_i  [\rho_z(\bm{r}_i)]^2,
\label{eq:Va}
\\
H_{\rm p} =& \frac{1}{2}E_{\rm G}d   \sum_i   \rho_z(\bm{r}_i) + \text{const.}
\end{align}
Here $H_0$ is the kinetic energy where $\bm k$ denotes momentum. 
The index $q$ labels the $Q$ electron flavors: $q=\uparrow,\downarrow$ representing the top and bottom layers for valleyless systems,  while $q=\uparrow K,\uparrow K',\downarrow K,\downarrow K'$ for the system with valley degrees of freedom.
The interaction, originating from the Coulomb repulsion, contains direct and exchange terms after being projected to the concerned bands~\cite{Hasebe2002, supplementary}. 
In \equa{eq:Vx}, we have kept the exchange terms relevant to the pseudospin ferromagnetic order, where  $\hat{\bm\Lambda}=(\hat\Lambda_1,\ldots,\hat\Lambda_{Q^2-1})$ is the pseudospin operator with  $\hat\Lambda_A(\bm r_j)=\sum_{q_1, q_2} a^\dagger_{q_1}(\bm r_j) \Lambda_A^{q_1 q_2} a_{q_2}(\bm r_j)$,
 $\Lambda_A$ being the generators of the group $SU(Q)$
 and $a_{q}(\bm r_j)$ being the electronic annihilation operator at location $\bm{r}_j$. The Fourier relation between $a_{q}(\bm r_j)$ and $a_{q,\bm k}$ reads 
 $a_{q}(\bm r_j) = \sum_{\bm k}$  $e^{i\bm k\cdot\bm r_j} a_{q,\bm k}/\sqrt{N}$ where $N$ is the number of lattice sites. Equation~\eqref{eq:Vx} and  the kinetic energy constitute the $SU(Q)$-symmetric part of the model, whose mean-field treatment yields $H^{\rm MF}$ introduced in the main text.  Equation~\eqref{eq:Va} is the easy-plane anisotropy term originating from the anisotropy between intralayer and interlayer interactions, where $g_{\rm a}$ is positive since the intralayer Coulomb repulsion is stronger than the interlayer one. We have defined  the local density imbalance $\rho_z(\bm r_j) \equiv \rho_\uparrow(\bm r_j)-\rho_\downarrow(\bm r_j)$ between the two layers where
$\rho_s(\bm r_j) \equiv \sum_{\alpha} a^\dagger_{s,\alpha}(\bm r_j)  a_{s,\alpha}(\bm r_j)$ is the population of layer $s=\uparrow,\downarrow$ with $\alpha=K,K'$ running over the valley index. Defining the uniform component of the exchange interaction by
$g\equiv\sum_{\bm R}J(\bm R)$, the easy-plane contribution to $E$ is
$2E_{\rm a}=\beta\Delta(n^z)^2$, where
$\beta=g_{\rm a}/(g-g_{\rm a})$ and we assume
$g>g_{\rm a}>0$~\cite{supplementary}.
The gate-field term $H_{\rm p}$ gives
$\langle H_{\rm p}\rangle/N
=D_Q(\Delta)E_{\rm G}d\,n^z/2+\mathrm{const.}$,
and hence contributes $2E_{\rm p}=E_{\rm G}d\,n^z$
to the normalized energy. 
The interlayer tunneling term $H_{\rm t}$ is given by \equa{tunneling_H_Q4}.
For $Q=2$, it reduces to
$
H_{\rm t}=-c_0 \sum_j  \left[
a^\dagger_{\uparrow}(\bm r_j) a_{\downarrow}(\bm r_j)
+{\rm h.c.}\right]
\label{eq:microscopic_tunneling}
$
and leads to the tunneling energy $E_{\rm t}= -c_0 n^x = -c_0\sin\theta\cos\phi$. 
For $Q=4$, it leads to the tunneling energy in \equa{E_tunnel}.

The layers are coupled to electronic reservoirs at chemical potentials $\mu_\uparrow$ and $\mu_\downarrow$, a coupling that cannot be captured by a subsystem Hamiltonian but is contained in the Keldysh path integral. 
This coupling produces a single-particle broadening $\eta$ and generates the damping term in the LLG equations, see Ref.~\cite{supplementary}.

\setcounter{equation}{0}
\renewcommand{\theequation}{B\arabic{equation}}
\emph{Appendix B: Generalized LLG equation for the order parameter.}
\phantomsection
\label{appe:EoM}
At zero interlayer bias, a general equation of motion for the order-parameter variables $(x^1, x^2, \ldots)$ takes the form 
\begin{align}
\Omega_{ab} \frac{d x^b}{dt}
=
\frac{\partial E}{\partial x^a}
+2\Gamma g_{ab}\frac{d x^b}{dt}.
\label{eq:general_coordinate_eom}
\end{align}
Equation~\eqref{eq:general_coordinate_eom} describes the deterministic mean-field dynamics, with reservoir-induced noise neglected. A similar form of dissipative dynamics was derived in
Ref.~\cite{PhysRevB.106.235154}; see Ref.~\cite{supplementary}
for a detailed derivation of \equa{eq:general_coordinate_eom} and comparison with Ref.~\cite{PhysRevB.106.235154}.
$\Omega_{ab}$ and $g_{ab}$ are the Berry curvature and
Fubini-Study metric, obtained from the imaginary and real parts of the quantum geometric tensor defined on the order-parameter manifold:
\begin{align}
\mathcal Q_{ab}
\equiv
\left\langle\partial_a\psi\middle|
(1-|\psi\rangle \langle \psi|)
\middle|
\partial_b \psi\right\rangle
=
g_{ab}-\frac{i}{2}\Omega_{ab}
\,,
\label{eq:projector_quantum_geometric_tensor}
\end{align}
where $|\psi\rangle$ is the lowest-energy
single-electron spinor wave function at each momentum, which is 
independent of momentum for the momentum-independent order parameter considered here.  It also parametrizes the order-parameter
matrix as $\Delta(Q|\psi\rangle\langle\psi|-1)$,
where $1$ denotes the identity matrix in the $Q$-dimensional flavor space and $Q=2$ ($4$) for systems without (with) valley degeneracy. 
In the valleyless case,
the spinor is $|\psi\rangle = e^{-i\phi/2}
\cos(\theta/2)\,
\ket{\uparrow}
+
e^{i\phi/2}
\sin(\theta/2)\,
\ket{\downarrow}$, which leads to the LLG equation in \equa{LLG_equili}.
For the layer-valley spinor wave function in \equa{four_spinor}, one can define 
\begin{align}
d\bm v_j
&=
\bm e_{j1}\,d\xi_j^1+\bm e_{j2}\,d\xi_j^2, \ \ \ 
\bm e_{j1}\times\bm e_{j2}
=\bm v_j , \nonum
\mathcal D\phi
&\equiv d\phi+i\langle 1|d1\rangle-i\langle 2|d2\rangle 
=
d\phi
+\bm a\cdot(d\bm v_1+d\bm v_2) 
\,,
\label{Dphi}
\end{align}
where $\bm a$ is defined in the main text. Taking 
\begin{align}
(dx^1,\ldots,dx^6)
={}&
\bigl(
d\theta,
\sin\theta\,\mathcal D\phi,
d\xi_1^1,
d\xi_1^2,
d\xi_2^1,
d\xi_2^2
\bigr),
\label{eq:Q4_tangent_variables}
\end{align} 
one obtains $
\mathcal Q
=
\frac{1}{4}\operatorname{diag}
\left(
B,
\cos^2\frac{\theta}{2}B,
\sin^2\frac{\theta}{2}B
\right).
$
Here $B = \sigma_0 - \sigma_y$. Substituting the Berry curvature and quantum metric into \equa{eq:general_coordinate_eom} yields the zero-bias forms of Eqs.~\eqref{LLG_eff} and ~\eqref{eq:Q4_LLG_v}. 
A nonzero interlayer bias is then incorporated by replacing $\mathcal D_t\bm n$ with $\mathcal D_t\bm n-\mu\hat{\bm z}\times\bm n$ in the damping term of \equa{LLG_eff}.

\begin{figure}
    \centering
    \includegraphics[width=0.8\columnwidth]{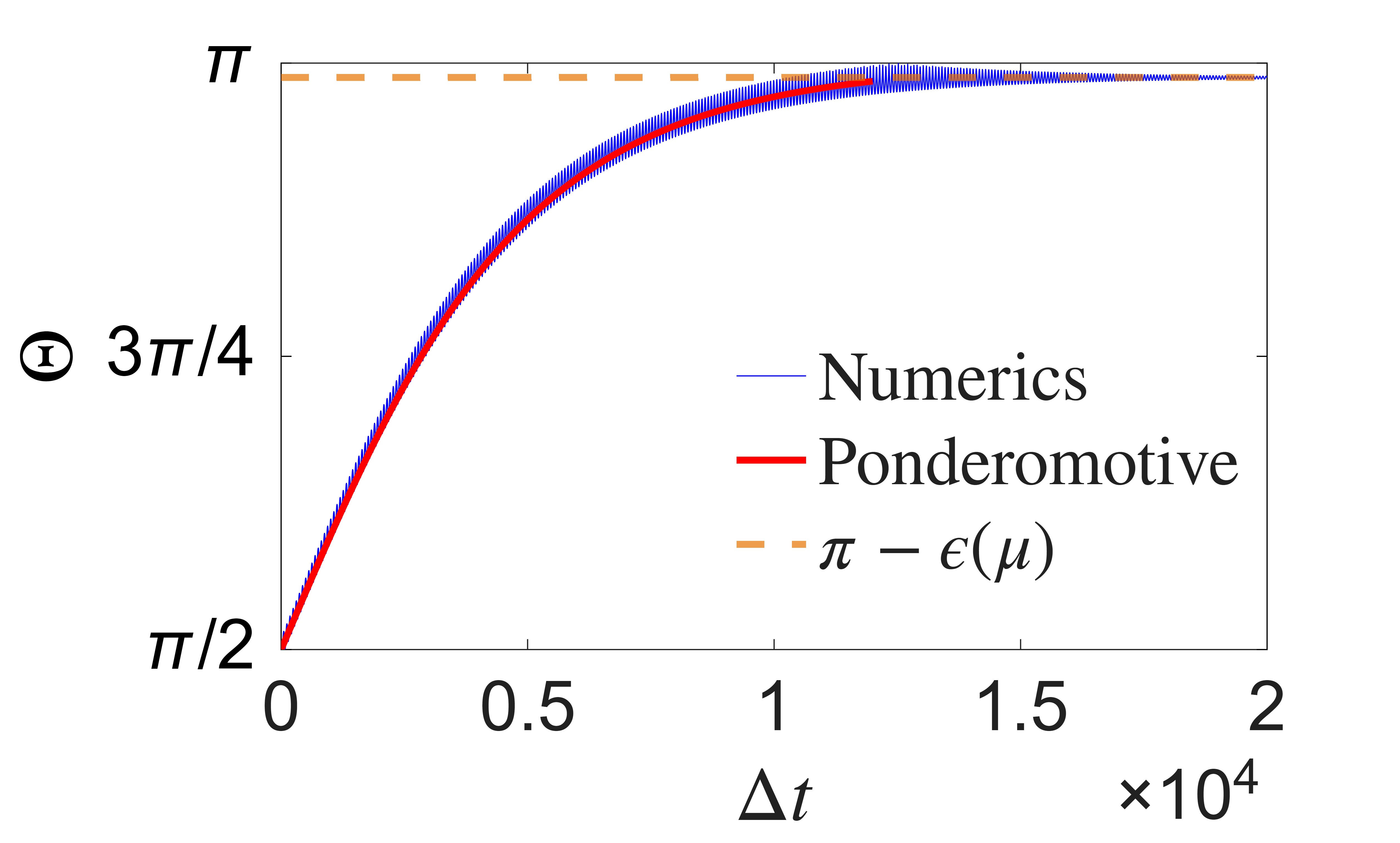}
    \caption{Time evolution of $\Theta$ toward the stable state with initial conditions $\Theta=\pi/2$, $\theta=\pi/2$, and $\phi=0$ at $\mu=4c_0/\Gamma$. 
     The blue curve is the numerical result, the red  curve is the analytical result from  \equa{eq_ponder}, and the orange dashed horizontal line is the stable value $\Theta=\pi-\epsilon$  in \equa{eq:blockade_static}.
     The  parameters are the same as those in Fig.~\ref{fig1}(c).
     }
    \label{fig_Theta}
\end{figure}

\setcounter{equation}{0}
\renewcommand{\theequation}{C\arabic{equation}}
\emph{Appendix C: Numerical results for the relaxation of $\Theta$.}
\phantomsection
\label{appe:Theta}
To examine the relaxation toward the high-bias state, we numerically solve Eqs.~\eqref{LLG_eff} and \eqref{eq:Q4_LLG_v} for the dynamics with the same parameters as in \fig{fig1}(c). Figure~\ref{fig_Theta} shows the time evolution of $\Theta$ at $\mu=4c_0/\Gamma$, starting from $\Theta=\pi/2$, $\theta=\pi/2$, and $\phi=0$. 
It exhibits fast oscillations on top of the slow relaxation toward the stationary value $\pi-\epsilon$ in \equa{eq:blockade_static}. 
Its slow relaxation agrees quantitatively with the analytical result obtained by solving \equa{eq_ponder} within the regime of validity of that equation.

\begin{figure}
    \centering
    \includegraphics[width=0.8\columnwidth]{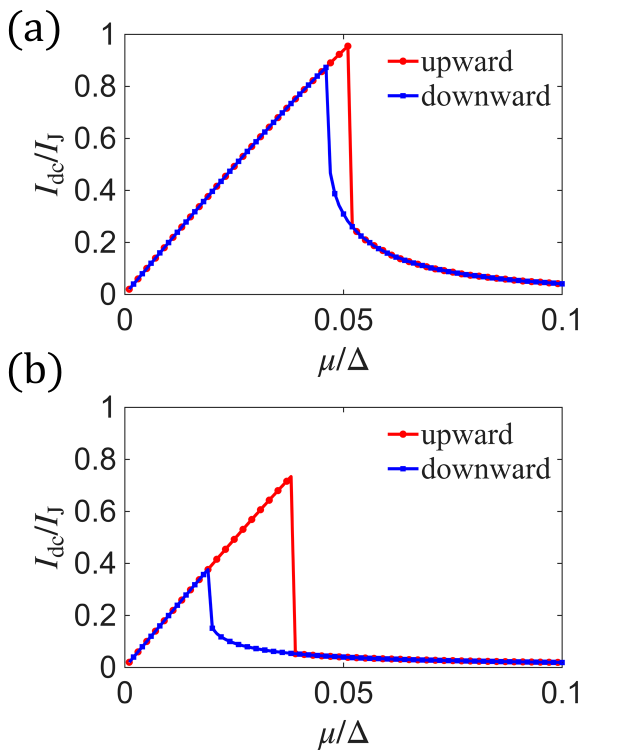}
    \caption{
    The dc interlayer current as a function of $\mu$ for systems (a) without and (b) with valley degeneracy.
    The red circles and blue squares show upward and downward bias sweeps, respectively. 
    The current density is expressed in units of $I_{\rm J}$. The parameters are the same as  those in Figs.~\ref{fig1}(b) and \ref{fig1}(c). 
    }
    \label{fig:hysteresis}
\end{figure}

\setcounter{equation}{0}
\renewcommand{\theequation}{D\arabic{equation}}
\emph{Appendix D: Hysteresis of the $I_{\rm dc}$--$\mu$ curves.}
\phantomsection
\label{appe:hysteresis}
At the mean-field level, the underdamped system has bistable steady states in a certain range of interlayer bias.
In the upward sweep shown in \fig{fig1}, the steady state obtained at each bias is used as the initial condition at the next higher bias, whereas the reverse protocol is used in a downward sweep. 
The system switches between the low- and high-bias branches, producing hysteresis between different critical biases.
This happens for systems both with and without valley degeneracy, as shown by the numerical result in Fig.~\ref{fig:hysteresis}.

\begin{figure}
    \centering
    \includegraphics[width=0.8\columnwidth]{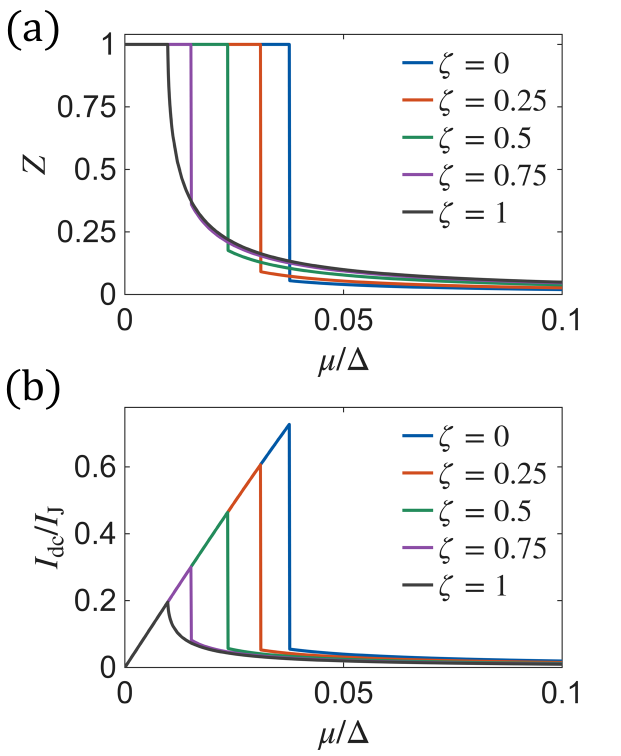}
    \caption{
    	(a) 
    	Tunneling blockade transitions for the generic  tunneling matrix  $X=\operatorname{diag}(1,1-\zeta)$.
    	Each curve plots the effective tunneling amplitude $Z\equiv|\langle1|X|2\rangle|$ for a given $\zeta$ as a function of the bias voltage during an upward sweep.
    	(b) Same as (a) but for the dc interlayer current density $I_{\rm dc}/I_{\rm J}$. 
    	Within numerical tolerance, both $Z$ and the current become time independent after relaxation. 
        The parameters $c_0,\Gamma$ and $\beta$ are the same as those in Fig.~\ref{fig1}(c). 
        }
    \label{fig:unequal_tunneling_Z}
\end{figure}

\setcounter{equation}{0}
\renewcommand{\theequation}{E\arabic{equation}}
\emph{Appendix E: Dynamical tunneling blockade with a generic form of tunneling.}
\phantomsection
\label{appe:unequal_tunneling}
To test the robustness of the dynamical tunneling blockade for a generic  tunneling matrix $X$, we take $X=\operatorname{diag}(1,1-\zeta)$ with $0\leq\zeta\leq1$, which covers the case of a translationally invariant tunneling barrier~\cite{supplementary}. 
For each value of $\zeta$, we solve the full dynamics 
during an upward bias sweep to obtain the steady states at each $\mu$. Figure~\ref{fig:unequal_tunneling_Z} shows that both the overlap $Z=|\langle1|X|2\rangle|$ and the dc current exhibit sharp transitions to suppressed values while the ac current is absent 
for every $\zeta<1$.
The limit $\zeta=1$
exhibits a continuous transition, see Ref.~\cite{supplementary} for details.

\setcounter{equation}{0}
\renewcommand{\theequation}{F\arabic{equation}}
\emph{Appendix F: The dynamical phase shift $\delta \phi(t)$ for \equa{j_dc_2}.}
\phantomsection 
\label{appe:varphi}
Substituting \equa{Substitute} into \equa{LLG_equili} and expanding to  linear order in the deviations $\delta\theta\equiv \theta - \pi/2$ and $\delta\phi \equiv \phi-\mu t-\phi_0$ from the unperturbed orbit, one obtains
\begin{equation}
\begin{aligned}
\frac{d\,\delta\theta}{dt}
&=
\Gamma\frac{d\delta\phi}{dt}
+2c_0\sin(\phi_0 +\mu t),
\\
\frac{d\delta\phi}{dt}
&= -2\beta\Delta\,\delta\theta
-\Gamma\frac{d\,\delta\theta}{dt}.
\end{aligned}
\label{lineared_SU2}
\end{equation}
The solution gives $ \delta\phi = C_{\phi} \cos(\mu t + \phi_0) + S_{\phi} \sin(\mu t + \phi_0)$ with 
\begin{align}
   C_{\phi} =  \frac{2c_0 \Gamma}{\mu}
\frac{4\beta^2\Delta^2+(1+\Gamma^2)\mu^2}
{4\Gamma^2\beta^2\Delta^2+(1+\Gamma^2)^2\mu^2}.
\end{align}
The time-averaged current density reads $I_{\rm dc} = I_{\rm J} C_{\phi} 
\langle \cos^2(\mu t + \phi_0) \rangle_t$, which gives the result in \equa{j_dc_2}.
\clearpage
\end{document}


\title{Supplemental Material for ``Dynamical Tunneling Blockade in Layer-Pseudospin Ferromagnets"}
	\author{Lingxian Kong}
	\affiliation{State Key Laboratory
	of Low-Dimensional Quantum Physics and Department of Physics, Tsinghua University, Beijing 100084, China}
	\affiliation{International Center for Quantum Materials, School of Physics, Peking University, Beijing 100871, China}
    \affiliation{Department of Physics, KTH Royal Institute of Technology, SE-106 91 Stockholm, Sweden}

    \author{Yuta Murakami}
    \affiliation{Institute for Materials Research, Tohoku University, Sendai 980-8577, Japan}
    \affiliation{RIKEN Center for Emergent Matter Science (CEMS), Wako 351-0198, Japan}
    
	\author{Ryuichi Shindou}
	\affiliation{International Center for Quantum Materials, School of Physics, Peking University, Beijing 100871, China}%
	
	\author{Zhiyuan Sun}
	\affiliation{State Key Laboratory
		of Low-Dimensional Quantum Physics and Department of Physics, Tsinghua University, Beijing 100084, China}
	\affiliation{Frontier Science Center for Quantum Information, Beijing 100084, People's Republic of China}
	\date{\today}
	\maketitle
	\vspace{-0.8cm}
	\tableofcontents
	\section{Microscopic Model Hamiltonian and the pseudospin description}
	The Hamiltonian of the double layer system is given by
	\begin{equation}
		H = H_0 + V + H_{\rm p} + H_{\rm t},		
	\end{equation}
	where $H_0$ is the kinetic and potential energy of electrons in the two layers, $V$ is the Coulomb interaction between electrons, $H_{\rm p}$ is the layer-polarization term generated by the gate field, and $H_{\rm t}$ is the interlayer tunneling term. $H_0$ and $H_{\rm p}$ are given by
	\begin{equation}		
		H_0 = \sum_{q = 1,\cdots Q} \sum_{\bm{k}} \epsilon_{\bm{k}} a^{\dagger}_{q,\bm{k}} a_{q,\bm{k}}, \ \ \ H_{\rm p} = \Phi_{\uparrow} \sum_i \rho_{\uparrow}(\bm{r}_i) + \Phi_{\downarrow} \sum_i \rho_{\downarrow}(\bm{r}_i).
	\end{equation}
    Here $q$ is the flavor index of the electron. For an electron gas in two dimensions (2D) under a strong magnetic field, $q = \uparrow, \downarrow$ is the layer-pseudospin index and $Q=2$. For double-layer graphene in the quantum Hall regime and double-moir\'e materials, $q = \uparrow K, \uparrow K^{\prime}, \downarrow K, \downarrow K^{\prime}$ with $K,K^{\prime}$ being the valley-pseudospin index and $Q=4$. $\epsilon_{\bm{k}}$ is the kinetic energy of the electrons with a momentum $\bm{k}$. In quantum Hall systems, $\bm{k}$ represents the magnetic translation quasi-momentum. $a^{\dagger}_{q,\bm{k}}$ ($a_{q,\bm{k}}$) is the creation (annihilation) operator of an electron with flavor $q$ and momentum $\bm{k}$. $\Phi_{\uparrow}$ and $\Phi_{\downarrow}$ are the potential energy for the up- and down-layers controlled by the top and bottom gates. They satisfy $\Phi_{\uparrow} - \Phi_{\downarrow} = E_{\rm G}d$ with $E_{\rm G}$ being vertical electric field and $d$ being the interlayer distance. 
	\begin{align}
		\rho_s(\bm{r}_i)= \sum_{\alpha}a^\dagger_{s\alpha}(\bm{r}_i)a_{s\alpha}(\bm{r}_i)
	\end{align}
	 is the population of layer $s=\uparrow,\downarrow$ at position $\bm{r}_i$, with the valley index $\alpha=K,K'$ absent for $Q=2$. Here $a^{\dagger}_{s\alpha}(\bm{r}_i)$ and $a_{s\alpha}(\bm{r}_i)$ are the creation and annihilation operators for an electron in layer $s$ with valley index $\alpha$ at site $i$, whose position is denoted by $\bm{r}_i$. These are discrete lattice operators. Since we focus on fixed total filling, $\rho_{\uparrow} + \rho_{\downarrow}$ is a constant and one can write the layer-polarization term as 
	\begin{align}
		H_{\rm p} = \frac{E_{\rm G}d}{2}\sum_i \rho_z(\bm{r}_i)  + \text{const}. \quad \rho_z(\bm{r}_i) \equiv \rho_{\uparrow}(\bm{r}_i) - \rho_{\downarrow}(\bm{r}_i)
	\end{align}
	The Coulomb interaction $V$ is firstly considered in the limit of zero inter-layer distance such that it has an $SU(Q)$ symmetry. $V$ can be separated into the direct part $V_{\rm d}$ and exchange part $V_{\rm x}$~\cite{PhysRevB.65.075311}.
	\begin{align}	
		V &= V_{\rm d} + V_{\rm x},  \nonum 
		V_{\rm d} &= \sum_{i \ne j} U(\bm{r}_i-\bm{r}_j) \Big( \sum_q a^{\dagger}_{q}(\bm{r}_i) a_q(\bm{r}_i) \Big) \Big( \sum_q a^{\dagger}_{q}(\bm{r}_j) a_q(\bm{r}_j) \Big) \equiv \sum_{i \ne j} U(\bm{r}_i-\bm{r}_j) n(\bm{r}_i) n(\bm{r}_j) , \nonum 
		V_{\rm x} &= -  \sum_{i \ne j} J(\bm{r}_i-\bm{r}_j)  \bm{\hat{\Lambda}}(\bm{r}_i) \cdot \bm{\hat{\Lambda}}(\bm{r}_j) -  \sum_{i \ne j} J(\bm{r}_i-\bm{r}_j)  n(\bm{r}_i) n(\bm{r}_j),
	\end{align}
	where $\bm{\hat{\Lambda}}(\bm{r}_i) = \Big(\hat{\Lambda}_{1}(\bm{r}_i), \cdots , \hat{\Lambda}_{Q^2 - 1}(\bm{r}_i)\Big)$ and 
    \begin{align}
		\hat{\Lambda}_A(\bm{r}_i) =  \sum_{q_1,q_2 =1,\cdots Q} a^{\dagger}_{q_1}(\bm{r}_i)  \Lambda_A^{q_1 q_2} a_{q_2}(\bm{r}_i), \quad A = 1, \cdots, Q^2 -1.
	\end{align}
    with $\Lambda_A$ being the $SU(Q)$ generators \footnote{For $Q=2$, the $SU(2)$ generators are given by the Pauli matrices $\sigma_x, \sigma_y, \sigma_z$. For $Q=4$, the $SU(4)$ generators are given by the following 15 matrices:
	\begin{align}
	\Lambda_{1/2/3}  \equiv \sigma_{x/y/z} \otimes \tau_0, \ \  \Lambda_{4/5/6} \equiv \sigma_0 \otimes \tau_{x/y/z}  , \ \  \Lambda_{7 \sim 15} = \sigma_{\alpha} \otimes \tau_{\beta} \quad \text{with } \alpha, \beta = x/y/z .
	\label{SU4_gen}
    \end{align}
    Here  $\tau_0$ and $\sigma_0$ are the $2\times2$ identity matrices. This definition gives a normalization and orthogonality condition of Tr$\Lambda_A \Lambda_B = 4\delta_{AB}$. }. $n(\bm{r}_i) = \sum_q a^{\dagger}_{q}(\bm{r}_i) a_q(\bm{r}_i)$ is the electron density operator at site $i$. $U(\bm{r}_i-\bm{r}_j)$ and $J(\bm{r}_i-\bm{r}_j)$ are the direct and exchange integrals determined by the local one-body wavefunction for each site~\cite{Altland_Simons_2010, PhysRevB.65.075311}. Since we focus on the pseudospin ferromagnetic order, the first term   in $V_{\rm x}$ with $J(\bm{r}_i-\bm{r}_j) > 0$, the pseudospin ferromagnetic coupling, is the relevant interaction. The second term in $V_{\rm x}$ can be absorbed into $V_{\rm d}$ by redefining $U = U- J$ and is irrelevant to our focus. 
	Written in the momentum space, the pseudospin ferromagnetic coupling term is given by
	\begin{equation}
		V_{\rm x} = - \frac{1}{N} \sum_{\bm{p}} J_{\bm{p}} \bm{\hat{\Lambda}}_{\bm{p}} \cdot \bm{\hat{\Lambda}}_{-\bm{p}},	    \quad \bm{\hat{\Lambda}}_{\bm{p}} = \Big(\hat{\Lambda}_{\bm{p},1}, \cdots , \hat{\Lambda}_{\bm{p},Q^2-1} \Big), 
		\label{eq:Vx}
	\end{equation}
    where $N$ is the number of sites, $J_{\bm{p}} = \sum_{\bm{R}} J(\bm{R}) e^{-i \bm{p} \cdot \bm{R}}$ is the Fourier transform of $J(\bm{R})$ with $J(\bm{R}=0)=0$. Furthermore, \footnote{In this work, we use the convention that the Fourier transform of a function $f(\bm{r})$ on a lattice is defined as $f_{\bm{p}} = \sum_i f(\bm{r}_i) e^{-i \bm{p} \cdot \bm{r}_i}$, where $\bm{r}_i$ is the position of site $i$. The inverse Fourier transform is given by $f(\bm{r}_i) = \frac{1}{N} \sum_{\bm{p}} f_{\bm{p}} e^{i \bm{p} \cdot \bm{r}_i}$. Moreover, for annihilation (creation) operators, we use $a_{q}(\bm{r}_i) = \frac{1}{\sqrt{N}} \sum_{\bm{p}} a_{q,\bm{p}} e^{i \bm{p} \cdot \bm{r}_i}$ ($a^{\dagger}_{q}(\bm{r}_i) = \frac{1}{\sqrt{N}} \sum_{\bm{p}} a^{\dagger}_{q,\bm{p}} e^{-i \bm{p} \cdot \bm{r}_i}$).} 
	\begin{align}
		\hat{\Lambda}_{\bm{p},A} = \sum_{\bm{k}} \sum_{q_1,q_2 =1,\cdots Q} a^{\dagger}_{q_1,\bm{k}} \Lambda_A^{q_1 q_2} a_{q_2,\bm{k}+\bm{p}}.
	\end{align} 
	In this work, we consider dynamics of a homogeneous pseudospin order, i.e., $\langle \hat{\Lambda}_{\bm{p}=0,A} \rangle$, which is an effective description for a single domain of the order parameter.  Therefore, we only consider the $\bm{p} = 0$ component of \equa{eq:Vx} and omit the momentum index $\bm{p}$ in the following. Moreover, due to the finite interlayer distance $d$, the interlayer Coulomb interaction is slightly weaker than the intralayer one. This leads to an easy-plane anisotropy in the pseudospin space~\cite{PhysRevLett.86.1825, PhysRevLett.86.1829, PhysRevLett.95.266804}. We include this effect by adding an anisotropic easy-plane term $V_{\rm a}$:
	\begin{align}
		V_{\rm a} = g_{\rm a} \sum_i [\rho_\uparrow(\bm{r}_i)-\rho_\downarrow(\bm{r}_i)]^2 =  g_{\rm a} \sum_i [\hat{\Lambda}_3(\bm{r}_i)]^2.
	\end{align}
	Therefore, the relevant interaction terms for the calculations below are given by the zero-momentum component of $V_{\rm x}$ and the easy-plane term $V_{\rm a}$:
	\begin{align}
		V_{\rm rel} = -\frac{g}{N} \hat{\bm{\Lambda}} \cdot \hat{\bm{\Lambda}} +  \frac{g_{\rm a}}{N} [\sum_i \rho_z(\bm{r}_i)]^2 = -\frac{g}{N} \hat{\bm{\Lambda}} \cdot \hat{\bm{\Lambda}} +  \frac{g_{\rm a}}{N} \hat{\Lambda}_3^2; \quad g \equiv  J_{\bm{p}=0}, \quad g_{\rm a} > 0.
		\label{Vx_relevant}
	\end{align}

	The interlayer tunneling term is given by
	\begin{align}	
		H_{\rm t} = - c_0 \sum_i \begin{cases}  a^{\dagger}_{\uparrow}(\bm{r}_i) a_{\downarrow}(\bm{r}_i) + \text{cc}.  , & Q = 2, \\ 
		\sum_{\alpha,\beta = K,K^{\prime}} a^{\dagger}_{\uparrow \alpha}(\bm{r}_i) X^{\alpha \beta} a_{\downarrow \beta}(\bm{r}_i) +  \text{cc}. , & Q=4. \end{cases}
		\label{eq:H_t}
	\end{align}
	Here $c_0 > 0 $ is the tunneling amplitude. In the case of $Q=4$, the simplest case has $X = \tau^0$, meaning that only valley-conserving tunneling is allowed and the tunneling strengths are equal in the two valley sectors. In the general case, one can always apply independent unitary transformations in the valley-pseudospin space for both layers to make $X = \text{diag}(1, 1-\zeta)$ with $0 \leq \zeta \leq 1$, which is nothing but a singular value decomposition of $X$. Here we absorb the overall factor of $X$ into $c_0$ so that $X$ is normalized. 

    \section{Mean field theory.}
	In this section, we derive the mean field solution for the order parameter.	To make it easy to derive the non-equilibrium equation in later sections, we use the Keldysh formalism.  
	The partition function of the system is given by 
    \begin{align}
	    \mathcal{Z} = \int \int \mathcal{D} a^{\dagger} \mathcal{D} a \ \ e^{iS_0 + iS_V}.
    \end{align}
Considering relevant interactions in \equa{Vx_relevant}, the interaction part of the action is given by
\begin{align}
	iS_V =  i\oint dt   \ \ & \frac{g}{N} \hat{\bm{\Lambda}} \cdot \hat{\bm{\Lambda}} -  \frac{g_{\rm a}}{N}  \hat{\Lambda}_3^2.
\end{align}
Here $\oint$ means a Keldysh time contour. Apply a Hubbard-Stratonovich transformation and obtain
\begin{align}
	e^{iS_V} \propto \int \mathcal{D} \bm{m} \ \ \exp \{ i\oint dt \ \ &-\frac{N}{4g}\bm{m}(t)\cdot \bm{m}(t) - \frac{\beta N}{4g} m^3(t) m^3(t)  +  \bm{m}(t) \cdot \hat{\bm{\Lambda}} 
	 \} .
\end{align}
Here, $\bm{m}$ is a $(Q^2-1)$-dimensional Hubbard-Stratonovich field, and $\beta=g_{\rm a}/(g-g_{\rm a})$. We restrict our analysis to $g>g_{\rm a}>0$, for which all interaction channels in the magnetic decomposition remain attractive and $\beta>0$.  We further apply the Keldysh-Larkin-Ovchinnikov rotation to obtain the action in the real-time contour
\footnote{ For the $\bm{m}$ field, the Keldysh-Larkin-Ovchinnikov rotation reads $\bm{m}_c = (\bm{m}_+ + \bm{m}_-)/2$, $\bm{m}_q = (\bm{m}_+ - \bm{m}_-)/2$. For the fermion field, the rotation reads 
\begin{align}
	\begin{bmatrix}
		a_c \\
		a_q 
	\end{bmatrix} = \frac{1}{\sqrt{2}} 
	\begin{bmatrix}
	1&1 \\
	1&-1 
	\end{bmatrix} 
	\begin{bmatrix}
		a_+ \\
		a_- 
	\end{bmatrix} 
	,   \   \  \  
	\begin{bmatrix}
		a^{\dagger}_c \\
		a^{\dagger}_q 
	\end{bmatrix} = \frac{1}{\sqrt{2}} 
	\begin{bmatrix}
	1&-1 \\
	1&1 
	\end{bmatrix} 
	\begin{bmatrix}
		a^{\dagger}_+ \\
		a^{\dagger}_- 
	\end{bmatrix} 
	\nonumber . 
\end{align} 
Here the subscript $+/-$ denotes the field defined on the forward/backward branch of the time contour. 
} 

\begin{align}
	e^{iS_V}
	={}&
	\int \mathcal{D} \bm{m}_c \int \mathcal{D} \bm{m}_q
	\exp \Bigg\{ i\int dt
	\Bigg[
	-\frac{N}{g}\bm{m}_c(t)\cdot \bm{m}_q(t)
	-\frac{\beta N}{g} m^{3}_c (t) m^{3}_q (t)
	\nonumber\\
	&\hspace{3.0cm}
	+ \sum_{\bm{k}}
	\begin{bmatrix}
		a^{\dagger}_{c,\bm{k}} (t) & a^{\dagger}_{q,\bm{k}} (t)
	\end{bmatrix}
	\begin{bmatrix}
		\bm{m}_{c} \cdot \bm{\Lambda} & \bm{m}_q \cdot \bm{\Lambda}  \\
		\bm{m}_{q} \cdot \bm{\Lambda} & \bm{m}_c \cdot \bm{\Lambda}
	\end{bmatrix}
	\begin{bmatrix}
		a_{c,\bm{k}} (t) \\
		a_{q,\bm{k}} (t)
	\end{bmatrix}
	\Bigg]
	\Bigg\} .
\end{align}
Here $\bm{m}_{c/q}$ is the classical/quantum order parameter, and $a_{c/q,\bm{k}} = [a_{c/q,1,\bm{k}}, \cdots, a_{c/q,Q,\bm{k}} ]^{\rm T}$ is a $Q$-component vector of the classical/quantum annihilation operator in the flavor space. $\bm{\Lambda} = (\Lambda_1, \Lambda_2, \cdots, \Lambda_{Q^2-1})$. Moreover,
\begin{align}
	iS_0 =  i\sum_{\bm{k}} \int dt \int dt^{\prime}  \ \ & 
	\begin{bmatrix}
		a^{\dagger}_{c,\bm{k}} (t) & a^{\dagger}_{q,\bm{k}} (t)
	\end{bmatrix}
	\begin{bmatrix}
		G_0^{R,-1}(\bm{k},t-t^{\prime}) + c_0(\sigma_x\blue{\otimes \tau_0}) & -\Sigma_{\mathrm{lab}}^K(t-t^{\prime}) \\
		0 & G_0^{A,-1}(\bm{k},t-t^{\prime}) + c_0(\sigma_x\blue{\otimes \tau_0})
	\end{bmatrix}
	\begin{bmatrix}
		a_{c,\bm{k}} (t^{\prime}) \\
		a_{q,\bm{k}} (t^{\prime})
	\end{bmatrix}
\end{align}
The $+c_0(\sigma_x\blue{\otimes \tau_0})$ terms in the inverse Green's functions arise from the lab-frame interlayer-tunneling Hamiltonian $-c_0(\sigma_x\blue{\otimes \tau_0})$, with the time-local factor $\delta(t-t^{\prime})$ understood. $G_0^{R/A,-1}$ and $\Sigma_{\mathrm{lab}}^K$ are the inverse retarded/advanced Green's function and the lab-frame Keldysh self-energy of the free electrons, respectively. They are given by
\begin{align}
	G_0^{R/A,-1}(\bm{k},t-t^{\prime}) = \delta(t-t^{\prime}) 
	\begin{bmatrix}
		i\partial_t - (\epsilon_{\bm{k}}+\Phi_{\uparrow}) +/- i\eta & 0 \\
		0 & i\partial_t - (\epsilon_{\bm{k}}+\Phi_{\downarrow}) +/- i\eta
	\end{bmatrix} \blue{\otimes \tau_0}
\end{align}
\begin{align}
	\Sigma_{\mathrm{lab}}^K(\omega)
	= -2i\eta
	\begin{bmatrix}
		\tanh\!\left(\dfrac{\omega-\mu_{\uparrow}}{2T}\right) & 0 \\
		0 & \tanh\!\left(\dfrac{\omega-\mu_{\downarrow}}{2T}\right)
	\end{bmatrix} \blue{\otimes \tau_0} .
\end{align}
Equivalently, in the time domain $
	\Sigma_{\mathrm{lab}}^K(t-t^{\prime})
	= \int \frac{d\omega}{2\pi} e^{-i\omega(t-t^{\prime})} \Sigma_{\mathrm{lab}}^K(\omega) .$
Here the $\blue{\otimes \tau_0}$ with a blue color is needed only for the case of $Q=4$, which denotes the valley space in each layer. We will keep using this notation later. $T$ is the temperature of the reservoir, $\mu_{\uparrow/\downarrow}$ is the chemical potential coupled to the two layers, which is controlled by the voltage of the leads, shown in Fig.~1 in the main text. $G_0^{R(A)}$ is the retarded(advanced) Green's function from $\uparrow/\downarrow$-layer electrons with the energy $\epsilon_{\bm{k}} + \Phi_{\uparrow/\downarrow}$. Here $\eta$ is the single-particle broadening induced by the system-reservoir coupling. The reservoir is taken to be energy independent for structureless reservoirs~\cite{Kamenev_2011}. The momentum-independent $\Sigma_{\mathrm{lab}}^{K}$ describes the coupling between the electrons and the reservoirs of the $\uparrow/\downarrow$ layer with chemical potentials $\mu_{\uparrow/\downarrow}$.

Since the two reservoirs have different chemical potentials, the action is interpreted as describing a non-equilibrium system. A convenient approach to deal with such a non-equilibrium system is to transform the action into the rotating frame~\cite{Altland_Simons_2010}. We define the unitary transformation matrix
\begin{equation}
 U(t)=\begin{pmatrix}
 e^{i\mu_\uparrow t}&0\\
 0&e^{i\mu_\downarrow t}
 \end{pmatrix}\blue{\otimes\tau_0}.
 \label{eq:rotating_frame_U}
\end{equation}
The fields transform as
\begin{equation}
 \Psi_{c/q}(t)=U(t)a_{c/q}(t),\qquad
 \bm M_{c/q}(t)\cdot\bm\Lambda
 =U(t)\bigl[\bm m_{c/q}(t)\cdot\bm\Lambda\bigr]U^\dagger(t).
 \label{eq:transformation}
\end{equation}
The full action is transformed as 
\begin{align}
	e^{iS_V}
	={}&
	\int \mathcal{D} \bm{M}_c \int \mathcal{D} \bm{M}_q
	\exp \Bigg\{ i\int dt
	\Bigg[
	-\frac{N}{g}\bm{M}_c(t)\cdot \bm{M}_q(t)
	-\frac{\beta N}{g} M^{3}_c (t) M^{3}_q (t)
	\nonumber\\
	&\hspace{3.0cm}
	+ \sum_{\bm{k}}
	\begin{bmatrix}
		\Psi^{\dagger}_{c,\bm{k}} (t) & \Psi^{\dagger}_{q,\bm{k}} (t)
	\end{bmatrix}
	\begin{bmatrix}
		\bm{M}_{c} \cdot \bm{\Lambda} & \bm{M}_q \cdot \bm{\Lambda}  \\
		\bm{M}_{q} \cdot \bm{\Lambda} & \bm{M}_c \cdot \bm{\Lambda}
	\end{bmatrix}
	\begin{bmatrix}
		\Psi_{c,\bm{k}} (t) \\
		\Psi_{q,\bm{k}} (t)
	\end{bmatrix}
	\Bigg]
	\Bigg\} .
\end{align}
\begin{align}
	iS_0 =  i\sum_{\bm{k}} \int dt \int dt^{\prime}  \ \ & 
	\begin{bmatrix}
		\Psi^{\dagger}_{c,\bm{k}} (t) & \Psi^{\dagger}_{q,\bm{k}} (t)
	\end{bmatrix}
	\begin{bmatrix}
		\mathcal{G}_0^{R,-1}(\bm{k},t-t^{\prime}) - \hat{h}(t,t^{\prime}) & -\Sigma_{\mathrm{rot}}^K(t-t^{\prime}) \\
		0 & \mathcal{G}_0^{A,-1}(\bm{k},t-t^{\prime}) - \hat{h}(t,t^{\prime})
	\end{bmatrix}
	\begin{bmatrix}
		\Psi_{c,\bm{k}} (t^{\prime}) \\
		\Psi_{q,\bm{k}} (t^{\prime})
	\end{bmatrix},
\end{align}
with 
\begin{align}
	\mathcal{G}_0^{R/A,-1}(\bm{k},t-t^{\prime}) &= \delta(t-t^{\prime}) 
	\begin{bmatrix}
		i\partial_t - \epsilon_{\bm{k}} +/- i\eta & 0 \\
		0 & i\partial_t - \epsilon_{\bm{k}} +/- i\eta
	\end{bmatrix} \blue{\otimes \tau_0} \nonum
    \hat{h}(t,t^{\prime}) &= \delta(t-t^{\prime})(\bm{h}(t) \cdot \bm{\sigma}) \blue{\otimes \tau_0}
    \label{eq:GRA_0}
\end{align}
Under the rotating-frame transformation, the Keldysh self-energy becomes 
\begin{align}
	\Sigma_{\mathrm{rot}}^K(\omega)
	&= -2i\eta f(\omega) 
	\sigma_0 \blue{\otimes \tau_0} , \quad f(\omega) \equiv \tanh\!\left(\dfrac{\omega}{2T}\right).
	\label{Sigma_rot}
\end{align}
In the rotating frame, the energies of the electrons in the two layers are shifted, and each layer is coupled to a reservoir at zero chemical potential. Therefore, the action is interpreted to describe an equilibrium system driven by a time-dependent Hamiltonian $\bm{h}\cdot \bm{\sigma}$ defined in \equa{eq:GRA_0}, which acts on the layer space and contains two parts:
\begin{align}
	\bm{h} \cdot \bm{\sigma} = \begin{pmatrix}
		\Phi_{\uparrow} - \mu_{\uparrow} & 0 \\
		0 & \Phi_{\downarrow} - \mu_{\downarrow}
	\end{pmatrix} - c_0
	 \begin{pmatrix}
		0 & e^{i\mu t} \\
		e^{-i\mu t} & 0
	\end{pmatrix} \equiv h^0 \sigma_0 +  h^z \sigma_z + \bm{c}(\mu t) \cdot \bm{\sigma}.
	\label{hz_ct}
\end{align} 
Here $h^z = [(\Phi_{\uparrow} - \mu_{\uparrow}) - (\Phi_{\downarrow} - \mu_{\downarrow})]/2 = (E_{\rm G} d - \mu)/2$. The constant term $h^0\sigma_0$ can be absorbed into the band energy $\epsilon_{\bm{k}}$. $h^z = 0$ corresponds to the charge balanced case with $\nu_{\uparrow} = \nu_{\downarrow} = 1/2$. $\bm{c} = -c_0(\cos(\mu t), -\sin(\mu t), 0)$ comes from the interlayer tunneling term in \equa{eq:H_t} with $\mu \equiv \mu_{\uparrow} - \mu_{\downarrow}$ being the chemical potential difference between the two layers. 

We first solve the mean-field equations without the external field $\bm{h}$ and then treat the external field as a perturbation for later analysis. Integrating out the fermion field, we obtain the effective action of the order parameter:
\begin{align}
	iS_{\rm eff} &= {\rm Tr} \ln  [i\mathcal{G}^{-1} - i\kappa_0 \otimes \hat{h} + i\kappa_x \otimes \bm{M}_q\cdot \bm{\Lambda}] - i\int dt \ [ \frac{N}{g} \bm{M}_c(t) \cdot \bm{M}_q(t) + \frac{\beta N}{g}M^{3}_c(t)M^{3}_q(t) ] 
	\label{iS_eff}
\end{align}
Here $\kappa_0$ and $\kappa_x$ are Pauli matrices acting in Keldysh space. The full Green's function $\mathcal{G}$ is given by 
\begin{align}
	\mathcal{G} = 	\begin{bmatrix}
		\mathcal{G}^{R}  & \mathcal{G}^{K}  \\
		0 & \mathcal{G}^{A} 
	\end{bmatrix} ,
\end{align} 
where $\mathcal{G}^{R/A}$ and $\mathcal{G}^K$ are the retarded/advanced Green's function and Keldysh Green's function of the electrons dressed by a static order parameter $\bm{M}_c$. They are given by 
\begin{align}
	\mathcal{G}^{R/A}(\bm{k},\omega)^{-1} &= \mathcal{G}_0^{R/A}(\bm{k}, \omega)^{-1} + \bm{M}_c \cdot \bm{\Lambda} ,  \nonum
	\mathcal{G}^K(\bm{k},\omega) & = \mathcal{G}^R(\bm{k},\omega) \Sigma_{\mathrm{rot}}^K(\omega) \mathcal{G}^A(\bm{k},\omega) = f(\omega)[ \mathcal{G}^R(\bm{k},\omega) - \mathcal{G}^A(\bm{k},\omega) ].
	\label{eq:Keldysh_GK}
\end{align}
The latter equation implies the fluctuation-dissipation theorem~\cite{Kamenev_2011}. The self-consistent gap equation of the order parameter is given by the saddle point $\delta S_{\rm eff}/\delta \bm{M}_q = 0$, which leads to
\begin{align}
	&\int \frac{d\omega}{2\pi} \frac{1}{N}\sum_{\bm{k}}  \text{Tr }\mathcal{G}^K (\bm{k}, \omega) \bm{\Lambda} = \int \frac{d\omega}{2\pi} \int \frac{d\nu}{2\pi} \rho(\nu) \text{Tr }\mathcal{G}^K (\nu, \omega) \bm{\Lambda} = i \frac{1}{g}  \bm{M}_c + i\frac{\beta}{g} M_c^3 \hat{\mathbf e}_3
	\label{saddle_M}
\end{align} 
Here $\hat{\mathbf e}_3$ denotes the unit basis vector along the third axis in order-parameter space; the corresponding component $M_c^3$ describes the electron density imbalance between the two layers. $\rho(\nu)$ is the density of states (DOS) of the band. For example, a typical DOS with a bandwidth $W$ reads $\rho(\nu) = 2W/(W^2 + \nu^2)$. 

At the mean-field level, the order-parameter matrix can be written as
\begin{align}
	\bm{M}_c\cdot\bm{\Lambda}
	=
	\Delta(Q P-1),
	\qquad
	P=|\Psi\rangle\langle\Psi| ,
    \label{eq_Mc_P}
\end{align}
where $P$ is a rank-one projector onto a normalized state $|\Psi\rangle$. Taking the trace with each generator and using $\operatorname{tr}(\Lambda_A\Lambda_B)=Q\delta_{AB}$ gives $\bm M_c=\Delta\langle\Psi|\bm\Lambda|\Psi\rangle$. The order-parameter matrix has eigenvalue $(Q-1)\Delta$ in the $P$ subspace and $-\Delta$ in its $(Q-1)$-dimensional complement. 
The Green's function has the spectral decomposition
\begin{align}
	\mathcal G^{R/A/K}
	=
	\mathcal G_+^{R/A/K} P+\mathcal G_-^{R/A/K}(1-P),
	\label{G_plus_minus}
\end{align}
with
\begin{align}
	\mathcal G_+^{R/A}(\nu,\omega)
	&=
	\frac{1}{\omega-\nu+(Q-1)\Delta\pm i\eta}, \ \ \
	\mathcal G_-^{R/A}(\nu,\omega)
	=
	\frac{1}{\omega-\nu-\Delta\pm i\eta},
	\nonumber\\
	\mathcal G_\pm^K(\nu,\omega)
	&=
	f(\omega)\left[
	\mathcal G_\pm^R(\nu,\omega)
	-
	\mathcal G_\pm^A(\nu,\omega)
	\right].
\end{align}
To evaluate the saddle-point equation~\eqref{saddle_M}, we integrate the Keldysh Green's function over frequency. For either branch $s=\pm$, with $E_+=\nu-(Q-1)\Delta$ and $E_-=\nu+\Delta$, this gives
\begin{align}
 \int\frac{d\omega}{2\pi}\,\mathcal G_s^K(\nu,\omega)
 &=\int\frac{d\omega}{2\pi}\,f(\omega)
 \left[\mathcal G_s^R(\nu,\omega)-\mathcal G_s^A(\nu,\omega)\right]
=\int\frac{d\omega}{2\pi}\,
 \frac{-2i\eta f(\omega)}{(\omega-E_s)^2+\eta^2}
 \equiv-if_\eta(E_s).
 \label{f_eta}
\end{align}
Here $f_\eta$ is the reservoir-broadened form of $f$ defined in \equa{Sigma_rot}. From Eqs.~\eqref{saddle_M}--\eqref{f_eta}, using $\bm M_c=\Delta\langle\Psi|\bm\Lambda|\Psi\rangle$, one obtains
\begin{align}
	\frac{1}{2g}\Delta &= \frac{1}{2}\int  \rho(\nu) [f_{\eta}(\nu+\Delta) - f_{\eta}(\nu-(Q-1
)\Delta)] \frac{d\nu}{2\pi}    \nonum
      &= \int  \rho(\nu) [ n_{\rm F}(\nu-(Q-1)\Delta) - n_{\rm F}(\nu+\Delta)] \frac{d\nu}{2\pi} \equiv D_Q(\Delta)
	  \label{eq:gap}
\end{align}
and $M_c^3 = 0$. Here 
\begin{align}
	n_{\rm F}(E) \equiv \int \frac{d\epsilon}{2\pi}\,
\frac{2\eta}{(\epsilon-E)^2+\eta^2}\, \frac{1}{e^{\epsilon/T} + 1}
\end{align} 
is the fermionic occupation function obtained by integrating the Fermi distribution over the reservoir-broadened spectral density~\cite{Datta_2005}. It reduces to the Fermi distribution function as $\eta\to0$. A solution with a non-zero $\Delta$ requires the condition $\frac{\partial D_{Q}}{\partial \Delta}|_{\Delta=0} > \frac{1}{2g}$. In the limit of $T,\eta \to 0$, this condition gives the Stoner criterion $Q \rho(\nu=0)/2\pi > 1/2g$. Therefore, in the flat-band limit, the enhanced density of states $\rho(\nu=0)$ allows the order parameter to develop even for a weak interaction $g$.

To parametrize the orientation of the order parameter for the subsequent dynamical analysis, we write the normalized wavefunction as
\begin{align}
	|\Psi\rangle
	=
	\begin{cases}
		e^{-i\varphi/2}\cos\frac{\theta}{2}\ket{\uparrow} + e^{i\varphi/2}\sin\frac{\theta}{2}\ket{\downarrow}, & Q=2; \\
		 e^{-i\varphi/2} \cos\frac{\theta}{2}\, \ket{\uparrow}\otimes\ket{1} + e^{i\varphi/2} \sin\frac{\theta}{2}\, \ket{\downarrow}\otimes\ket{2}. & Q=4.
		 \label{macro_wf}
	\end{cases}
\end{align}
Here $\ket{\uparrow}$ and $\ket{\downarrow}$ denote the layer-up and layer-down states. The angles $\theta$ and $\varphi$ define the layer-pseudospin unit vector $\bm N=(\sin\theta\cos\varphi,\sin\theta\sin\varphi,\cos\theta)$ in the rotating frame. For $Q=4$, the normalized spinors $|1\rangle$ and $|2\rangle$ describe the valley states in the two layers, with valley-pseudospin unit vectors $\bm v_j=\langle j|\bm\tau|j\rangle$ for $j=1,2$. Thus the order parameter is parametrized by $\bm N$, $\bm v_1$, and $\bm v_2$. The relative valley angle $\Theta$ between the two valleys satisfies $\bm v_1\cdot\bm v_2=\cos\Theta$, and we choose the relative spinor phase so that $\langle1|2\rangle=\cos(\Theta/2)$ is real and nonnegative. The lab-frame wavefunction $|\psi\rangle$ is obtained by replacing $\varphi$ with $\phi=\varphi+\mu t$ in \equa{macro_wf}. 

Following \equa{eq:transformation}, the lab-frame layer-pseudospin vector $\bm n(t)$ is related to $\bm N(t)$ by $N^x - iN^y = e^{i\mu t} (n^x - in^y)$ (or equivalently $\phi = \varphi + \mu t$) and $N^z = n^z$. Here $\mu \equiv \mu_{\uparrow} - \mu_{\downarrow}$ is the bias voltage between the two layers. In the following sections, we will use the notation of 
\begin{align}
	\bm{n} = (n^x, n^y, n^z) = (\sin\theta \cos\phi, \sin\theta \sin\phi, \cos\theta), \ \ \ \bm{N} = (N^x, N^y, N^z) = (\sin\theta \cos\varphi, \sin\theta \sin\varphi, \cos\theta).
\end{align}
Here $\phi = \varphi + \mu t$. According to \equa{eq_Mc_P}, the order parameter in the lab frame can be written as $\bm{m}_c \cdot \bm{\Lambda} = \Delta (Q \ket{\psi} \bra{\psi}  - 1)$.

The notation used in the two frames throughout Secs.~II--IV is summarized in Table~\ref{tab:frame_notation}.
\begin{table}[htbp]
\caption{Notation in the lab and rotating frames used in Secs.~II--IV.}
\label{tab:frame_notation}
\centering
\small
\setlength{\tabcolsep}{4pt}
\renewcommand{\arraystretch}{1.35}
\begin{tabular}{llll}
\hline\hline
\parbox[t]{0.23\textwidth}{\raggedright Quantity\strut} & \parbox[t]{0.13\textwidth}{\raggedright Lab frame\strut} & \parbox[t]{0.13\textwidth}{\raggedright Rotating frame\strut} & \parbox[t]{0.42\textwidth}{\raggedright Relation \strut} \\
\hline
\parbox[t]{0.23\textwidth}{\raggedright Fermionic field in layer $s$\strut} & \parbox[t]{0.13\textwidth}{\raggedright $a_s(t)$\strut} & \parbox[t]{0.13\textwidth}{\raggedright $\Psi_s(t)$\strut} & \parbox[t]{0.42\textwidth}{\raggedright $\Psi_s=e^{i\mu_s t}a_s$ \strut} \\
\parbox[t]{0.23\textwidth}{\raggedright Order-parameter field\strut} & \parbox[t]{0.13\textwidth}{\raggedright $\bm m_{c/q}$\strut} & \parbox[t]{0.13\textwidth}{\raggedright $\bm M_{c/q}$\strut} & \parbox[t]{0.42\textwidth}{\raggedright $\bm M_{c/q}\!\cdot\!\bm\Lambda=U(\bm m_{c/q}\!\cdot\!\bm\Lambda)U^\dagger$ \strut} \\
\parbox[t]{0.23\textwidth}{\raggedright Macroscopic spinor\strut} & \parbox[t]{0.13\textwidth}{\raggedright $|\psi\rangle$\strut} & \parbox[t]{0.13\textwidth}{\raggedright $|\Psi\rangle$\strut} & \parbox[t]{0.42\textwidth}{\raggedright $|\Psi\rangle=U(t)|\psi\rangle$, up to an overall phase \strut} \\
\parbox[t]{0.23\textwidth}{\raggedright Layer-pseudospin vector\strut} & \parbox[t]{0.13\textwidth}{\raggedright $\bm n$\strut} & \parbox[t]{0.13\textwidth}{\raggedright $\bm N$\strut} & \parbox[t]{0.42\textwidth}{\raggedright $N^x-iN^y=e^{i\mu t}(n^x-in^y)$, $N^z=n^z$ \strut} \\
\parbox[t]{0.23\textwidth}{\raggedright Layer azimuthal angle\strut} & \parbox[t]{0.13\textwidth}{\raggedright $\phi$\strut} & \parbox[t]{0.13\textwidth}{\raggedright $\varphi$\strut} & \parbox[t]{0.42\textwidth}{\raggedright $\varphi=\phi-\mu t$ \strut} \\
\parbox[t]{0.23\textwidth}{\raggedright Layer-polarization energy\strut} & \parbox[t]{0.13\textwidth}{\raggedright $E_{\rm p}=\frac{E_{\rm G}d}{2}n^z$\strut} & \parbox[t]{0.13\textwidth}{\raggedright $E_h=h^zN^z$\strut} & \parbox[t]{0.42\textwidth}{\raggedright $h^z=(E_{\rm G}d-\mu)/2$ \strut} \\
\parbox[t]{0.23\textwidth}{\raggedright Energy in the LLG equation\strut} & \parbox[t]{0.13\textwidth}{\raggedright $E$\strut} & \parbox[t]{0.13\textwidth}{\raggedright $\mathcal E$\strut} & \parbox[t]{0.42\textwidth}{\raggedright $\mathcal E=E-\mu N^z/2$ \strut} \\
\parbox[t]{0.23\textwidth}{\raggedright Keldysh self-energy\strut} & \parbox[t]{0.13\textwidth}{\raggedright $\Sigma^K_{\mathrm{lab}}$\strut} & \parbox[t]{0.13\textwidth}{\raggedright $\Sigma^K_{\mathrm{rot}}$\strut} & \parbox[t]{0.42\textwidth}{\raggedright $\Sigma_{\mathrm{rot}}^K(t,t')=U(t)\Sigma_{\mathrm{lab}}^K(t,t')U^\dagger(t')$ \strut} \\
\hline\hline
\end{tabular}
\par\smallskip
\begin{minipage}{\linewidth}
\small
Here $\mu=\mu_\uparrow-\mu_\downarrow$ and $U(t)=\operatorname{diag}(e^{i\mu_\uparrow t},e^{i\mu_\downarrow t}) \blue{\otimes \tau_0} $.
The quantities $\Delta$, $\theta$, $\bm v_{1,2}$, and $\Theta$ are unchanged by the frame transformation.
\end{minipage}
\end{table}
\FloatBarrier
\section{Derivation of the equation of motion in the tangent space.}
\label{sec:derive_LLG}
In this section, we derive the equation of motion for the order parameter in the tangent space in the rotating frame. Let $\delta\mathcal M_c \equiv \delta (\bm{M}_c \cdot \bm{\Lambda})$ and $\mathcal M_q \equiv \bm{M}_q \cdot \bm{\Lambda}$ denote the
classical and quantum variations of the order-parameter matrix. From Eq.~\eqref{iS_eff}, keeping the terms that are linear in $\mathcal M_q$ and first order in either
$\delta\mathcal M_c$ or the external field $\hat h$ gives
\begin{align}
iS_{\rm eff}
\approx {}&
iS_{\rm dyn}-iS_E.
\label{S_eff_parts}
\end{align}
The dynamic part is
\begin{align}
iS_{\rm dyn}
\equiv{}&
-\operatorname{Tr}
\left[
\mathcal G(\kappa_0\otimes\delta\mathcal M_c)
\mathcal G(\kappa_x\otimes\mathcal M_q)
\right] -
i\frac{N}{g}\int dt\,
\delta\bm{M}_c(t) \cdot \bm{M}_q(t)
\label{S_dyn}
\end{align}
The energy part is
\begin{align}
iS_E
={}&
-\operatorname{Tr}
\left[
\mathcal G(\kappa_0\otimes\hat h)
\mathcal G(\kappa_x\otimes\mathcal M_q)
\right] 
+i\frac{\beta N}{g}\int dt\,
\delta M_c^3(t)M_q^3(t).
\label{S_e}
\end{align}
Here 
$\operatorname{tr}$ is the internal $Q$-dimensional trace, and
$\operatorname{Tr}$ includes all space-time, internal and Keldysh indices. It is useful to define the retarded response function $\mathcal P_\omega^R$ by
\begin{align}
-\operatorname{Tr}[\cdots]
&\equiv
i\int\frac{d\omega}{2\pi}\,
\mathcal P_\omega^R
\left(
\mathcal M_q(-\omega),
\delta\mathcal M_c(\omega)
\right),
\nonumber\\
\mathcal P_\omega^R(\mathcal M_q(-\omega),\delta\mathcal M_c(\omega))
&\equiv
i\sum_{\bm k}\int\frac{d\epsilon}{2\pi}\,
\operatorname{tr}
\left[
\mathcal G^R(\bm k,\epsilon+\omega)\delta\mathcal M_c
\mathcal G^K(\bm k,\epsilon) \mathcal M_q
+
\mathcal G^K(\bm k,\epsilon+\omega)\delta\mathcal M_c
\mathcal G^A(\bm k,\epsilon) \mathcal M_q
\right].
\label{GNGM_general_Q}
\end{align}

At fixed order-parameter amplitude, we take the two tangent variations as
$\delta\mathcal M_c=Q\Delta X$ and $\mathcal M_q=Q\Delta Y$. Here $X$ and $Y$ are tangent matrices satisfying
$X=P X\bar P+\bar PXP$ and $Y=P Y\bar P+\bar PYP$, with
$\bar P\equiv 1-P = 1-|\Psi\rangle\langle\Psi|$.  Therefore, they satisfy the following identity:
\begin{align}
\operatorname{tr}(YX)
=
\operatorname{tr}(PY\bar PX)
+
\operatorname{tr}(PX\bar PY).
\label{eq:tangent_trace_identity}
\end{align}

Combining Eqs.~\eqref{G_plus_minus},\eqref{GNGM_general_Q},\eqref{eq:tangent_trace_identity}, and the calculation in Appendix~\ref{app:green_integrals_inter}, the retarded response function can be calculated as
\begin{align}
\mathcal P_\omega^R(Q\Delta Y,Q\Delta X)
={}&
\left\{
i(Q\Delta)^2\sum_{\bm k}\int\frac{d\epsilon}{2\pi}
\left[
\mathcal G_-^R(\bm k,\epsilon+\omega)\mathcal G_+^K(\bm k,\epsilon)
+
\mathcal G_-^K(\bm k,\epsilon+\omega)\mathcal G_+^A(\bm k,\epsilon)
\right]
\right\}
\operatorname{tr}(PY\bar PX)
\nonumber\\
+&
\left\{
i(Q\Delta)^2\sum_{\bm k}\int\frac{d\epsilon}{2\pi}
\left[
\mathcal G_+^R(\bm k,\epsilon+\omega)\mathcal G_-^K(\bm k,\epsilon)
+
\mathcal G_+^K(\bm k,\epsilon+\omega)\mathcal G_-^A(\bm k,\epsilon)
\right]
\right\} \operatorname{tr}(PX\bar PY)
\nonumber\\
\approx {}&
\left(
\frac{NQ\Delta^2}{g}
\right)
\operatorname{tr}(YX) +
\mathcal S
(\omega+i\Gamma\omega)
\operatorname{tr}(PY\bar PX) 
-\mathcal S
(\omega-i\Gamma\omega)
\operatorname{tr}(PX\bar PY)
+\mathcal{O}(\omega^2).
\label{eq:projector_response_two_channels}
\end{align}
Here $D_Q(\Delta)$ is defined in \equa{eq:gap}. $\mathcal{S}$ and $\Gamma$ are calculated in \appe{app:green_integrals_inter}. $\mathcal S \simeq 2N D_Q(\Delta) $ in the leading order of $\eta/\Delta$. $\Gamma$ is expressed in \equa{eq:newapp_interband_real_imaginary_slopes}. $\Gamma$ approaches zero in the limit of $\eta/\Delta \to 0$, meaning that the damping here comes from the finite system-reservoir coupling. Damping can also arise from disorder~\cite{PhysRevLett.87.196802}, which is not considered in this work.

For further analysis, we introduce the complex structure on the tangent space,
\begin{align}
\mathcal J_{\rm c} X
&\equiv-i[P,X],
\qquad
\mathcal J_{\rm c}^2X=-X.
\label{eq:projector_response_complex_structure}
\end{align}
The same tangent decomposition gives
\begin{align}
\operatorname{tr}\left( Y\mathcal J_{\rm c}X \right)
&=
i\left(
\operatorname{tr}(PY\bar PX)
-
\operatorname{tr}(PX\bar PY)
\right).
\label{eq:projector_response_trace_algebra}
\end{align}
Together with Eq.~\eqref{eq:tangent_trace_identity},
Eq.~\eqref{eq:projector_response_two_channels} can be written
as
\begin{align}
\mathcal P_\omega^R(Q\Delta Y,Q\Delta X)
&= 
\frac{NQ\Delta^2}{g} \operatorname{tr}(YX)
+
 i\mathcal S\Gamma\omega\operatorname{tr}(YX)
-i\mathcal S\omega\operatorname{tr}
\left( Y\mathcal J_{\rm c}X \right)+\mathcal O(\omega^2).
\label{eq:projector_response_bilinear}
\end{align}

We now pull this general response back to a set of real collective coordinates
$x_c^a/x_q^a$ in the classical/quantum components. The two tangent matrices in
Eq.~\eqref{eq:projector_response_bilinear} can be expressed in terms of the
coordinate variations by
\begin{align}
X
&=
\delta P
=
\frac{\partial P}{\partial x_c^b}\delta x_c^b,
\ \ \ 
Y
=
P_q
=
\frac{\partial P}{\partial x_c^a}x_q^a.
\label{eq:projector_coordinate_pullback}
\end{align}
Equations~\eqref{GNGM_general_Q}--\eqref{eq:projector_response_bilinear}
give the fermionic-bubble response contained in $S_{\rm dyn}$.  The second term
of $S_{\rm dyn}$ contributes exactly the opposite static term,
which cancels the first term in Eq.~\eqref{eq:projector_response_bilinear}.  The remaining finite-frequency part is
the dynamical response.
Therefore, the dynamic part of the coordinate action is
\begin{align}
S_{\rm dyn}
={}&
\int\frac{d\omega}{2\pi}\,
x_q^a(-\omega)K^R_{ab}(\omega)\delta x_c^b(\omega).
\label{eq:projector_coordinate_dynamic_action}
\end{align}
Here $K^R_{ab}(\omega)$ denotes only the dynamical kernel.  Substituting
Eq.~\eqref{eq:projector_coordinate_pullback} into
Eq.~\eqref{eq:projector_response_bilinear} gives
\begin{align}
K^R_{ab}(\omega)
={}&
i\mathcal S\Gamma\omega\operatorname{tr}
\left(\partial_aP\partial_bP\right) -i\mathcal S\omega\operatorname{tr}
\left[
\partial_aP\mathcal J_{\rm c}(\partial_bP)
\right]+\mathcal O(\omega^2),
\label{eq:projector_coordinate_response_traces}
\end{align}
Equation
\eqref{eq:projector_coordinate_response_traces} naturally defines the Fubini--Study metric and Berry curvature on the $CP^{Q-1}$ manifold~\cite{Cheng2010},
\begin{align}
g_{ab}
&\equiv
\frac{1}{2}\operatorname{tr}
\left(\partial_aP\partial_bP\right)
=\mathrm{Re}
\left\langle\partial_a\Psi\middle|
(1-P)\middle|\partial_b\Psi\right\rangle ,
\nonumber\\
\Omega_{ab}
&\equiv
\operatorname{tr}
\left[
\partial_aP\mathcal J_{\rm c}(\partial_bP)
\right]
=
i\operatorname{tr}
\left(P[\partial_aP,\partial_bP]\right)
=-2\mathrm{Im}
\left\langle\partial_a\Psi\middle|
(1-P)\middle|\partial_b\Psi\right\rangle
=\partial_a\mathcal A_b-\partial_b\mathcal A_a.
\label{eq:projector_symplectic_metric}
\end{align}
Here $P=|\Psi\rangle\langle\Psi|$ and
$\mathcal A_a=i\langle\Psi|\partial_a\Psi\rangle$. It is convenient to define the quantum geometric tensor $\mathcal Q_{ab}$ as
\begin{align}
\mathcal Q_{ab}
\equiv
\left\langle\partial_a\Psi\middle|
(1-P)\middle|\partial_b\Psi\right\rangle
=
g_{ab}-\frac{i}{2}\Omega_{ab}.
\label{eq:projector_quantum_geometric_tensor}
\end{align}
The dynamic kernel can be written as
\begin{align}
K^R_{ab}(\omega)
 = \mathcal S\left[
-i\omega\Omega_{ab}
+2i\omega\Gamma g_{ab}
\right]
+\mathcal{O}(\omega^2).
\label{eq:projector_coordinate_response_kernel}
\end{align}

We now derive the energy term in Eq.~\eqref{S_e} directly in terms of the
layer-pseudospin vector.  According to the definition in
Eq.~\eqref{GNGM_general_Q} and substituting $M^3=\Delta N^z = \Delta\cos\theta$, the energy term reads
\begin{align}
iS_E
&=
i\int\frac{d\omega}{2\pi}\,
\mathcal P_\omega^R
\left(
\mathcal M_q(-\omega),\hat h(\omega)
\right) + i\beta N \Delta 2D_Q(\Delta) \int dt N^z_c(t) N^z_q(t).
\label{eq:energy_response_frequency}
\end{align}
In the first term, we keep the first order contribution in $h^z$, $c_0$ and neglect the contribution in the order $\mu c_0$ by taking the static limit of the retarded response
function $\mathcal P_\omega^R\to\mathcal P_0^R$. Taking
$\mathcal M_q= Q\Delta x_q^a \partial P/\partial x_c^a$, this gives
\begin{align}
iS_E
&=
i\mathcal S \int\frac{d\omega}{2\pi}\, x_q^a(-\omega)  \frac{\partial (E_h + E_{\rm t} + E_{\rm a})}{\partial x_c^a}(\omega)  = i\mathcal S \int dt \ \  x_q^a(t)  \frac{\partial (E_h + E_{\rm t} + E_{\rm a})}{\partial x_c^a}(t),
\label{eq:energy_static_response}
\end{align}
with 
\begin{align}
	E_h + E_{\rm t}
	&\equiv \operatorname{tr}(P\hat h)
	= \langle \Psi|(h^z \sigma_z + \bm{c}(\mu t) \cdot \bm{\sigma})
	\blue{\otimes \tau_0} |\Psi\rangle ,  \nonum
	E_{\rm a} &\equiv \frac{\beta}{2} \Delta (N^z)^2 = \frac{\beta}{2} \Delta (\langle\Psi| \sigma_z \blue{\otimes \tau_0} |\Psi \rangle )^2.
\end{align}
Here $E_{\rm t}$ is the interlayer tunneling energy, $E_{\rm a}$ is the easy-plane anisotropy energy of the layer pseudospin, and $E_h=h^zN^z=E_{\rm p}-\mu N^z/2$ is the rotating-frame counterpart of the gate-induced layer-polarization energy $E_{\rm p} = (E_{\rm G}d/2)N^z$. Using Eqs.~\eqref{macro_wf} and \eqref{hz_ct}, the tunneling energy becomes
\begin{align}
	E_{\rm t} &= 
	\begin{cases}
	-c_0\sin\theta\cos(\varphi+\mu t),
	& Q=2,\\[3pt]
	-c_0\sin\theta\cos\dfrac{\Theta}{2}
	\cos(\varphi+\mu t),
	& Q=4.
	\end{cases}
\label{eq:collective_energy_components}
\end{align}
For $Q=4$, we used
$\langle1|2\rangle \equiv \cos(\Theta/2)$.

Combining Eqs.~\eqref{S_eff_parts}, \eqref{eq:projector_coordinate_dynamic_action},
\eqref{eq:projector_coordinate_response_kernel}, and
\eqref{eq:energy_static_response}, and imposing the saddle-point condition
$\delta S_{\rm eff}/\delta x_q^a=0$, we obtain the general equation of motion
in the collective-coordinate space,
\begin{align}
\Omega_{ab}\frac{d x_c^b}{dt}
=
\frac{\partial \mathcal E}{\partial x_c^a}
+2\Gamma g_{ab}\frac{d x_c^b}{dt} .
\label{eq:general_coordinate_eom}
\end{align}
Here $\mathcal E \equiv E_h+E_{\rm t}+E_{\rm a} $ is the energy in the rotating frame. It is related to the energy in the lab frame $E$ by $\mathcal E = E - \mu N^z / 2$. This equation represents all $2(Q-1)$ real equations of motion and is valid
for both $Q=2$ and $Q=4$. In principle, the nonequilibrium description also includes thermal noise arising from coupling to the reservoirs~\cite{Kamenev_2011,RevModPhys.49.435}, which we neglect here under the assumption of low temperatures and weak thermal fluctuations.

\equa{eq:general_coordinate_eom} is  expressed in collective coordinates on the order-parameter manifold, while it can also be expressed in terms of the wavefunction $|\Psi\rangle$:
\begin{align}
(i-\Gamma)\bar{P}|\dot\Psi\rangle
=\bar{P} H_{\rm eff}|\Psi\rangle,
\qquad \bar{P} =1-|\Psi\rangle\langle\Psi|,
\label{eq:rtg_full_projected_spinor}
\end{align}
where $H_{\rm eff}$ is defined through $H_{\rm eff}|\Psi\rangle
=\frac{\partial\mathcal E}{\partial\langle\Psi|}$. Taking twice the real part of the inner product with $\langle\partial_a\Psi|$ produces
\begin{align}
	2\operatorname{Re}\!\left[
(i-\Gamma)\mathcal Q_{ab}
\right]\dot{x}^{b}
=
\left(\Omega_{ab}-2\Gamma g_{ab}\right)\dot{x}^{b}
=
\frac{\partial\mathcal E}{\partial x^{a}},
\end{align}
which is equivalent to \equa{eq:general_coordinate_eom}. A similar form as \equa{eq:rtg_full_projected_spinor} is also obtained in Eq.~(8) of Ref.~\cite{PhysRevB.106.235154} for $SU(Q)$ coherent states.

\subsection{$Q=2$ case.}
For $Q=2$, the wavefunction is given in \equa{macro_wf}. We use the local tangent variables
\begin{align*}
dx^1&=d\theta,
&
dx^2&=\sin\theta\,d\varphi,
&
\partial_1&=\partial_\theta,
&
\partial_2&=\frac{1}{\sin\theta}\partial_\varphi.
\end{align*}
In this basis, one has the quantum geometric tensor defined in \equa{eq:projector_quantum_geometric_tensor} 
\begin{align}
\mathcal Q
 = 
\frac{1}{4}
\begin{pmatrix}
1 & i\\
-i & 1
\end{pmatrix}.
\end{align}
Correspondingly, $ g_{ab} = \frac{1}{4}\delta_{ab}$, and $ \Omega_{ab} = -\frac{1}{2}\varepsilon_{ab}$ with $ \varepsilon_{12}=1 $, $\varepsilon_{21}=-1$. Therefore, \equa{eq:general_coordinate_eom} gives the LLG equation 
\begin{align}
	\frac{1}{2} \bm{N} \times \frac{d{\bm N}}{dt}
= 
\frac{1}{2}\Gamma\frac{d{\bm N}}{dt}
+
\frac{\partial \mathcal E}{\partial\bm N}, \text{or equivalently,} \ \ \frac{d{\bm N}}{dt} = -\bm{N} \times \left[ \Gamma \frac{d{\bm N}}{dt} + 2\frac{\partial \mathcal E}{\partial\bm N} \right].
	\label{LLG_Q2}
\end{align}
Here $\mathcal E = E_h+E_{\rm t}+E_{\rm a}$. Transforming back to the lab frame gives 
\begin{align}
	\frac{d{\bm n}}{dt}= -\bm{n} \times \left[ \Gamma (\frac{d{\bm n}}{dt} -\mu\hat{\bm z}\times\bm n) + 2\frac{\partial E}{\partial\bm n} \right].
\end{align}
Here $E = \mathcal E + n^z\mu/2 = E_{\rm p} + E_{\rm t} + E_{\rm a}$ with $E_{\rm p} = E_h+\mu n^z/2 = n^z E_{\rm G} d/2$. This is the LLG equation in the main text.

\subsection{$Q=4$ case.}
For $Q=4$, the wavefunction given in \equa{macro_wf} contains two valley spinors. We use the local tangent variables for the valley spinors $\bm{v}_1$ and $\bm{v}_2$ as
\begin{align}
d\bm v_j
&=
\bm e_{j1}\,d\xi_j^1+\bm e_{j2}\,d\xi_j^2, \ \ \ 
\bm e_{j1}\times\bm e_{j2}
=\bm v_j.
\end{align}
Moreover, we define
\begin{align}
\mathcal A_j
\equiv i\langle j|dj\rangle, \ \ \ 
\mathcal D\varphi
&\equiv d\varphi+\mathcal A_1-\mathcal A_2 \equiv 
d\varphi
+\bm a\cdot(d\bm v_1+d\bm v_2) 
\label{Dvarphi}
\end{align}
Here\footnote{To derive \equa{internal_a}, we use
$d|j\rangle=-i\mathcal A_j|j\rangle+\frac12(d\bm v_j\cdot\bm\tau)|j\rangle$.
Consequently, the overlap changes according to
\begin{align*}
d\langle1|2\rangle
&=\langle d1|2\rangle+\langle1|d2\rangle =i(\mathcal A_1-\mathcal A_2)\langle1|2\rangle
+\frac12\langle1|(d\bm v_1+d\bm v_2)\cdot\bm\tau|2\rangle.
\end{align*}
The Pauli-matrix algebra gives
$\langle1|\bm\tau|2\rangle/\langle1|2\rangle
=(\bm v_1+\bm v_2-i\bm v_1\times\bm v_2)/(1+\bm v_1\cdot\bm v_2)$.
Since we absorb the phase of $\langle1|2\rangle$ into $\varphi$ and choose the overlap to be real and nonnegative, its imaginary part must vanish:
\begin{align*}
0=\Im d\langle1|2\rangle
=\langle1|2\rangle\left[
\mathcal A_1-\mathcal A_2
-\frac{(\bm v_1\times\bm v_2)\cdot(d\bm v_1+d\bm v_2)}
{2(1+\bm v_1\cdot\bm v_2)}
\right].
\end{align*}
This gives \equa{Dvarphi} and \equa{internal_a}.}
\begin{align}
	\bm a 
&=
\frac{\bm v_1\times\bm v_2}
{2(1+\bm v_1\cdot\bm v_2)} = \frac{\bm v_1\times\bm v_2}{4\cos^2(\Theta/2)}.  \label{internal_a}
\end{align}
The six local tangent variables are chosen as
\begin{align}
(dx^1,\ldots,dx^6)
={}&
\bigl(
d\theta,
\sin\theta\,\mathcal D\varphi,
d\xi_1^1,
d\xi_1^2,
d\xi_2^1,
d\xi_2^2
\bigr).
\label{eq:Q4_tangent_variables}
\end{align}
Direct substitution of Eq.~\eqref{macro_wf} gives the quantum geometric tensor
\begin{align}
\mathcal Q_{ab}
\equiv
\left\langle\partial_a\Psi\middle|
(1-P)\middle|\partial_b\Psi\right\rangle
=
\frac{1}{4}\operatorname{diag}
\left(
B,
\cos^2\frac{\theta}{2}B,
\sin^2\frac{\theta}{2}B
\right),
\qquad
B\equiv
\begin{pmatrix}
1&i\\
-i&1
\end{pmatrix}.
\label{eq:Q4_quantum_geometric_tensor_blocks}
\end{align}
Taking the real and imaginary parts of Eq.~\eqref{eq:Q4_quantum_geometric_tensor_blocks}
gives
\begin{align}
g_{ab}
&=
\frac14\operatorname{diag}\left(
\sigma_0,
\cos^2\frac{\theta}{2}\sigma_0,
\sin^2\frac{\theta}{2}\sigma_0
\right),
\nonumber\\
\Omega_{ab}
&=
-\frac12\operatorname{diag}\left(
\varepsilon,
\cos^2\frac{\theta}{2}\varepsilon,
\sin^2\frac{\theta}{2}\varepsilon
\right),
\qquad
\varepsilon_{12}=+1.
\label{eq:Q4_metric_curvature_blocks}
\end{align}
Here $\sigma_0$ denotes the $2\times2$ identity matrix.
Define the temporal geometric gauge potential~\cite{PhysRevB.80.024420}
\begin{align}
a_t
&\equiv
\bm a\cdot\left(\frac{d\bm v_1}{dt}+\frac{d\bm v_2}{dt}\right),
\end{align}
and the covariant time derivative of $\bm{N}$
\begin{align}
\mathcal D_t\bm N
&\equiv
\frac{d\theta}{dt}\,\bm e_\theta
+
\sin\theta\left(
\frac{d\varphi}{dt}+a_t
\right)\bm e_\varphi = 
\frac{d\bm N}{dt}
+a_t\hat{\bm z}\times\bm N.
\label{eq:Q4_covariant_layer_velocity}
\end{align}
Substituting Eq.~\eqref{eq:Q4_metric_curvature_blocks} into
Eq.~\eqref{eq:general_coordinate_eom} gives the three LLG equations
\begin{align}
\mathcal D_t\bm N
&=
-\bm N\times\left[
\Gamma\mathcal D_t\bm N
+2\frac{\partial\mathcal E}{\partial\bm N}
\right],
\label{eq:Q4_LLG_N}
\\
\frac{d\bm v_1}{dt}
&=
-\bm v_1\times\left[
\Gamma\frac{d\bm v_1}{dt}
+\frac{4}{1 + \cos\theta}
\mathcal D_{\bm v_1}\mathcal E
\right],
\label{eq:Q4_LLG_v1}
\\
\frac{d\bm v_2}{dt}
&=
-\bm v_2\times\left[
\Gamma\frac{d\bm v_2}{dt}
+\frac{4}{1 - \cos\theta}
\mathcal D_{\bm v_2}\mathcal E
\right].
\label{eq:Q4_LLG_v2}
\end{align}
Here we define
\begin{align}
\mathcal D_{\bm v_i}\mathcal E
\equiv
\frac{\partial\mathcal E}{\partial\bm v_i}
-\bm a\frac{\partial\mathcal E}{\partial\varphi},
\qquad i=1,2.
\label{eq:Q4_covariant_energy_derivative}
\end{align}
This means the energy differential can be expressed as
\begin{align}
d\mathcal E
=&
\frac{\partial\mathcal E}{\partial\theta}\,d\theta
+
\frac{\partial\mathcal E}{\partial\varphi}\,
d\varphi
+
\frac{\partial\mathcal E}{\partial\bm v_1}
\cdot d\bm v_1
+
\frac{\partial\mathcal E}{\partial\bm v_2}
\cdot d\bm v_2 \nonum
=&
\frac{\partial\mathcal E}{\partial\theta}\,d\theta
+
\frac{\partial\mathcal E}{\partial\varphi}\,
\mathcal D\varphi +
\mathcal D_{\bm v_1}\mathcal E\cdot d\bm v_1 +
\mathcal D_{\bm v_2}\mathcal E\cdot d\bm v_2.
\end{align}
The coefficients of $\mathcal D\varphi$, $d\bm v_1$, and $d\bm v_2$
can therefore be read off directly.
The valley derivatives in Eqs.~\eqref{eq:Q4_LLG_v1} and
\eqref{eq:Q4_LLG_v2} are the gauge-covariant derivatives dual to the
one-form $\mathcal D\varphi$. 

We next evaluate the three energy derivatives in
Eqs.~\eqref{eq:Q4_LLG_N}--\eqref{eq:Q4_LLG_v2}.
The energy in Eq.~\eqref{eq:collective_energy_components} is
\begin{align}
\mathcal E
={}&
h^zN^z+\frac{\beta\Delta}{2}(N^z)^2
-
c_0\sin\theta
\cos\frac{\Theta}{2}
\cos(\varphi+\mu t).
\label{eq:Q4_explicit_energy}
\end{align}

The derivative with respect to the physical layer pseudospin denotes its
tangential gradient on the unit sphere.  Explicitly,
\begin{align}
\frac{\partial\mathcal E}{\partial\bm N}
&\equiv
\bm e_\theta\frac{\partial\mathcal E}{\partial\theta}
+\bm e_\varphi
\frac{1}{\sin\theta}
\frac{\partial\mathcal E}{\partial\varphi}
\nonumber\\
&=
-\Biggl[
\sin\theta
\left(h^z+\beta\Delta N^z\right)
+c_0N^z\cos\frac{\Theta}{2}
\cos(\varphi+\mu t)
\Biggr]\bm e_\theta 
+c_0\cos\frac{\Theta}{2}
\sin(\varphi+\mu t)\,\bm e_\varphi.
\label{eq:Q4_energy_derivative_N}
\end{align}

For the valley derivatives, since
$d\bm v_i\cdot\bm v_i=0$, the corresponding coefficients are defined only
up to terms parallel to $\bm v_i$.  Such terms also vanish inside the
cross products in Eqs.~\eqref{eq:Q4_LLG_v1} and \eqref{eq:Q4_LLG_v2}.
We therefore use the simplest representatives.  Using
\begin{align*}
d\cos\frac{\Theta}{2}
&=
\frac{\bm v_2\cdot d\bm v_1+\bm v_1\cdot d\bm v_2}
{4\cos(\Theta/2)}
\end{align*}
gives
\begin{align}
\mathcal D_{\bm v_1}\mathcal E
&=
-\frac{
c_0\sin\theta\cos(\varphi+\mu t)}
{4\cos(\Theta/2)}
\bm v_2 
-c_0\sin\theta
\cos\frac{\Theta}{2}
\sin(\varphi+\mu t)\,\bm a,
\label{eq:Q4_energy_derivative_v1}
\\
\mathcal D_{\bm v_2}\mathcal E
&=
-\frac{
c_0\sin\theta\cos(\varphi+\mu t)}
{4\cos(\Theta/2)}
\bm v_1 
-c_0\sin\theta
\cos\frac{\Theta}{2}
\sin(\varphi+\mu t)\,\bm a.
\label{eq:Q4_energy_derivative_v2}
\end{align}

Substituting Eqs.~\eqref{eq:Q4_energy_derivative_N},
\eqref{eq:Q4_energy_derivative_v1}, and
\eqref{eq:Q4_energy_derivative_v2} into the three LLG equations gives
\begin{align}
\mathcal D_t\bm N
={}&
-\bm N\times\Bigl\{
\Gamma\mathcal D_t\bm N
-2\Biggl[
\sin\theta
\left(h^z+\beta\Delta N^z\right)
+c_0N^z\cos\frac{\Theta}{2}
\cos(\varphi+\mu t)
\Biggr]\bm e_\theta
\nonumber\\
&\qquad\qquad
+2c_0\cos\frac{\Theta}{2}
\sin(\varphi+\mu t)\,\bm e_\varphi
\Bigr\},
\label{eq:Q4_explicit_LLG_N}
\\
\frac{d\bm v_1}{dt}
={}&
-\bm v_1\times\Biggl\{
\Gamma\frac{d\bm v_1}{dt}
-\frac{
c_0\sin\theta\cos(\varphi+\mu t)}
{\cos(\Theta/2)(1+N^z)}
\bm v_2 
-\frac{
4c_0\sin\theta\cos(\Theta/2)
\sin(\varphi+\mu t)}
{1+N^z}\bm a
\Biggr\},
\label{eq:Q4_explicit_LLG_v1}
\\
\frac{d\bm v_2}{dt}
={}&
-\bm v_2\times\Biggl\{
\Gamma\frac{d\bm v_2}{dt}
-\frac{
c_0\sin\theta\cos(\varphi+\mu t)}
{\cos(\Theta/2)(1-N^z)}
\bm v_1 
-\frac{
4c_0\sin\theta\cos(\Theta/2)
\sin(\varphi+\mu t)}
{1-N^z}\bm a
\Biggr\}.
\label{eq:Q4_explicit_LLG_v2}
\end{align}
For numerical calculations, the first of these equations can equivalently
be written in the local coordinates
$(d\theta,\sin\theta\,\mathcal D\varphi)$ as
\begin{equation}
\begin{aligned}
\frac{d\theta}{dt}
&=
2c_0
\cos\frac{\Theta}{2}
\sin(\varphi+\mu t)
 +\Gamma\sin\theta
\left(\frac{d\varphi}{dt}+a_t\right), \\
\sin\theta\left(\frac{d\varphi}{dt}+a_t\right)
&=
2\sin\theta\left(
h^z+\beta\Delta\cos\theta
\right)
+2c_0\cos\theta\cos\frac{\Theta}{2}
\cos(\varphi+\mu t)
-\Gamma\frac{d\theta}{dt}.
\end{aligned}
\label{eq:Q4_explicit_layer_scalar_eom}
\end{equation}

The layer equations in \equa{eq:Q4_explicit_layer_scalar_eom}
depend on the valley dynamics only through $\Theta$ and $a_t$.  These two
quantities can be extracted directly from
\equa{eq:Q4_explicit_LLG_v1} and \equa{eq:Q4_explicit_LLG_v2}.  For brevity in the
following projection, define
\begin{align*}
u_1&=
\frac{c_0\sin\theta\cos(\varphi+\mu t)}
{\cos(\Theta/2)(1+N^z)},
&
u_2&=
\frac{c_0\sin\theta\cos(\varphi+\mu t)}
{\cos(\Theta/2)(1-N^z)},
\\
w_1&=
\frac{4c_0\sin\theta\cos(\Theta/2)\sin(\varphi+\mu t)}
{1+N^z},
&
w_2&=
\frac{4c_0\sin\theta\cos(\Theta/2)\sin(\varphi+\mu t)}
{1-N^z}.
\end{align*}
The two valley equations then take the common form
\begin{align*}
\frac{d\bm v_i}{dt}
=-\bm v_i\times
\left(\Gamma\frac{d\bm v_i}{dt}-u_i\bm v_{\bar i}-w_i\bm a\right),
\qquad
\bar 1=2,\quad \bar 2=1.
\end{align*}
Inverting the Gilbert term gives
\begin{align*}
\frac{d\bm v_i}{dt}
=\frac{1}{1+\Gamma^2}
\Bigl\{
\bm v_i\times\left(u_i\bm v_{\bar i}+w_i\bm a\right)
-\Gamma\bm v_i\times
\left[
\bm v_i\times\left(u_i\bm v_{\bar i}+w_i\bm a\right)
\right]
\Bigr\}.
\end{align*}
Since $\cos\Theta=\bm v_1\cdot\bm v_2$, the two required projections are
\begin{align*}
-\sin\Theta\,\frac{d\Theta}{dt}
&=
\frac{d\bm v_1}{dt}\cdot\bm v_2
+\bm v_1\cdot\frac{d\bm v_2}{dt},
&
a_t&=\bm a\cdot
\left(\frac{d\bm v_1}{dt}+\frac{d\bm v_2}{dt}\right).
\end{align*}
Using
\begin{align*}
\bm a\cdot\bm v_1
&=\bm a\cdot\bm v_2=0,
&
\bm a\cdot(\bm v_1\times\bm v_2)
&=\sin^2\frac{\Theta}{2},
&
\bm a^2&=\frac14\tan^2\frac{\Theta}{2},
\end{align*}
one obtains, before substituting the coefficients,
\begin{align}
-\sin\Theta\,\frac{d\Theta}{dt}
&=
\frac{1}{1+\Gamma^2}
\left[
\sin^2\frac{\Theta}{2}(w_2-w_1)
+\Gamma\sin^2\Theta\,(u_1+u_2)
\right],
\\
a_t
&=
\frac{1}{1+\Gamma^2}
\left[
\sin^2\frac{\Theta}{2}(u_1-u_2)
+\Gamma\bm a^2(w_1+w_2)
\right].
\label{Theta_dyn}
\end{align}
Here the combinations of coefficients reduce to
\begin{align*}
w_2-w_1
&=
\frac{8c_0N^z\cos(\Theta/2)}
{\sin\theta}\sin(\varphi+\mu t),
&
u_1+u_2
&=
\frac{2c_0}
{\sin\theta\cos(\Theta/2)}
\cos(\varphi+\mu t),
\\
u_1-u_2
&=
-\frac{2c_0N^z}
{\sin\theta\cos(\Theta/2)}
\cos(\varphi+\mu t),
&
w_1+w_2
&=
\frac{8c_0\cos(\Theta/2)}
{\sin\theta}\sin(\varphi+\mu t).
\end{align*}
Therefore, the valley equations needed to close
\equa{eq:Q4_explicit_layer_scalar_eom} are
\begin{equation}
\begin{aligned}
\frac{d\Theta}{dt}
&=
-\frac{4c_0}{1+\Gamma^2}
\frac{\sin(\Theta/2)}{\sin\theta}
\left[
N^z\sin(\varphi+\mu t)
+\Gamma\cos(\varphi+\mu t)
\right], \\
a_t
&= \frac{2c_0}{1+\Gamma^2} \frac{\sin(\Theta/2)}{\sin\theta} \tan \frac{\Theta}{2}
\left[
-N^z\cos(\varphi+\mu t)
+\Gamma\sin(\varphi+\mu t)
\right].
\end{aligned}
\label{eq:Q4_reduced_valley_scalars}
\end{equation}
Transforming \equa{eq:Q4_LLG_N} into the lab frame and combining \equa{eq:Q4_reduced_valley_scalars}, one obtains a complete set of EoMs for the system. 

\section{Solution of the LLG equation of N.}
In this section, we solve \equa{eq:Q4_explicit_layer_scalar_eom} under the condition $h^z = 0$, or equivalently $E_{\rm G}d = \mu$. \equa{LLG_Q2} in the $Q=2$ case can be seen as a special case of \equa{eq:Q4_explicit_layer_scalar_eom} with $a_t = 0$ and $\Theta = 0$. Define
\begin{align}
c_{\rm eff}(t)
&\equiv
c_0\cos\frac{\Theta(t)}{2}, \ \ \ 
\widetilde{\varphi}(t)
\equiv
\varphi(t)+\int^t dt'\,a_{t},
\end{align}
so that
$\frac{d\widetilde{\varphi}}{dt}
=\frac{d\varphi}{dt}+a_t$. In this section, we solve the dynamics of $\bm{N}$ in the weak-tunneling regime, assuming that $c_0$ is sufficiently small that all relevant dimensionless expansion parameters proportional to $c_0$ are much smaller than unity.

\subsection{Dynamical state}
In this section, we solve the dynamical state with $\bm{N}$ undergoing small oscillation, which means $\bm{n}$ precessing around the equator. For weak tunneling,
we set $\theta=\pi/2+\delta\theta$ with
$\delta\theta=\mathcal{O}(c_0)$ and $\varphi = \phi_0 + \delta \phi $, $\delta\phi \equiv \phi-\mu t - \phi_0 = \mathcal{O}(c_0)$. Moreover, according to \equa{eq:Q4_reduced_valley_scalars}, $a_t$ also belongs to $\mathcal{O}(c_0)$. Therefore, keeping the terms in the linear order of $c_0$ in Eq.~\eqref{eq:Q4_explicit_layer_scalar_eom} gives
\begin{equation}
\begin{aligned}
\frac{d\,\delta\theta}{dt}
&=
\Gamma\frac{d\widetilde{\varphi}}{dt}
+2c_{\rm eff}\sin(\phi_0 +\mu t),
\\
\frac{d\widetilde{\varphi}}{dt}
&= -2\beta\Delta\,\delta\theta
-\Gamma\frac{d\,\delta\theta}{dt}.
\label{lineared_theta_phi}
\end{aligned}
\end{equation}
The term
$2c_{\rm eff}\cos\theta\cos(\varphi+\mu t)$ in the second equation of
Eq.~\eqref{eq:Q4_explicit_layer_scalar_eom} is of order $c_0^2$ and has
therefore been neglected.  Eliminating
$\frac{d\,\delta\theta}{dt}$ gives
\begin{align}
\delta\theta
=
-\frac{1}{2\beta\Delta}
\left[
(1+\Gamma^2)\frac{d\widetilde{\varphi}}{dt}
+2\Gamma c_{\rm eff}\sin(\phi_0+\mu t)
\right].
\end{align}
Taking a time derivative and using the first equation in \equa{lineared_theta_phi} again,
we obtain
\begin{align}
&\frac{1+\Gamma^2}{2\beta\Delta}
\frac{d^2\widetilde{\varphi}}{dt^2}
+\Gamma\frac{d\widetilde{\varphi}}{dt}
+2c_{\rm eff}\sin(\phi_0+\mu t) 
+\frac{\Gamma}{\beta\Delta}
\left[
\frac{dc_{\rm eff}}{dt}\sin(\phi_0+\mu t)
+c_{\rm eff} \mu 
\cos(\phi_0+\mu t)
\right]
=0.
\label{eq:Q4_covariant_phase_general}
\end{align}
For weak tunneling, $d\Theta/dt = \mathcal{O}(c_0)$. Thus,
$d c_{\rm eff}/dt=\mathcal{O}(c_0^2)$ and is neglected below.
We then obtain the linearized equation for $\widetilde{\varphi}$:
\begin{align}
M\frac{d^2\widetilde{\varphi}}{dt^2}
+\Gamma\frac{d\widetilde{\varphi}}{dt}
+2c_{\rm eff}
\sqrt{1+\left(\frac{\Gamma \mu}{2\beta \Delta}\right)^2}
\sin(\phi_0 + \mu t + \varphi_1) =0.
\label{eq:Q4_covariant_RCSJ}
\end{align}
Here $M=\frac{1+\Gamma^2}{2\beta\Delta}, c_{\rm eff}=c_0\cos\frac{\Theta_0}{2}$. $\varphi_1 = \arctan\left(\frac{\Gamma \mu}{2\beta \Delta}\right)$ is a constant phase shift. The solution of the equation is given by
\begin{align}
	\widetilde{\varphi}(t) - \phi_0 &= 2c_{\rm eff} \sqrt{1+(\frac{\Gamma \mu}{2\beta \Delta})^2} \frac{1}{\mu} \frac{1}{\sqrt{\Gamma^2 + \mu^2 M^2}} \sin(\phi_0 + \mu t + \varphi_1 + \frac{\pi}{2} - \varphi_2) \nonum
	&= \frac{2c_{\rm eff}\Gamma\bigl[4\beta^2\Delta^2+\mu^2(1+\Gamma^2)\bigr]}
	{\mu\bigl[4\beta^2\Delta^2\Gamma^2+\mu^2(1+\Gamma^2)^2\bigr]}
	\cos(\phi_0+\mu t)
	+\frac{4c_{\rm eff}\beta\Delta}
	{4\beta^2\Delta^2\Gamma^2+\mu^2(1+\Gamma^2)^2}
	\sin(\phi_0+\mu t); \nonum
	\delta \theta(t) &= \frac{2c_{\rm eff}}{2\beta \Delta \Gamma } \cos \varphi_2 \sin(\phi_0 + \mu t - \varphi_2) \nonum
	&= -\frac{2c_{\rm eff}\mu(1+\Gamma^2)}
	{4\beta^2\Delta^2\Gamma^2+\mu^2(1+\Gamma^2)^2}
	\cos(\phi_0+\mu t)
	+\frac{4c_{\rm eff}\beta\Delta\Gamma}
	{4\beta^2\Delta^2\Gamma^2+\mu^2(1+\Gamma^2)^2}
	\sin(\phi_0+\mu t).  
	\label{LLG_Q2_sol}
\end{align}
Here $\varphi_2 = \arctan\left(\frac{\mu M}{\Gamma}\right) = \arctan\left(\frac{\mu(1+\Gamma^2)}{2\beta \Delta \Gamma}\right)$ is a constant phase shift. 

In the $Q=2$ case one has $c_{\rm eff} = c_0$ and $a_t = 0$. According to \appe{Appe:current}, the interlayer-current density reads
\begin{align}	
    I(t) = I_{\rm J} \Big [\sin(\phi_0 + \mu t) + \cos(\phi_0 + \mu t)\delta\phi + \mathcal{O}(c_0^2)\Big ].
	\label{j_dynamical}
\end{align}
Here the leading term gives an ac Josephson current oscillating with a frequency $\mu$. The time-averaged dc Josephson current comes from the second term, which reads 
\begin{align}
	I_{\rm dc} \equiv \langle I(t) \rangle_t
	&=  I_{\rm J}\frac{c_0\Gamma}{\mu}
	\frac{4\beta^2\Delta^2+(1+\Gamma^2)\mu^2}
	{4\Gamma^2\beta^2\Delta^2+(1+\Gamma^2)^2\mu^2}.
	\label{j_dc}
\end{align}
For the $Q=4$ case, we will show in the next section that the tunneling current in the dynamical state is not obtained by the solution in \equa{LLG_Q2_sol} since $\Theta$ is always driven toward $\pi$. 

\subsection{Static solution.}
In this section, we solve the stationary state with $\bm{n}$ being static, which means $\bm{N}$ precessing. Setting $d\bm{n}/dt = 0$, or equivalently $d\phi/dt = d(\varphi+\mu t)/dt = 0$ and $d\theta/dt = 0$, \equa{eq:Q4_explicit_layer_scalar_eom} becomes
\begin{align}
0 & =
2c_{\rm eff} \sin \phi
 -\Gamma\sin\theta \mu_{\rm eff},
\nonumber\\
-\sin\theta \mu_{\rm eff}
&=
2\sin\theta\left(
h^z+\beta\Delta\cos\theta
\right)
+2 \cos\theta c_{\rm eff}
\cos \phi.
\label{eq:Q4_explicit_layer_scalar_static}
\end{align}
Here $\mu_{\rm eff} \equiv \mu - a_t$, $c_{\rm eff} \equiv c_0 \cos(\Theta/2)$. In the valleyless case, $\mu_{\rm eff} = \mu$ and $c_{\rm eff} = c_0$. Keeping terms to linear order in $c_0$ in both equations, we obtain the solution for $h^z = 0$
\begin{align}
	\mu_{\rm eff} = \frac{2c_{\rm eff}}{\Gamma} \sin \phi = -2\beta \Delta \cos\theta.
	\label{eq:Q4_static_leading_solution}
\end{align}
Thus, one solution is
$\phi=\arcsin(\mu_{\rm eff}/\mu_{\rm eff,c})$, where
$\mu_{\rm eff,c}=2c_{\rm eff}/\Gamma$, and the stationary solution exists
only for $|\mu_{\rm eff}|\leq\mu_{\rm eff,c}$. 

There is another solution with $\phi=\pi - \arcsin(\mu_{\rm eff}/\mu_{\rm eff,c})$, which is unstable. We prove this through writing \equa{eq:Q4_explicit_layer_scalar_eom} explicitly in the lab frame with $h^z = 0$.  Using
$d\varphi/dt+a_t=d\phi/dt-\mu_{\rm eff}$ gives
\begin{align}
(1+\Gamma^2)\frac{d\theta}{dt}
={}&
2c_{\rm eff}\sin\phi
+2\Gamma\left[
\sin\theta \beta\Delta\cos\theta 
+c_{\rm eff}\cos\theta\cos\phi
\right],
\\
(1+\Gamma^2)\sin\theta\frac{d\phi}{dt}
={}&
(1+\Gamma^2)\sin\theta\,\mu_{\rm eff}
+2\sin\theta \beta\Delta\cos\theta 
+2c_{\rm eff}\cos\theta\cos\phi
-2\Gamma c_{\rm eff}\sin\phi .
\end{align}
We now write
$\theta=\theta_*+\delta\theta$ and
$\phi=\phi_*+\delta\phi$ with $(\theta_*, \phi_*)$ being a static solution. Here 
$\cos\theta_*=-\mu_{\rm eff}/(2\beta\Delta)=\mathcal{O}(c_{\rm eff}/\beta\Delta)$.
Consequently, to the same order used above, one may set
$\sin\theta_*=1$ in the following stability matrix.  Linearization then yields
\begin{align}
\frac{d}{dt}
\begin{pmatrix}
\delta\theta\\
\delta\phi
\end{pmatrix}
=
\frac{2}{1+\Gamma^2}
\begin{pmatrix}
-\Gamma\left(\beta\Delta+c_{\rm eff}\cos\phi_*\right)
&
c_{\rm eff}\cos\phi_*
\\
-\left(\beta\Delta+c_{\rm eff}\cos\phi_*\right)
&
-\Gamma c_{\rm eff}\cos\phi_*
\end{pmatrix}
\begin{pmatrix}
\delta\theta\\
\delta\phi
\end{pmatrix}.
\end{align}
Denoting this matrix by $J$, its determinant and trace are
\begin{align*}
\det J
&=
\frac{
4c_{\rm eff}\cos\phi_*
\left(\beta\Delta+c_{\rm eff}\cos\phi_*\right)
}{1+\Gamma^2},  \ \ \ 
\operatorname{Tr}J
=
-\frac{2\Gamma}{1+\Gamma^2}
\left(\beta\Delta+2c_{\rm eff}\cos\phi_*\right).
\end{align*}
The stability requires
$\det J>0$ and $\operatorname{Tr}J<0$, which gives
$\cos\phi_*>0$, i.e., $-\pi/2 < \phi_* < \pi/2$. Therefore, only the branch $\phi_*=\arcsin(\mu_{\rm eff}/\mu_{\rm eff,c})$ is stable against layer-pseudospin perturbations.

For the parallel state with $\Theta = 0$, the analysis above gives a critical bias $\mu_{\rm c}=2c_0/\Gamma$ for the static solution. However, the instability associated with the valley dynamics can also give a critical bias, which turns out to be smaller than $\mu_{\rm c}$ as shown in the numerical results in the main text. To explain this, we write the following linearized dynamical equation for $\Theta$ near $\Theta = 0$ using \equa{eq:Q4_reduced_valley_scalars}:
\begin{equation}
 \dot\Theta=-\frac{2c_0}{(1+\Gamma^2)\sin\theta}
 \left(\Gamma\cos\phi + n^z\sin\phi \right)\Theta  \equiv -
 \kappa_{\rm v} \Theta,
 \label{eq:vai_linear_rate}
\end{equation}
where $n^z$ and $\phi$ are evaluated on the static branch. The branch is
stable against this perturbation for $\kappa_{\rm v}>0$ and unstable for
$\kappa_{\rm v}<0$. The instability boundary therefore satisfies
\begin{equation}
 \Gamma\cos\phi + n^z\sin\phi =0.
 \label{eq:vai_boundary}
\end{equation}
This condition can be reached while $\cos\phi>0$. Using the approximated static solution in \equa{eq:Q4_static_leading_solution} with $c_{\rm eff} = c_0$, we obtain the threshold $\mu_{\rm v}$ from
\begin{equation}
 \Gamma\sqrt{1-\frac{\mu_{\rm v}^2}{\mu_{\rm c}^2}}
 =\frac{\mu_{\rm v}^2}{2\beta\Delta\,\mu_{\rm c}}.
 \label{eq:vai_threshold_condition}
\end{equation}
Within this approximation, $\mu_{\rm v}<\mu_{\rm c}$, consistent with the numerical results for the parameters used in the main text. 

\section{Ponderomotive force for $\bar{\Theta}$.}
In the dynamical state, the precession of $\phi = \varphi + \mu t$ generates an averaged force acting on the slow, cycle-averaged relative valley angle $\bar{\Theta}\equiv\langle\Theta\rangle_t$. Combining \equa{eq:Q4_reduced_valley_scalars} and the solution in
\equa{LLG_Q2_sol}, one obtains the ponderomotive force to second order in $c_0$:
\begin{align}
	\frac{1}{\Gamma}\frac{d\bar{\Theta}}{dt}
	\equiv \frac{1}{\Gamma}\left\langle \frac{d\Theta}{dt} \right\rangle_t 
	&\approx -\frac{4c_0}{1+\Gamma^2} \sin\frac{\bar{\Theta}}{2} \Big(  -\frac{1}{\Gamma} \langle \delta\theta(t) \sin(\varphi+\mu t) \rangle_t +  \langle \cos(\varphi+\mu t) \rangle_t  \Big) \nonum
	&\approx -\frac{4c_0}{1+\Gamma^2} \sin\frac{\bar{\Theta}}{2} \Big(  -\frac{1}{\Gamma} \langle \delta\theta(t) \sin(\phi_0 + \mu t) \rangle_t + \langle \cos(\widetilde{\varphi} - \int^t dt'\,a_{t} +\mu t) \rangle_t \Big) \nonum
	&= -\frac{4c_0}{1+\Gamma^2} \sin\frac{\bar{\Theta}}{2} \cos\frac{\bar{\Theta}}{2} \Big( -\frac{c_0}{2\beta \Delta \Gamma} \frac{1}{\Gamma} \cos^2 \varphi_2 + \frac{c_0}{ \Gamma \mu }  \frac{\cos \varphi_2}{\cos \varphi_1} \sin(\varphi_1  - \varphi_2) \Big) \nonum
	&= \frac{2c_0^2}{1+\Gamma^2} \frac{\cos^2 \varphi_2}{\beta \Delta \Gamma^2}  \sin\bar{\Theta}   \equiv -\frac{8}{1+\Gamma^2}\frac{\partial V_{\rm eff}}{\partial \bar{\Theta}}.
	\label{P-force}
\end{align}
At first order in $c_0$, \equa{eq:Q4_reduced_valley_scalars} gives $\Theta-\bar\Theta\propto c_0\sin(\phi_0+\mu t)$, so that $\langle(\Theta-\bar\Theta)\cos(\phi_0+\mu t)\rangle_t=0$. This vanishing average allows us to neglect the oscillatory part of $\Theta$ in the first approximate equality of \equa{P-force}, where the prefactor $\sin(\Theta/2)$ is replaced by $\sin(\bar\Theta/2)$. Here $\int^t dt'\,a_{t}$ only has a $\cos(\phi_0 + \mu t)$-component to the first order in $c_0$ according to \equa{eq:Q4_reduced_valley_scalars}. Therefore, it does not contribute to the time-average $\langle \cos(\widetilde{\varphi} - \int^t dt'\,a_{t} +\mu t) \rangle_t$ to the first order in $c_0$. From \equa{P-force}, one obtains the ponderomotive potential 
\begin{align}
	V_{\rm eff}(\bar{\Theta}) = c_0^2 \frac{\cos^2 \varphi_2}{4\beta \Delta \Gamma^2}  (\cos\bar{\Theta}+1) = c_0^2 \frac{\beta \Delta}{(2\beta \Delta \Gamma)^2 + \mu^2 (1+\Gamma^2)^2 } (\cos\bar{\Theta}+1) = \langle E_{\rm t} \rangle_t.
	\label{P-force-expr}
\end{align}
In Fig.~2 shown in the main-text end matter, we plot the numerical evolution of $\Theta$ obtained by numerically solving the coupled equations in \equa{eq:Q4_explicit_layer_scalar_eom} and \equa{eq:Q4_reduced_valley_scalars}. The slow dynamics are well described by the \equa{P-force} obtained above.

\section{The steady state as $\Theta \to \pi$.}
As $\Theta \to \pi$, the factor $\tan(\Theta/2)$ in
\equa{eq:Q4_reduced_valley_scalars} makes $a_t$ large and invalidates the
perturbative precessing solution and leads to a static state.  To determine the resulting steady state,
we write \equa{eq:Q4_LLG_N} in the lab frame as
\begin{align}
	\frac{d{\bm n}}{dt}-\mu_{\rm eff}\hat{\bm z}\times\bm n = -\bm{n} \times \left[ \Gamma (\frac{d{\bm n}}{dt} -\mu_{\rm eff} \hat{\bm z}\times\bm n) + 2\frac{\partial \mathcal E}{\partial\bm n} \right].
	\label{LLG_eff}
\end{align}
Here $\mu_{\rm eff}\equiv\mu-a_t$.  For $d\Theta/dt=0$, \equa{eq:Q4_reduced_valley_scalars} gives the exact
constraint
\begin{align}
	\cos\theta\sin\phi+\Gamma\cos\phi=0.
	\label{eq:Q4_static_theta_constraint}
\end{align}
Using the leading-order static relation in Eq.~\eqref{eq:Q4_static_leading_solution}, together with Eq.~\eqref{eq:Q4_reduced_valley_scalars}, we obtain
\begin{align}
	\mu - \frac{2c_0}{1+\Gamma^2} \frac{\sin(\Theta/2)}{\sin\theta} \tan \frac{\Theta}{2}
\left(
-\cos\theta \cos\phi
+\Gamma\sin \phi \right) &= -2\beta \Delta \cos\theta =      \frac{2c_0 \cos(\Theta/2)}{\Gamma} \sin\phi .
	\label{eq:Q4_static_three_angles}
\end{align}
Equations~\eqref{eq:Q4_static_theta_constraint} and
\eqref{eq:Q4_static_three_angles} jointly determine the values of the three
angles $\theta$, $\phi$, and $\Theta$.
For either sign of the nonzero bias, we take $c_0\to0$ at fixed $\mu$,
$\Gamma$, and $\beta\Delta$, and write
\begin{align}
	\Theta=\pi-\epsilon,
	\qquad
	\theta=\frac{\pi}{2}+\delta\theta,
	\qquad
	\phi=\operatorname{sgn}(\mu)\frac{\pi}{2}+\delta\phi,
	\qquad \epsilon\ll1.
\end{align}
The last equality in Eq.~\eqref{eq:Q4_static_three_angles}, together with
Eq.~\eqref{eq:Q4_static_theta_constraint}, gives to leading order
\begin{align}
	\delta\theta
	&\simeq \operatorname{sgn}(\mu)
	\frac{c_0\epsilon}{2\beta\Delta\Gamma}, \ \ \ 
	\delta\phi
	\simeq-\frac{\delta\theta}{\Gamma}.
\end{align}
Substituting these relations and
$\sin(\Theta/2)\simeq1$, $\tan(\Theta/2)\simeq2/\epsilon$ into the first
equality in Eq.~\eqref{eq:Q4_static_three_angles} yields
\begin{align}
	\mu
	&=\operatorname{sgn}(\mu)
	\frac{4c_0\Gamma}{(1+\Gamma^2)\epsilon}
	+ \mathcal{O}(c_0\epsilon), \ \ \ 
	\epsilon
	=\frac{4c_0\Gamma}{|\mu|(1+\Gamma^2)}
	\left[1+\mathcal O\!\left(\frac{c_0^2}{\mu^2}\right)\right].
	\label{eq:Q4_blockade_epsilon_leading}
\end{align}
Consequently,
\begin{align}
	\delta\theta
	&\simeq
	\operatorname{sgn}(\mu)
	\frac{2c_0^2}{\beta\Delta|\mu|(1+\Gamma^2)}, \ \ \
	\delta\phi
	\simeq
	-\operatorname{sgn}(\mu)
	\frac{2c_0^2}{\beta\Delta|\mu|\Gamma(1+\Gamma^2)}, \ \ \ 
	\mu_{\rm eff}=\mu-a_t
	\simeq \operatorname{sgn}(\mu)\frac{c_0\epsilon}{\Gamma}
	=\frac{4c_0^2}{\mu(1+\Gamma^2)}.
\end{align}
Thus $\theta$ and $\phi$ receive no correction at linear order in $c_0$,
while $\Theta$ does.  Therefore, to leading order,
$a_t\simeq\mu$, $\phi\simeq\operatorname{sgn}(\mu)\pi/2$, and
$\Theta=\pi-4c_0\Gamma/[|\mu|(1+\Gamma^2)]$, which is precisely the
solution quoted in the main text. This solution is consistent with the numerical results shown in Fig.~2 of the main-text end matter.

\section{Amplitude fluctuations.}
We derive the dynamics of the amplitude fluctuations for $Q=2$ and $Q=4$ together. Using $P\equiv|\Psi\rangle\langle\Psi|$, we define
\begin{align}
	L\equiv QP-1=(Q-1)P-\bar P,\qquad \bar P\equiv1-P,
	\label{eq:amplitude_generator}
\end{align}
so that the order-parameter matrix and its amplitude variations are
\begin{align}
	\mathcal M_c(t)
	={}&[\Delta+\delta\Delta(t)]L,
	&
	\delta\mathcal M_c(t)
	={}&\delta\Delta(t)L,
	&
	\mathcal M_q(t)
	={}&\Delta_q(t)L .
	\label{eq:amplitude_variations}
\end{align}
At quadratic order in fluctuations about the mean-field saddle point, the longitudinal and tangent sectors decouple. To see this, note that $L=(Q-1)P-\bar P$ and the Green's functions in Eq.~\eqref{G_plus_minus} are block diagonal in the $P$ and $\bar P$ subspaces, whereas the tangent matrices $X$ and $Y$ have only off-diagonal blocks in the $P$ and $\bar P$ subspaces. For any tangent matrix $X=PX\bar P+\bar PXP$ (and likewise for $Y$), the mixed longitudinal-tangent response and trace vanish:
\begin{align}
\mathcal P_\omega^R(L,X)
=\mathcal P_\omega^R(X,L)
&=0,\qquad
\operatorname{tr}(LX)=0.
\label{eq:longitudinal_tangent_decoupling}
\end{align}
Thus, the amplitude and directional fluctuations decouple at quadratic order in the effective action, and their linearized dynamics can be treated independently. In the following, we analyze the amplitude fluctuation and its contribution to the interlayer tunneling current. 

Applying similar decomposition as Eqs.~\eqref{S_eff_parts}-\eqref{S_e}, we keep
the terms that are linear in $\Delta_q$ and first order in either
$\delta\Delta$ or the external field.  This part of the effective action can
be separated into a dynamical part and an energy part:
\begin{align}
	iS_{\rm eff}^{\Delta}
	\simeq
	iS_{\rm dyn}^{\Delta}
	-iS_E^{\Delta}.
	\label{eq:amplitude_action_parts}
\end{align}
The dynamical part contains the fermionic bubble and the
Hubbard--Stratonovich term,
\begin{align}
	iS_{\rm dyn}^{\Delta}
	\equiv{}&
	-\operatorname{Tr}
	\left[
	\mathcal G(\kappa_0\otimes\delta\Delta L)
	\mathcal G(\kappa_x\otimes\Delta_q L)
	\right]
	-i\frac{N(Q-1)}{g}\int dt\,
	\Delta_q(t)\delta\Delta(t).
	\label{eq:amplitude_dynamic_action}
\end{align}
Here the factor $Q-1$ in the second term follows from
$\operatorname{Tr}(L^2)/Q=Q-1$.  The energy part contains the response to
the external field
\begin{align}
	iS_E^{\Delta}
	\equiv{}&
	-\operatorname{Tr}
	\left[
	\mathcal G(\kappa_0\otimes\hat h)
	\mathcal G(\kappa_x\otimes\Delta_q L)
	\right] = -\operatorname{Tr}
	\left[
	\mathcal G(\kappa_0\otimes h_{\Delta,Q} L)
	\mathcal G(\kappa_x\otimes\Delta_q L)
	\right].
	\label{eq:amplitude_energy_action}
\end{align}
Here $\hat h$ is the external-field matrix defined in Eq.~\eqref{eq:GRA_0}. $h_{\Delta,Q}$ represents the projection coefficient of $\hat h$ onto the subspace spanned by $L$, whose explicit form in real time is given by
\begin{align}
	h_{\Delta,Q}(t)
	&\equiv
	\frac{\operatorname{Tr}[L\hat h(t)]}
	{\operatorname{Tr}(L^2)}
	=
	\frac{\operatorname{Tr}[P\hat h(t)]}{Q-1}
	=
	\frac{E_h(t)+E_{\rm t}(t)}{Q-1} .
\end{align}  Both the first term in \equa{eq:amplitude_dynamic_action} and \equa{eq:amplitude_energy_action} can be expressed by the retarded response function $\mathcal P_\omega^R$, evaluated for the amplitude channel as
\begin{align}
	\mathcal P_{\omega}^{R} (L,  L) \equiv 
	\sum_{\bm k}&\int\frac{d\epsilon}{2\pi}
	i  \Big\{
	(Q-1)^2
	\left[
	\mathcal G_+^R(\epsilon+\omega)\mathcal G_+^K(\epsilon)
	+\mathcal G_+^K(\epsilon+\omega)\mathcal G_+^A(\epsilon)
	\right]
	\nonum
	&+(Q-1)
	\left[
	\mathcal G_-^R(\epsilon+\omega)\mathcal G_-^K(\epsilon)
	+\mathcal G_-^K(\epsilon+\omega)\mathcal G_-^A(\epsilon)
	\right]
	\Big\} \nonum
	\equiv 2N(Q-1) & \Pi^R_{\Delta, Q}(\omega) .
	\label{eq:amplitude_projector_bubble}
\end{align}

Here the factors $(Q-1)^2$ and $(Q-1)$ arise, respectively, from the
eigenvalues of $L$ in the rank-one $P$ sector and from the degeneracy of the
$\bar P$ sector. The integrals over $\epsilon$ are analyzed in Appendix~\ref{app:green_integrals_intra}. Based on the results in Appendix~\ref{app:green_integrals_intra}, the response function $\Pi^R_{\Delta,Q}(\omega)$ has a low-frequency form
\begin{align}
\Pi_{\Delta,Q}^R(\omega) \approx \Pi_{\Delta,Q}^R(0) + \alpha_Q(\Delta) \frac{i\omega}{2\eta}.
	\label{eq:amplitude_projector_bubble_low_freq}
\end{align}
Here $\alpha_Q(\Delta)$ is given in \equa{alphaQ_def}, and
\begin{align}
	\Pi_{\Delta,Q}^R(0)
    &=\frac{1}{2}\int\frac{d\nu}{2\pi}\rho(\nu)
    \left[(Q-1)f_\eta'(\nu-(Q-1)\Delta)+f_\eta'(\nu+\Delta)\right]
    =\frac{\partial D_Q(\Delta)}{\partial\Delta},
	\label{Pi_Delta_0}
\end{align}
where $D_Q(\Delta)$ is defined in \equa{eq:gap}. 
Transforming back to real time, with $i\omega\to-\partial_t$, we obtain 
\begin{align}
 iS_{\rm dyn}^{\Delta}
 &\approx
 -i2N(Q-1)\int dt\,\Delta_q(t)
 \left[
 \frac{\alpha_Q(\Delta)}{2\eta}\partial_t\delta\Delta(t)
 +\left(\frac{1}{2g}-\Pi_{\Delta,Q}^R(0)\right)\delta\Delta(t)
 \right],\nonumber\\
 iS_E^{\Delta}
 &\approx
 i2N(Q-1)\Pi_{\Delta,Q}^R(0)\int dt\,
 \Delta_q(t)h_{\Delta,Q}(t)
 +i\frac{\beta N}{g}\int dt\,
 [N^z(t)]^2\Delta_q(t)\delta\Delta(t).
 \label{eq:amplitude_action_evaluated}
\end{align}
Here we only consider the leading-order effect in the driving field by neglecting the contribution in the order $\omega c_0$ in $S_E^{\Delta}$, or equivalently, taking the static approximation $\Pi^R_{\Delta,Q}(\omega) \approx \Pi^R_{\Delta,Q}(0)$. Combining \equa{eq:amplitude_action_parts} and \equa{eq:amplitude_action_evaluated}, we obtain the linearized equation of motion for the amplitude fluctuation around the mean-field saddle point with amplitude $\Delta$ and $N^z = 0$: 
\begin{align}
	\frac{\alpha_Q(\Delta)}{2\eta}\partial_t\delta\Delta(t)
	+\left(\frac{1}{2g}-\Pi_{\Delta,Q}^R(0)\right)\delta\Delta(t)
	=-\Pi_{\Delta,Q}^R(0)h_{\Delta,Q}(t).
	\label{eq:amplitude_eom_general_Q}
\end{align}
Here $h_{\Delta,Q}(t) \approx -c_{\rm eff}\cos\phi(t)/(Q-1)$. Combining the gap equation~\eqref{eq:gap} with Eq.~\eqref{Pi_Delta_0}, we identify $1/(2g)-\Pi_{\Delta,Q}^R(0)$ as the amplitude stiffness, which must be positive for the mean-field ordered state to be stable. Following \equa{eq:amplitude_eom_general_Q}, $\delta \Delta$ satisfies
\begin{align}
\frac{|\delta\Delta(\omega)|}{\Delta}
&=
\frac{\Pi_{\Delta,Q}^R(0)|h_{\Delta,Q}(\omega)|}
{\Delta\sqrt{
\left[1/2g-\Pi_{\Delta,Q}^R(0)\right]^2
+\left[\omega\alpha_Q(\Delta)/2\eta\right]^2
}} \leq
\frac{\Pi_{\Delta,Q}^R(0)}
{1/2g -\Pi_{\Delta,Q}^R(0)}
\frac{|h_{\Delta,Q}(\omega)|}{\Delta}.
\label{delta_Delta_small}
\end{align}
Therefore, as long as the amplitude fluctuations are not very soft, the condition 
\begin{align*}
\frac{\Pi_{\Delta,Q}^R(0)}{1/(2g)-\Pi_{\Delta,Q}^R(0)} \lesssim \mathcal{O}(1)
\end{align*}
is satisfied, which guarantees $\delta \Delta /\Delta \ll 1$ since $c_{\rm eff} \ll \Delta$. Near amplitude-fluctuation softening, this condition is not satisfied and amplitude response can become important.

To establish the contribution to the interlayer current from amplitude fluctuations, we write the amplitude correction to the dc current as $ I^{\Delta}_{\rm dc ,Q}=c_{\rm eff}\Pi_{\Delta,Q}^R(0) \langle \delta\Delta \sin\phi \rangle_t /a^2 $, which is obtained by combining \equa{eq:jt} and \equa{Pi_Delta_0}. $ I^{\Delta}_{\rm dc , Q} $ is calculated below for two cases:

\textit{(i) Precessing state of $\phi$.} This state only exists for $Q=2$, $c_{\rm eff}=c_0$. For $\phi\simeq\phi_0+\mu t$, we solve Eq.~\eqref{eq:amplitude_eom_general_Q} with its static driving coefficient $\Pi_{\Delta,2}^R(0)$ and take the cycle average of the amplitude-induced current correction, obtaining
\begin{align}
 |I_{{\rm dc},2}^{\Delta}|
 &=I_{\rm J}\frac{c_0}{2\Delta}
 [2g\Pi_{\Delta,2}^R(0)]
 \left|\operatorname{Im}\frac{\Pi_{\Delta,2}^R(0)}
 {1/2g - \Pi_{\Delta,2}^R(0)-i\, \alpha_2(\Delta)\mu/2\eta}\right|
 \nonumber\\
 &\leq I_{\rm J}\frac{\Gamma c_0}{2|\mu|}
 \Big[ \Big( \frac{ \Pi_{\Delta,2}^R(0) }{1/2g -\Pi_{\Delta,2}^R(0)} \Big)^2
 g\alpha_2(\Delta) \Big] \frac{\mu^2}{\Gamma\eta \Delta} 
 .
 \label{eq:amplitude_dc_current_low_frequency}
\end{align}
Here we used the gap equation \equa{eq:gap}. The bound in \equa{alpha_leq_Pi} and stability imply $0\leq\alpha_2\leq2\Pi_{\Delta,2}^R(0)<1/g$, which, together with the bounded amplitude response specified above, ensures that the factor in square brackets is at most $\mathcal O(1)$. In the low-frequency overdamped regime $|\mu|\ll\min\{\eta,\Delta\}$, the amplitude-induced dc current is negligible compared with the directional contribution in \equa{j_dc} as long as $\mu^2/\Gamma \eta \Delta \ll 1$.

\textit{(ii) Static state of $\phi$.} In a static state, including the low-bias state for $Q=2$ and both low-bias and high-bias states for $Q=4$, the dc current is proportional to
$D_Q(\Delta)$ at fixed angular variables. Using
$\partial_\Delta D_Q=\Pi_{\Delta,Q}^R(0)$ and the saddle-point
relation $D_Q(\Delta)=\Delta/(2g)$, the relative amplitude
correction is
\begin{align}
 \left|\frac{I^{\Delta}_{\rm dc,Q}}{I_{\rm dc}}\right|
 \simeq
 \frac{\Pi_{\Delta,Q}^R(0)}{D_Q(\Delta)}|\delta\Delta|
 =
 2g\Pi_{\Delta,Q}^R(0)\frac{|\delta\Delta|}{\Delta}
 \ll 1.
 \label{eq:amplitude_dc_current_suppression}
\end{align}
Here $0\leq 2g\Pi_{\Delta,Q}^R(0)<1$ follows from amplitude
stability, while $|\delta\Delta|/\Delta\ll1$ follows from
Eq.~\eqref{delta_Delta_small}. 

The above analysis justifies neglecting the amplitude fluctuations in the LLG equation.

\section{Radiation emission due to the ac current.}
As shown in Eq.~\eqref{j_dynamical}, the dynamical state of the order parameter gives an ac Josephson current with frequency $\mu$. Radiation emission is a direct physical consequence of the ac interlayer current in the dynamical state and can serve as a complementary probe of this current. In the dynamical state, the initial phase $\phi_0$ in \equa{j_dynamical} could be arbitrary and forms certain spatial order $\phi_0(\bm{r})$, which determines the radiation power calculated below. The ac current distribution in the sample corresponds to a time-dependent electric dipole distribution, which can be treated as a source of electromagnetic radiation. 
\begin{align}
	p(\bm{r},t) \hat{z} =  d c_0 D_2(\Delta) \hat{z} \sum_i \delta(\bm{r} - \bm{r}_i)  \frac{-1}{\mu} \cos(\mu t + \phi_0(\bm{r}_i)).
\end{align}
In this section, we calculate the radiation emission in the far-field limit, i.e. the observer is far away from the 2D sample. Moreover, we consider the long-wavelength limit $L\ll \lambda_{\mu}$, where $L$ is the linear size of the sample and $\lambda_{\mu}$ is the wavelength corresponding to the oscillation frequency $\mu$. This limit is relevant for typical experimental parameters, such as $\mu \sim 2\pi\times 10^2\,{\rm GHz}$ and $L\sim 10^2\,\mu{\rm m}$. In this regime, the whole two-dimensional sample can be treated as a macroscopic dipole. The total dipole moment is given by 
\begin{align}
	\mathsf{p}\hat{z}  = \hat{z} \int  p(\bm{r},t) d^2 \bm{r} =  d c_0 D_2(\Delta) \frac{-1}{\mu} \hat{z} \sum_i  \cos(\mu t + \phi_0(\bm{r}_i)) \equiv  \hat{z} (\frac{1}{2} \mathsf{p}_{\mu} e^{i\mu t} + \frac{1}{2} \mathsf{p}^*_{\mu} e^{-i\mu t} ).
\end{align}
Here we have defined the Fourier component of the dipole moment $\mathsf{p}_{\mu}$ as
\begin{align}
	\mathsf{p}_{\mu} \equiv  d c_0 D_2(\Delta) \frac{-1}{\mu} \sum_i e^{i\phi_0(\bm{r}_i)}.
\end{align}
The radiation power per solid angle in the far-field limit is given by~\cite{jackson2021classical}
\begin{align}
	\frac{d P}{d \Omega}=\frac{\mu_0 \mu^4}{32 \pi^2 c}\langle |\mathsf{p}_{\mu}|^2 \rangle \sin ^2 \theta_r ,
	\label{dPdOmega}
\end{align} 
Here $\theta_r$ is the polar angle of the observation direction measured from the $z$ axis, assuming that the distance between the sample and the observer is much larger than the sample size. 
The constants $\mu_0$ and $c$ are the vacuum permeability and the speed of light, respectively.  $\langle \cdots \rangle$ denotes the ensemble average. $\langle |\mathsf{p}_{\mu}|^2 \rangle$ is given by 
\begin{align}
	\langle |\mathsf{p}_{\mu}|^2 \rangle = ( d c_0 D_2(\Delta) \frac{-1}{\mu})^2 \sum_{i,j} \langle e^{i(\phi_0(\bm{r}_i) - \phi_0(\bm{r}_j))} \rangle \equiv ( d c_0 D_2(\Delta) \frac{-1}{\mu})^2 \sum_{i,j} C(\bm{r}_i - \bm{r}_j). 
	\label{p_mu_sum}
\end{align}
Here $C(\bm{r}_i - \bm{r}_j) = \langle e^{i\phi_0(\bm{r}_i)} e^{-i\phi_0(\bm{r}_j)} \rangle$ is the correlation function of the $U(1)$ phase $\phi_0$ with $\langle \cdots \rangle$ being the ensemble average. In the $0_{\rm th}$ order of interlayer tunneling, the two-dimensional system has quasi-long-range order (QLRO) at low temperatures, with $C(\bm{r}_i,\bm{r}_j)=(a/|\bm{r}_i-\bm{r}_j|)^b$ and $b<1/4$, while it becomes disordered with $C(\bm{r}_i - \bm{r}_j) = \delta_{ij}$ at high temperatures as a consequence of the $U(1)$ symmetry and the Berezinskii-Kosterlitz-Thouless (BKT) transition~\cite{berezinskii1971destruction, Kosterlitz_1973}. For weak interlayer tunneling, the same QLRO and disordered correlation functions can be used to evaluate the radiation power to leading order in $c_0$ in the dynamical state.

Without loss of generality, one can consider a square lattice with a size $L \times L$ and lattice constant $a$. Therefore, one has $\sum_{i,j} = \sum_{\bm{r}_i - \bm{r}_j = \bm{U}L} (L/a)^2 A(\bm{U})$ with 
\begin{align}
	A(\bm{U}) =
	\begin{cases}
		 (1 - |U_x|)(1 - |U_y|), \ \ \ |U_{x}| < 1 \ \ \& \ \  |U_y| < 1, \\
		 0, \ \ \ \text{otherwise}. 
	\end{cases}
\end{align}
$L^2 A(\bm{U})$ denotes the area of the overlap between the sample and the sample shifted by a vector $\bm{U}L$. Dividing this area by $a^2$ gives the number of ordered lattice-site pairs with that displacement, $(L/a)^2 A(\bm{U})$. Therefore, 
\begin{align}
	\langle |\mathsf{p}_{\mu}|^2 \rangle = ( d c_0 D_2(\Delta) \frac{-1}{\mu})^2  (\frac{L}{a})^4 \int A(\bm{U}) C(\bm{U}L)  d^2 \bm{U} .
	\label{p_mu_integral}
\end{align}
The integral depends on the correlation function $C(\bm{U}L)$. We discuss relevant correlations. 

$\bullet$ The disordered case: $C(\bm{r}_i - \bm{r}_j) = \delta_{ij}$. In this case, the summation in Eq.~\eqref{p_mu_sum} gives $(\frac{L}{a})^2$. Therefore, the radiation power per solid angle reads
\begin{align}
	\frac{dP}{d\Omega} = \frac{\mu_0 \mu^2}{32\pi^2 c}  (\sin \theta_r  d c_0 D_2(\Delta))^2 (\frac{L}{a})^2  .	
\end{align}

$\bullet$ The quasi-long-range order case: $C(\bm{r}_i - \bm{r}_j) = (\frac{a}{|\bm{r}_i - \bm{r}_j|})^b$ with $b < \frac{1}{4}$. In this case, the integral in \equa{p_mu_integral} gives \begin{align}
	\int  (\frac{a}{L})^b \frac{1}{U^b} A(\bm{U}) d^2 \bm{U} = (\frac{a}{L})^b \int_0^1 dU_x \int_0^1 dU_y \ \   4\frac{(1-U_x)(1-U_y)}{U^b}.
\end{align}
For $b < \frac{1}{4}$, the integral over $U_x$ and $U_y$ is convergent and gives a constant independent of $L$. Therefore, the radiation power per solid angle reads
\begin{align}
    \frac{d P}{d \Omega} \sim \frac{\mu_0 \mu^2}{32\pi^2 c} (\sin \theta_r  d c_0 D_2(\Delta))^2 (\frac{L}{a})^{4-b} . 
	\label{Radiation_power}
 \end{align}
 
This result shows that, if the system undergoes a BKT transition from the QLRO phase to the disordered phase, the transition can be detected through the size dependence of the radiation power. Note that we have used natural units in the derivation. Under SI units, there is a prefactor of $e^2/\hbar^4$ in the result of \equa{Radiation_power}. Using $\mu/\hbar \sim 2\pi\times10^2\,{\rm GHz}$, $L\sim 10^2\,\mu{\rm m}$, $a\sim10\,{\rm nm}$, $d=5\,{\rm nm}$, and $c_0D_2(\Delta)\sim0.1\mu$, we estimate the radiation power $dP/d\Omega$ at $\sin\theta_r=1$ to be of order $10^{-18}\,{\rm W/sr}$ in the disordered phase and $10^{-10-4b}\,{\rm W/sr}$ in the QLRO phase.

\section{Unequal tunneling amplitudes: equations and numerical results}
\label{sec:rtg}
In this section, we test the robustness of dynamical tunneling blockade (DTB) for
$X=\operatorname{diag}(1,1-\zeta)$ and $0\leq\zeta \leq 1$, which cover general cases according to the analysis below \equa{eq:H_t}. We choose a gauge so that
$Z\equiv \langle 1|X|2\rangle$ is real and nonnegative. In this case, the energy in the rotating frame is still given by \equa{eq:Q4_explicit_energy} with a replacement of $\cos(\Theta/2)$ by $Z$. Therefore, the LLG equation for $\bm N$ has the same structure as
Eq.~\eqref{eq:Q4_explicit_LLG_N}, with $\cos(\Theta/2)$ replaced by $Z$. However, the dynamical equations \equa{eq:Q4_reduced_valley_scalars} do not apply, and we need to derive the valley-spinor equations for $|1\rangle$ and $|2\rangle$ to make numerical simulations. To make convenient numerical calculations, we start from \equa{eq:rtg_full_projected_spinor} in the numerics, in which $H_{\rm eff}$ has the following explicit form: 
\begin{align}
H_{\rm eff}
&=
\begin{pmatrix}
(h^z+\beta\Delta N^z)\,\tau_0 & -c_0 e^{i\mu t}X \\
-c_0 e^{-i\mu t}X^\dagger & -(h^z+\beta\Delta N^z)\,\tau_0
\end{pmatrix}.
\end{align}  
For the calculation below, we define the projectors in the valley space by $
 \bar P_1=1-|1\rangle\langle1|,
 \bar P_2=1-|2\rangle\langle2|. $ Substituting the wavefunction in \equa{macro_wf} into
\equa{eq:rtg_full_projected_spinor}, we apply
$\langle\uparrow|\otimes\bar P_1$ and
$\langle\downarrow|\otimes\bar P_2$ to its two sides to obtain 
\begin{equation}
\begin{aligned}
 \bar P_1|\dot 1\rangle
 &=-(i-\Gamma)^{-1}c_0\sqrt{\frac{1-N^z}{1+N^z}}e^{i\phi}\bar P_1X|2\rangle,\\
 \bar P_2|\dot 2\rangle
 &=-(i-\Gamma)^{-1}c_0\sqrt{\frac{1+N^z}{1-N^z}}e^{-i\phi}\bar P_2X|1\rangle.
\end{aligned}
\label{eq:rtg_projected_spinors}
\end{equation}

The projected equations determine the motion of $|1\rangle$ and $|2\rangle$. Meanwhile, the evolution of $Z$ is determined by 
\begin{align}
 \dot Z
 =\langle\dot1|X|2\rangle+\langle1|X|\dot2\rangle &= \langle\dot1|\bar P_1X|2\rangle
 +\langle1|X\bar P_2|\dot2\rangle
 +\bigl(\langle\dot1|1\rangle+\langle2|\dot2\rangle\bigr)Z \nonum
 &=\langle\dot1|\bar P_1X|2\rangle
 +\langle1|X\bar P_2|\dot2\rangle+i a_tZ.
 \label{eq:rtg_Z_at}
\end{align}
Here we introduce the gauge potential $a_t=i\langle 1|\dot 1\rangle-i\langle 2|\dot 2\rangle$, whose value is determined from the requirement that $Z$ remain real. Substituting Eq.~\eqref{eq:rtg_projected_spinors} into
Eq.~\eqref{eq:rtg_Z_at},
its real part gives $\dot Z$, while its imaginary part, set to zero,
determines the gauge choice of $a_t$ for $Z>0$. Thus an initially real $Z$ remains real as
long as the trajectory does not reach $Z=0$. In the equal-amplitude
limit $\zeta=0$, $Z=\cos(\Theta/2)$;
Eq.~\eqref{eq:rtg_Z_at} then reduces to
\equa{eq:Q4_reduced_valley_scalars}.

The dynamics of the layer-pseudospin vector $\bm N$ can be written as \equa{eq:Q4_explicit_layer_scalar_eom} with 
$\cos(\Theta/2)$ replaced by $Z$. Moreover, $Z$ and the gauge potential $a_t$ are determined by \equa{eq:rtg_Z_at}. Together with \equa{eq:rtg_projected_spinors}, these equations form a closed set for a numerical solution of the full dynamics.

To obtain the numerical plot with $\zeta \ne 0$ shown in the end matter of the main text, we numerically integrate the coupled spinor and layer equations. We start at zero bias from the aligned state $|1\rangle=|2\rangle=(1,0)^T$. In the upward bias sweep,
the final state at each bias serves as the initial state for the
next, with a small transverse spinor perturbation added at the
start of each bias interval to seed possible instabilities of the parallel-valley configuration in the low-bias static state. At each bias, the system is evolved until the monitored observables are stationary within the prescribed numerical
tolerance over a finite time window. 

\begin{figure}[htbp]
    \centering
    \includegraphics[width=0.32\linewidth]{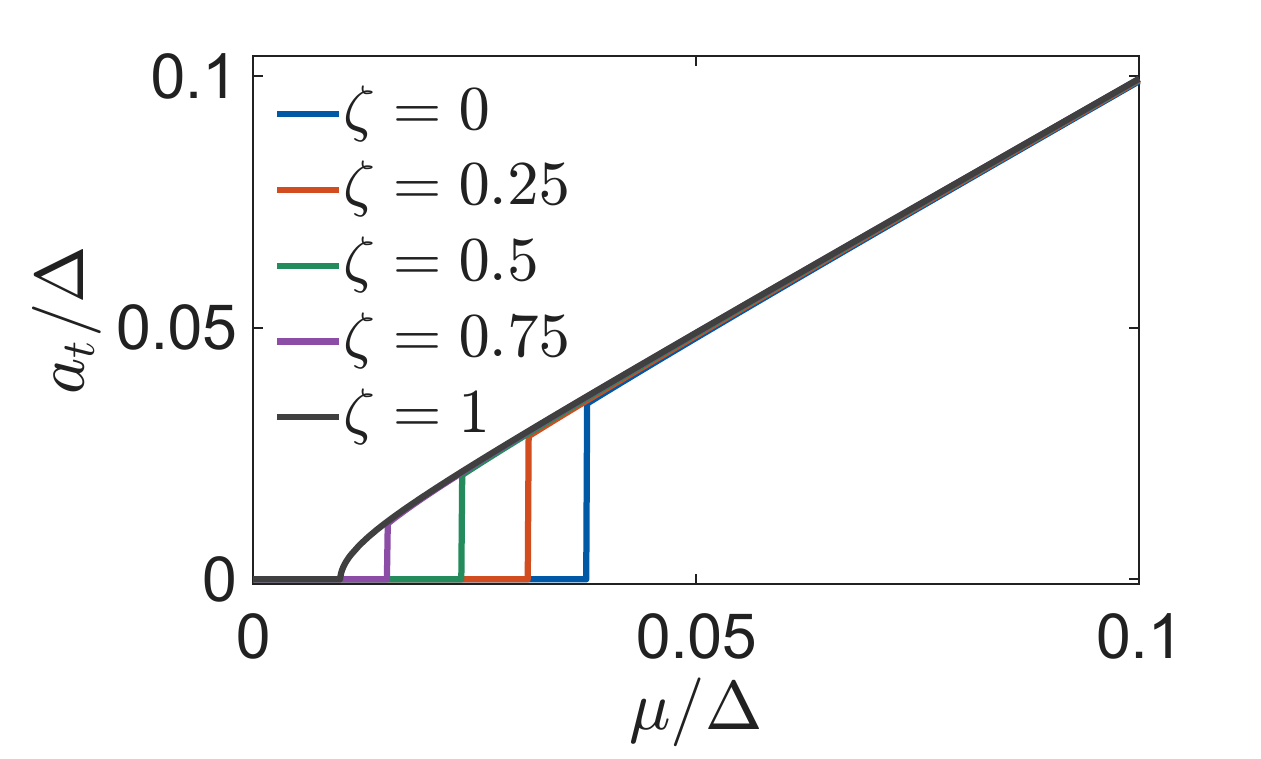}
    \caption{Effective gauge potential $a_t/\Delta$ versus bias
    $\mu/\Delta$ for $X=\operatorname{diag}(1,1-\zeta)$ and
    $\zeta=0,0.25,0.5,0.75,1$. The upward sweeps use the same converged
    states and color convention as Fig.~4 of the main-text end matter,
    with $c_0/\Delta=0.005$, $\beta=0.1$, and $\Gamma=0.2$. For all curves, the gauge potential $a_t$ becomes time independent after relaxation, within numerical tolerance.}
    \label{fig:rtg_gauge_potential}
\end{figure}

As a supplement to Fig.~4 in the main-text end matter, \fig{fig:rtg_gauge_potential} shows the effective gauge potential
$a_t$ in the real-$Z$ gauge,
evaluated from the imaginary part of Eq.~\eqref{eq:rtg_Z_at} at the
converged final spinor state of each bias interval. It is negligible on
the low-bias parallel branch and approaches the applied bias on the
high-bias DTB branch for all five values of $\zeta$. Thus the valley
dynamics compensates most of the applied bias, leaving a small effective
bias $\mu_{\rm eff}=\mu-a_t$, also for unequal tunneling amplitudes. This figure, together with Fig.~4 of the main-text end matter, shows that both $Z$ and $a_t$ change discontinuously at the onset of the DTB state for $\zeta=0,0.25,0.5,0.75$, but continuously for $\zeta=1$. 

\section{Dynamical tunneling blockade with valley anisotropy}
\label{sec:valley_anisotropy}
In this section, we investigate the effects of valley anisotropy by adding the following term to the energy entering the LLG equations
\begin{equation}
 E_v=\frac{K\Delta}{4}\left[(1+N^z)^2(v_1^z)^2+(1-N^z)^2(v_2^z)^2\right].
 \label{eq:valley_anisotropy_energy}
\end{equation}
Here $K$ is the dimensionless valley-anisotropy strength. The two identical layers share the same valley anisotropy. $K<0$ favors easy-axis valley polarization, whereas $K>0$ favors easy-plane valley order, corresponding to intervalley coherence~\cite{Jin2026}. To ensure a layer-balanced equilibrium state at $h^z=0$ in the weak-tunneling limit, we require $\beta+K>0$. The layer easy-plane anisotropy arises from the finite separation $d$ imposed by the insulating barrier, whereas the two valleys reside within the same layer. This distinction motivates us to consider the physically plausible regime $|K|<\beta$ in the following analysis.

In the absence of interlayer tunneling, the anisotropy term does not
lift the degeneracy between parallel and antiparallel valley
configurations. 
For valley-conserving tunneling, $X=\tau_0$, the
static tunneling energy favors parallel alignment at low bias. At high bias, the contribution to the time-averaged tunneling energy arising from the fast layer-pseudospin dynamics favors antiparallel valley alignment, as discussed in Eqs.~\eqref{P-force} and \eqref{P-force-expr} and argued in the main text. This mechanism remains operative for weak valley anisotropy. 
We therefore expect DTB to persist at least for sufficiently weak anisotropy, and test its robustness numerically below.

We first derive the dynamical equations in the presence of $E_v$.
As in Sec.~\ref{sec:rtg},
the normalized valley spinors are chosen so that
$Z=\langle1|X|2\rangle$ is real and nonnegative. The full projected spinor equation~\eqref{eq:rtg_full_projected_spinor}
remains applicable, but the effective Hamiltonian is replaced by
\begin{equation}
 H_{\rm eff}=
 \begin{pmatrix}
 (h^z+\beta\Delta N^z)\tau_0+K\Delta(1+N^z)v_1^z\tau_z
 &-c_0e^{i\mu t}X\\
 -c_0e^{-i\mu t}X^\dagger
 &-(h^z+\beta\Delta N^z)\tau_0+K\Delta(1-N^z)v_2^z\tau_z
 \end{pmatrix}.
 \label{eq:anisotropy_Heff}
\end{equation}
The projected
valley-spinor equations are modified to
\begin{equation}
 \begin{aligned}
 (i-\Gamma)\bar P_1|\dot1\rangle&=-c_0\sqrt{\frac{1-N^z}{1+N^z}}
 e^{i\phi}\bar P_1X|2\rangle
 +K\Delta(1+N^z)v_1^z\bar P_1\tau_z|1\rangle,\\
 (i-\Gamma)\bar P_2|\dot2\rangle&=-c_0\sqrt{\frac{1+N^z}{1-N^z}}
 e^{-i\phi}\bar P_2X|1\rangle
 +K\Delta(1-N^z)v_2^z\bar P_2\tau_z|2\rangle.
 \end{aligned}
 \label{eq:anisotropy_projected_spinors}
\end{equation}
Substituting Eq.~\eqref{eq:anisotropy_projected_spinors} into
Eq.~\eqref{eq:rtg_Z_at} gives $\dot Z$ from the real part.
Requiring the imaginary part to vanish fixes the gauge potential
$a_t$ for $Z>0$. The layer-pseudospin equations, Eqs.~\eqref{eq:Q4_explicit_LLG_N}
and \eqref{eq:Q4_explicit_layer_scalar_eom}, also acquire an additional
effective field because $E_v$ depends explicitly on $N^z$. We define
\begin{equation}
 h_v^z=\frac{\partial E_v}{\partial N^z}
 =\frac{K\Delta}{2}\left[(1+N^z)(v_1^z)^2-(1-N^z)(v_2^z)^2\right].
 \label{eq:anisotropy_layer_field}
\end{equation}
The modified equations retain the form of
Eqs.~\eqref{eq:Q4_explicit_LLG_N} and
\eqref{eq:Q4_explicit_layer_scalar_eom}, with
$\cos(\Theta/2)\to Z$ and $h^z\to h^z+h_v^z$.
We set $h^z=0$ throughout the calculations.
The current remains $I/I_{\rm J}=Z\sin\theta\sin\phi$.
We integrate Eq.~\eqref{eq:anisotropy_projected_spinors} together with
the modified layer-pseudospin equations using the upward-bias protocol
of Sec.~\ref{sec:rtg}, with $c_0/\Delta=0.005$, $\beta=0.1$,
$\Gamma=0.2$, and $\zeta=0$. The initial state has balanced layer
populations and valley spinors nearly parallel to the positive $z$ axis;
a transverse perturbation of $10^{-3}$ is added at each bias step.
We consider $K=-0.03,-0.01,0.01,0.03$.

\begin{figure}[htbp]
 \centering
 \begin{minipage}{0.32\linewidth}\centering
 \includegraphics[width=\linewidth]{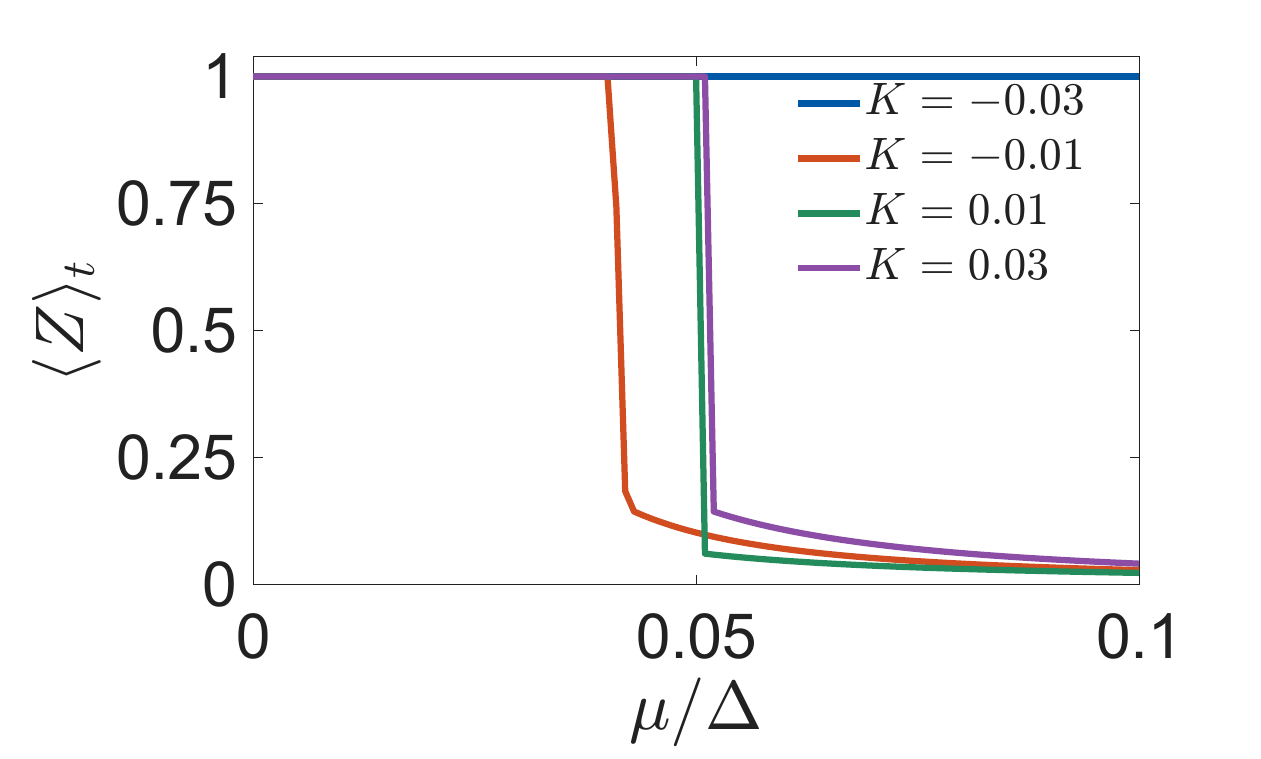}\\(a)
 \end{minipage}\hfill
 \begin{minipage}{0.32\linewidth}\centering
 \includegraphics[width=\linewidth]{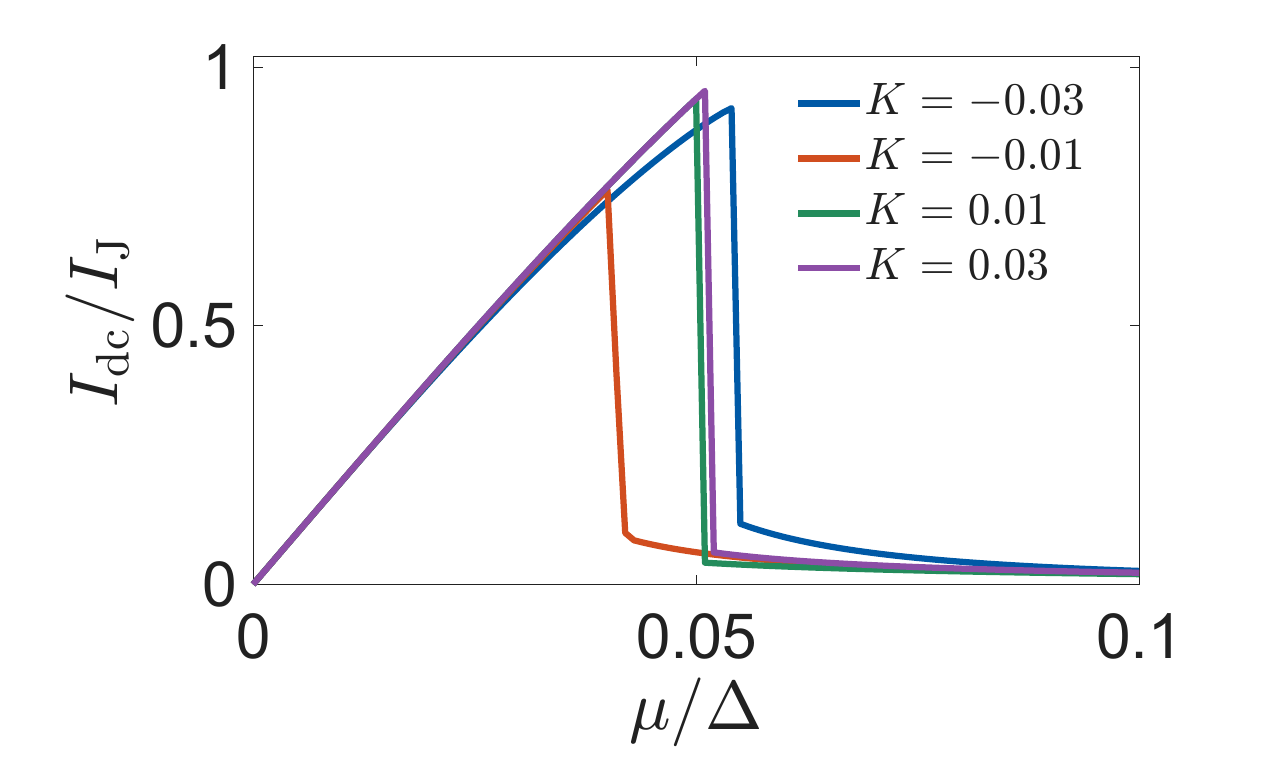}\\(b)
 \end{minipage}\hfill
 \begin{minipage}{0.32\linewidth}\centering
 \includegraphics[width=\linewidth]{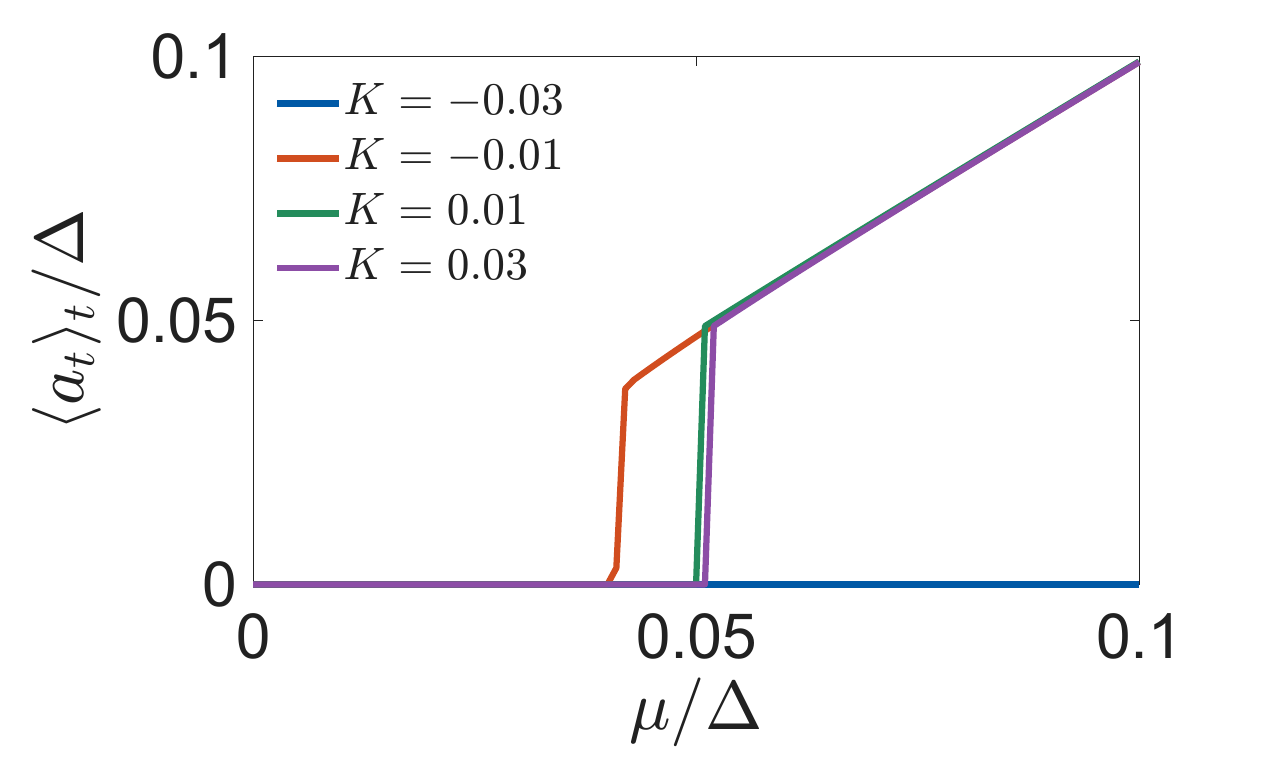}\\(c)
 \end{minipage}
 \caption{Upward bias sweeps with valley anisotropy.
 (a) the mean valley overlap, (b) the dc current, and (c) the mean
 gauge potential. The four curves have fixed
 $K=-0.03,-0.01,0.01,0.03$.
 The remaining parameters are $c_0/\Delta=0.005$, $\beta=0.1$,
 $\Gamma=0.2$, and $\zeta = 0$. Time averages are used for time-dependent states.}
 \label{fig:valley_anisotropy_sweep}
\end{figure}
\FloatBarrier

Figure~\ref{fig:valley_anisotropy_sweep} shows the time-averaged valley
overlap, dc current, and gauge potential. For
$K=-0.01,0.01$, and $0.03$, increasing the bias beyond a
critical value induces DTB: both $Z$ and $I_{\rm dc}$ drop, while
$\langle a_t\rangle_t$ approaches $\mu$. 

For $K=-0.03$,
the DTB state is not reached within the bias range shown.
The absence of DTB  in this sweep can be
understood in terms of an anisotropy barrier.
For $K<0$, the low-bias static state has parallel valley polarizations
along the easy axis, $\bm v_1=\bm v_2=\pm\hat{\bm z}$.
In the high bias regime, the oscillating tunneling term from the layer pseudospin precession is expected to generate a ponderomotive potential favoring
antiparallel valley alignment. However, once the easy-axis anisotropy exceeds a certain strength, 
there is a barrier for the valley pseudospins to leave the easy axis, and reaching this low energy configuration requires crossing the barrier. 
In our deterministic dynamics, the system remains trapped in the metastable minimum as the bias is swept up.
In practice, if the bias is swept up slowly enough, thermal or quantum noise would enable the escape from the metastable minimum to the global minimum of antiparallel valley alignment, so that DTB would occur. 
The exact time scale requires treatment of fluctuations, which we leave for further study.

\begin{appendices}
\renewcommand{\thesubsection}{\Roman{subsection}}

\section{Evaluation of the Green's-function integrals.}
In this section, we evaluate the Green's-function integrals in Eq.~\eqref{eq:projector_response_two_channels} and Eq.~\eqref{eq:amplitude_projector_bubble}. The Green's functions are
\begin{align}
E_+&=\nu-(Q-1)\Delta,\qquad E_-=\nu+\Delta,
\qquad E_--E_+=Q\Delta,\nonumber\\
\mathcal G_\pm^{R/A}(\nu,\epsilon)
&=\frac{1}{\epsilon-E_\pm\pm i\eta},\qquad
\mathcal G_\pm^K(\nu,\epsilon)
=f(\epsilon)\left[\mathcal G_\pm^R-\mathcal G_\pm^A\right]
=\frac{-2i\eta f(\epsilon)}{(\epsilon-E_\pm)^2+\eta^2}.
\label{eq:newapp_green_functions}
\end{align}
Here $f(\epsilon)=\tanh(\epsilon/2T)$. 

\subsection{Integrals in Eq.~\eqref{eq:projector_response_two_channels}}
\label{app:green_integrals_inter}
Denote the first channel in
Eq.~\eqref{eq:projector_response_two_channels} by
\begin{align}
\mathcal B_1(\omega)
&\equiv iN(Q\Delta)^2
\int\frac{d\nu}{2\pi}\rho(\nu)
\int\frac{d\epsilon}{2\pi}
\left[
\mathcal G_-^R(\epsilon+\omega)\mathcal G_+^K(\epsilon)
+\mathcal G_-^K(\epsilon+\omega)\mathcal G_+^A(\epsilon)
\right].
\label{eq:newapp_interband_channel}
\end{align}
Taking a shift of $\epsilon + \omega \to \epsilon$ in the second term and substituting Eq.~\eqref{eq:newapp_green_functions} give
\begin{align}
\frac{\mathcal B_1(\omega)}{N(Q\Delta)^2}
={}&\int\frac{d\nu}{2\pi}\rho(\nu)
\int\frac{d\epsilon}{2\pi}
\Bigg[
\frac{2\eta f(\epsilon)}{(\epsilon-E_+)^2+\eta^2}
\frac{1}{\epsilon+\omega-E_-+i\eta} 
+\frac{2\eta f(\epsilon)}{(\epsilon-E_-)^2+\eta^2}
\frac{1}{\epsilon-\omega-E_+-i\eta}
\Bigg].
\label{eq:newapp_interband_expanded}
\end{align}
The second channel, denoted by $\mathcal B_2(\omega)$, is obtained by
interchanging the band labels and satisfies
$\mathcal B_2(\omega)=\mathcal B_1(-\omega)^*$.

At $\omega=0$, using the broadened occupation $f_\eta$ defined before
Eq.~\eqref{eq:gap}, we therefore obtain
\begin{align}
\mathcal B_1(0)=\mathcal B_2(0)
&=iNQ\Delta\int\frac{d\nu}{2\pi}\rho(\nu)
\int\frac{d\epsilon}{2\pi}f(\epsilon)
\left[(\mathcal G_-^R-\mathcal G_-^A)
-(\mathcal G_+^R-\mathcal G_+^A)\right]
\nonumber\\
&=NQ\Delta\int\frac{d\nu}{2\pi}\rho(\nu)
\left[f_\eta(E_-)-f_\eta(E_+)\right]
=\frac{NQ\Delta^2}{g}.
\label{eq:newapp_interband_static}
\end{align}
The equality involving $g$ uses the gap equation
\eqref{eq:gap}.

For the term linear in frequency, shift the integration variable in the second
term of Eq.~\eqref{eq:newapp_interband_expanded} so that both occupation factors
are $f(\epsilon)$. Differentiating the resulting denominators at $\omega=0$ gives
\begin{align}
\left.\frac{\partial\mathcal B_1(\omega)}{\partial\omega}\right|_{\omega=0}
={}&N(Q\Delta)^2\int\frac{d\nu}{2\pi}\rho(\nu)
\int\frac{d\epsilon}{2\pi}\,f(\epsilon)  \nonum
&\Bigg[
-\frac{2\eta}{(\epsilon-E_+)^2+\eta^2}
\frac{1}{(\epsilon-E_-+i\eta)^2}
+\frac{2\eta}{(\epsilon-E_-)^2+\eta^2}
\frac{1}{(\epsilon-E_+-i\eta)^2}
\Bigg] \nonum
&\equiv \mathcal{S} + i \mathcal{S}\Gamma.
\label{eq:newapp_interband_frequency_derivative}
\end{align}
The real and imaginary parts read
\begin{align}
\mathcal S
&=4N(Q\Delta)^3\eta^3
\int\frac{d\nu}{2\pi}\rho(\nu)
\int\frac{d\epsilon}{2\pi}
\frac{f(\epsilon)(2\epsilon-E_+-E_-)}
{\big[(\epsilon-E_+)^2+\eta^2\big]^2
 \big[(\epsilon-E_-)^2+\eta^2\big]^2},
\nonumber\\
\mathcal S\Gamma
&=2N(Q\Delta)^2\eta^2
\int\frac{d\nu}{2\pi}\rho(\nu)
\int\frac{d\epsilon}{2\pi}
\frac{f'(\epsilon)}
{\big[(\epsilon-E_+)^2+\eta^2\big]
 \big[(\epsilon-E_-)^2+\eta^2\big]}.
\label{eq:newapp_interband_real_imaginary_slopes}
\end{align}
The second expression follows by integration by parts in $\epsilon$.
For $\eta\ll\Delta$, the leading contribution to $\mathcal S$ comes from
$\epsilon\simeq E_\pm$. Since $E_--E_+=Q\Delta$, the numerator
$2\epsilon-E_+-E_-$ is approximately $\mp Q\Delta$ near $E_\pm$, while
the squared denominator associated with the other peak is approximately
$(Q\Delta)^4$. Thus, we obtain the expression of $\mathcal{S}$ in the equilibrium limit $\eta/\Delta \to 0$:
\begin{align}
\frac{\mathcal S}{N}
&\simeq\int\frac{d\nu}{2\pi}\rho(\nu)
\int\frac{d\epsilon}{2\pi}\,f(\epsilon)
\left[
\frac{4\eta^3}{\big[(\epsilon-E_-)^2+\eta^2\big]^2}
-\frac{4\eta^3}{\big[(\epsilon-E_+)^2+\eta^2\big]^2}
\right]
\nonum
&\simeq\int\frac{d\nu}{2\pi}\rho(\nu)
\big[f_{\eta}(E_-)-f_{\eta}(E_+)\big]
=2D_Q(\Delta).
\end{align}
The second step uses the normalized squared Lorentzian,
$\int d\epsilon\,4\eta^3/\{2\pi[(\epsilon-E)^2+\eta^2]^2\}=1$. Here $D_Q(\Delta)$ is defined in Eq.~\eqref{eq:gap}.
Consequently, $\mathcal S\simeq 2ND_Q(\Delta)$ in the leading order of $\eta/\Delta$, with the bandwidth and
temperature dependence retained in $D_Q$. The dissipative coefficient \(\mathcal S\Gamma\) vanishes as \(\eta/\Delta \to0\). To retain a nonzero dissipative response, we therefore keep \(\eta\) finite and use the full integral in \equa{eq:newapp_interband_real_imaginary_slopes}.

\subsection{Integrals in Eq.~\eqref{eq:amplitude_projector_bubble}}
\label{app:green_integrals_intra}
For either band $s=+,-$, define the response per site appearing in
Eq.~\eqref{eq:amplitude_projector_bubble} as
\begin{align}
\mathcal I_s(\omega)
&\equiv i\int\frac{d\nu}{2\pi}\rho(\nu)
\int\frac{d\epsilon}{2\pi}
\left[
\mathcal G_s^R(\epsilon+\omega)\mathcal G_s^K(\epsilon)
+\mathcal G_s^K(\epsilon+\omega)\mathcal G_s^A(\epsilon)
\right].
\label{eq:newapp_intraband_definition}
\end{align}
Substituting the explicit Green's functions and shifting the integration
variable in the second term gives
\begin{align}
\mathcal I_s(\omega)
&=\int\frac{d\nu}{2\pi}\rho(\nu)
\int\frac{d\epsilon}{2\pi}
\frac{2\eta f(\epsilon)}{(\epsilon-E_s)^2+\eta^2}
\left[\frac{1}{\epsilon-E_s+\omega+i\eta}
+\frac{1}{\epsilon-E_s-\omega-i\eta}\right].
\label{eq:newapp_intraband_expanded}
\end{align}
This expression is exact for the Green's functions used here.

At zero frequency, integration by parts gives
\begin{align}
\mathcal I_s(0)
&=\int\frac{d\nu}{2\pi}\rho(\nu)
\int\frac{d\epsilon}{2\pi}
\frac{4\eta(\epsilon-E_s)f(\epsilon)}{[(\epsilon-E_s)^2+\eta^2]^2}
=\int\frac{d\nu}{2\pi}\rho(\nu)
\int\frac{d\epsilon}{2\pi}
\frac{2\eta f'(\epsilon)}{(\epsilon-E_s)^2+\eta^2}
=\int\frac{d\nu}{2\pi}\rho(\nu)f_\eta'(E_s).
\label{eq:newapp_intraband_static}
\end{align}
Here $f_\eta$ is defined before Eq.~\eqref{eq:gap}. According to the definition $\Pi^R_{\Delta,Q}(0) \equiv [(Q-1)\mathcal I_+(0) + \mathcal I_-(0)]/2$, one obtains \equa{Pi_Delta_0}.

Differentiating Eq.~\eqref{eq:newapp_intraband_expanded} with respect to
$\omega$ at zero frequency gives
\begin{align}
\left.\partial_\omega\mathcal I_s(\omega)\right|_{\omega=0}
&=i \frac{1}{\eta} 2\int\frac{d\nu}{2\pi}\rho(\nu)
\int\frac{d\epsilon}{2\pi}
\frac{\eta^3 f'(\epsilon)}{[(\epsilon-E_s)^2+\eta^2]^2}.
\label{eq:newapp_intraband_frequency_derivative}
\end{align}
Therefore, one can define a coefficient 
\begin{align}
\alpha_Q(\Delta) &\equiv (Q-1)\int\frac{d\nu}{2\pi}\rho(\nu)
\int\frac{d\epsilon}{2\pi}
\frac{2\eta^3 f'(\epsilon)}{[(\epsilon-E_+)^2+\eta^2]^2} + \int\frac{d\nu}{2\pi}\rho(\nu)
\int\frac{d\epsilon}{2\pi}
\frac{2\eta^3 f'(\epsilon)}{[(\epsilon-E_-)^2+\eta^2]^2}
\label{alphaQ_def}
\end{align}
to represent the amplitude response function in the low-energy regime by  Eq.~\eqref{eq:amplitude_projector_bubble_low_freq}.

Comparing the integrands in the definition of $\Pi_{\Delta,Q}^R(0)$ and $\alpha_Q(\Delta)$, one obtains
\begin{align}
0\leq\alpha_Q(\Delta) < 2\Pi_{\Delta,Q}^R(0).
\label{alpha_leq_Pi}
\end{align}

\section{Derivation of the interlayer current.}
\label{Appe:current}
To derive the interlayer current, one needs to consider a $z$-direction vector potential $A^z$ between the double layers, and the tunneling Hamiltonian acquires an extra term to the first order in $dA^z$ 
\begin{align}
		H_{\rm t}^{'} &=  -c_0 \sum_i  \ \ ( idA^z  a^{\dagger}_{\uparrow  \alpha}(\bm{r}_i) X^{\alpha \beta} a_{\downarrow \beta}(\bm{r}_i) + cc. )  \nonumber \\
		&= -c_0 \sum_i \ \ ( idA^z e^{i\mu t} \Psi^{\dagger}_{\uparrow \alpha}(\bm{r}_i) X^{\alpha \beta} \Psi_{\downarrow \beta}(\bm{r}_i) + cc. ) 
\end{align}
For $Q=4$, $X$ is the interlayer tunneling matrix acting in valley space, and the blue tensor-product factor $\blue{\otimes X}$ is included in the following expressions. For $Q=2$, this factor and the valley indices are absent.
Note that the classical component of the current is coupled to the quantum component of the vector potential. Thus, we set $(dA_{z,+} - dA_{z,-})/2 \equiv dA_q$ with the subscript $+/-$ being the variable defined in the forward/backward time contour. 

The Lagrangian acquires the following term to the first order in  $dA^z$, 
\begin{align}
	iS^{\prime}_{\rm t} \equiv \bar{\Psi} T' \Psi
	&= c_0
	i\sum_i \int dt  \begin{bmatrix}
		\bar{\Psi}_+(\bm{r}_i,t) & \bar{\Psi}_-(\bm{r}_i,t) 
	\end{bmatrix}  dA_q(t)
	\begin{bmatrix}
		\sigma(t) \blue{\otimes X} & 0 \\
		0 & \sigma(t) \blue{\otimes X}
	\end{bmatrix}
	\begin{bmatrix}
		\Psi_+(\bm{r}_i,t) \\
		\Psi_-(\bm{r}_i,t)
	\end{bmatrix} \nonumber \\
	&=  c_0
	i\sum_i \int dt  \begin{bmatrix}
		\bar{\Psi}_c(\bm{r}_i,t) & \bar{\Psi}_q(\bm{r}_i,t) 
	\end{bmatrix}  dA_q(t)
	\begin{bmatrix}
		0 & \sigma(t) \blue{\otimes X} \\
		\sigma(t) \blue{\otimes X} & 0
	\end{bmatrix}
	\begin{bmatrix}
		\Psi_c(\bm{r}_i,t) \\
		\Psi_q(\bm{r}_i,t)
	\end{bmatrix}
\end{align}
Here $\sigma(t) = (\cos(-\mu t + \pi/2), \sin(-\mu t + \pi/2), 0) \cdot \bm{\sigma}$. After integrating out the fermions, we get the following expansion of the action to the order of $\mathcal{O}(dA_q)$
\begin{align}
	iS^{\prime}_{\rm t} = {\rm Tr}[\mathcal{G} T'] = c_0 \sum_{\bm{k}} \int dt \int dt^{\prime} \ \ {\rm tr}[\mathcal{G}^K(\bm{k},t^{\prime} - t) (\sigma(t)\blue{\otimes X})] dA_q(t) \delta(t-t^{\prime})  
\end{align}
The interlayer tunneling current gives
\begin{align*}
I(t)
&\equiv
\frac{1}{a^2N}
\frac{\delta S^{\prime}_{\rm t}}{2\delta dA_q(t)}
=
-\frac{ic_0}{2a^2N}
\sum_{\bm{k}}\int\frac{d\omega}{2\pi}\,
\operatorname{Tr}\!\left[
\mathcal G^K(\bm{k},\omega)(\sigma(t)\blue{\otimes X})
\right].
\end{align*}
Using the spectral decomposition in Eq.~\eqref{G_plus_minus} and the
tracelessness of $(\sigma(t)\blue{\otimes X})$, we obtain
\begin{align*}
\operatorname{Tr}\!\left(\mathcal G^K(\sigma(t)\blue{\otimes X})\right)
&=
\mathcal G_+^K\operatorname{Tr}(P(\sigma(t)\blue{\otimes X}))
+\mathcal G_-^K\operatorname{Tr}[(1-P)(\sigma(t)\blue{\otimes X})]
=
\left(\mathcal G_+^K-\mathcal G_-^K\right)
\operatorname{Tr}(P(\sigma(t)\blue{\otimes X})).
\end{align*}
Since $P=\ket{\Psi}\bra{\Psi}$, one has $
\operatorname{Tr}(P(\sigma(t)\blue{\otimes X}))
=
\braket{\Psi|(\sigma(t)\blue{\otimes X})|\Psi}$. Restoring $e$ and $\hbar$ and combining the gap equation \equa{eq:gap} and the wavefunction \equa{macro_wf}, the current density is calculated as
\begin{align}
I(t)
&=
I_{\rm J}
\braket{\Psi|(\sigma(t)\blue{\otimes X})|\Psi}
=
I_{\rm J}
\begin{cases}
\sin\theta\sin\phi,
&Q=2,\\[3pt]
Z\sin\theta\sin\phi,
&Q=4.
\end{cases}
\label{eq:jt}
\end{align}
Here $I_{\rm J}\equiv \frac{c_0 D_Q(\Delta)e}{\hbar a^2}$ is the ``Josephson critical current density'' and $Z=\langle1|X|2\rangle$ is chosen real and nonnegative, as in Sec.~\ref{sec:rtg}. $N$ is the number of sites and $a$ is the lattice constant or the
magnetic length in quantum Hall systems. 
\end{appendices}	

\bibliography{supp}